\documentclass[preprint]{ptephy}

\preprintnumber{%
J-PARC-TH-0138,
KYUSHU-HET-190,
RIKEN-QHP-385
}

\usepackage{amssymb}
\usepackage{amsthm}
\usepackage{amsmath}
\usepackage{booktabs}
\usepackage{bbm}
\usepackage{mathtools}
\usepackage{subcaption}
\usepackage{braket}
\usepackage{here}

\numberwithin{equation}{section}

\usepackage{url}

\begin{document}

\title{Thermodynamics in quenched QCD: energy--momentum tensor with two-loop
order coefficients in the gradient flow formalism}

\author{%
\name{\fname{Takumi} \surname{Iritani}}{1},
\name{\fname{Masakiyo} \surname{Kitazawa}}{2,3},
\name{\fname{Hiroshi} \surname{Suzuki}}{4,\ast},
\name{\fname{Hiromasa} \surname{Takaura}}{4}
}

\address{%
\affil{1}{Theoretical Research Division, Nishina Center, RIKEN, Wako 351-0198,
Japan}
\affil{2}{Department of Physics, Osaka University, Toyonaka, Osaka 560-0043,
Japan}
\affil{3}{J-PARC Branch, KEK Theory Center, Institute of Particle and Nuclear
Studies, KEK, 203-1, Shirakata, Tokai, Ibaraki, 319-1106, Japan}
\affil{4}{Department of Physics, Kyushu University 744 Motooka, Nishi-ku,
Fukuoka, 819-0395, Japan}
\email{hsuzuki@phys.kyushu-u.ac.jp}
}

\date{\today}

\begin{abstract}
Recently, Harlander et al.\ [Eur.\ Phys.\ J.\ C {\bf 78}, 944 (2018)] have
computed the two-loop order (i.e., NNLO) coefficients in the gradient-flow
representation of the energy--momentum tensor (EMT) in vector-like gauge
theories. In this paper, we study the effect of the two-loop order corrections
(and the three-loop order correction for the trace part of the EMT, which is
available through the trace anomaly) on the lattice computation of
thermodynamic quantities in quenched QCD. The use of the two-loop order
coefficients generally reduces the $t$~dependence of the expectation values of
the EMT in the gradient-flow representation, where $t$~is the flow time. With
the use of the two-loop order coefficients, therefore, the $t\to0$
extrapolation becomes less sensitive to the fit function, the fit range, and
the choice of the renormalization scale; the systematic error associated with
these factors is considerably reduced.
\end{abstract}

\subjectindex{B01, B31, B32, B38}
\maketitle

\section{Introduction}
\label{sec:1}
The energy--momentum tensor (EMT)~$T_{\mu\nu}(x)$ is a fundamental physical
observable in quantum field theory. It has been pointed out
in~Refs.~\cite{Suzuki:2013gza,Makino:2014taa} that a ``universal''
representation of the EMT can be written down by utilizing the so-called
gradient flow~\cite{Narayanan:2006rf,Luscher:2009eq,Luscher:2010iy,%
Luscher:2011bx,Luscher:2013cpa} and its small flow-time
expansion~\cite{Luscher:2011bx}. This representation of the EMT is universal in
the sense that it is independent of the adopted regularization. The
representation can thus be employed in particular with the lattice
regularization that makes nonperturbative computations possible. An advantage
of this approach to the lattice EMT is that the expression of the EMT is known
a priori and it is not necessary to compute the renormalization constants
involved in the lattice EMT~\cite{Caracciolo:1989pt}.\footnote{See Ref.~\cite{Suzuki:2016ytc}
and references cited therein. In particular,
in~Refs.~\cite{DelDebbio:2013zaa,Capponi:2015ucc}, the gradient flow is applied
to the construction of the EMT in the conventional
approach~\cite{Caracciolo:1989pt}.} This approach instead requires the limit
$t\to0$, where $t$ is the flow time (see below), because the representation is
obtained in the small flow-time limit. In actual lattice simulations, however,
since $t$ is limited as~$t\gtrsim a^2$ by the lattice spacing~$a$, the $t\to0$
limit has to be obtained by the extrapolation from the range of~$t$
satisfying~$t\gtrsim a^2$; this $t\to0$ extrapolation can be a source of
systematic error. By employing this gradient-flow representation of the EMT,
expectation values and correlation functions of the EMT have been computed to
study various physical questions~\cite{Asakawa:2013laa,Taniguchi:2016ofw,%
Kitazawa:2016dsl,Ejiri:2017wgd,Kitazawa:2017qab,Kanaya:2017cpp,%
Taniguchi:2017ibr,Yanagihara:2018qqg,Hirakida:2018uoy,Shirogane:2018zbp}.

One important application of the lattice EMT is the thermodynamics of gauge
theory at finite temperature; see
Refs.~\cite{Boyd:1996bx,Okamoto:1999hi,Borsanyi:2012ve,Borsanyi:2013bia,%
Bazavov:2014pvz} and more recent works
in~Refs.~\cite{Shirogane:2016zbf,Giusti:2014ila,Giusti:2015daa,Giusti:2016iqr,%
DallaBrida:2017sxr,Caselle:2018kap} on this problem. Two independent
thermodynamic quantities, such as the energy density~$\varepsilon$ and the
pressure~$p$, can be computed from the finite-temperature expectation value of
the traceless part and the trace part of the EMT, respectively, as\footnote{For
simplicity of expression, here and in what follows we omit the subtraction of
the vacuum expectation value of an expression; this is always assumed.}
\begin{align}
   \varepsilon+p
   &=-\frac{4}{3}\left\langle T_{00}(x)-\frac{1}{4}T_{\mu\mu}(x)\right\rangle,
\label{eq:(1.1)}
\\
   \varepsilon-3p
   &=-\left\langle T_{\mu\mu}(x)\right\rangle.
\label{eq:(1.2)}
\end{align}
In the gradient-flow approach, moreover, the computation of
isotropic/anisotropic Karsch coefficients (i.e., the lattice $\beta$ function)
is not necessary~\cite{Engels:1999tk}, because the expression of the EMT is a
priori known.

In this paper, we investigate the thermodynamics in quenched QCD (quantum
chromodynamics), i.e., the pure Yang--Mills theory, in the gradient-flow
approach. The EMT in the gradient-flow representation is obtained as follows.
Assuming dimensional regularization, the EMT in the pure Yang--Mills theory is
given by
\begin{equation}
   T_{\mu\nu}(x)
   =\frac{1}{g_0^2}\left[
   F_{\mu\rho}^a(x)F_{\nu\rho}^a(x)
   -\frac{1}{4}\delta_{\mu\nu}F_{\rho\sigma}^a(x)F_{\rho\sigma}^a(x)
   \right],
\label{eq:(1.3)}
\end{equation}
where $g_0$ is the bare gauge coupling and
$F_{\mu\nu}^a(x)\equiv%
\partial_\mu A_\nu^a(x)-\partial_\nu A_\mu^a(x)+f^{abc}A_\mu^b(x)A_\nu^c(x)$ is
the field strength.\footnote{$f^{abc}$ denote the structure constants of the
gauge group~$G$.} Note that this is an expression in
$D\equiv4-2\epsilon$-dimensional spacetime and is not generally traceless.

One can express any composite operator in gauge theory such as the
EMT~\eqref{eq:(1.3)} as a series of flowed composite operators through the
small flow-time expansion~\cite{Luscher:2011bx}. That is, one can
write\footnote{Note that our convention for~$c_2(t)$ differs from that
of~Ref.~\cite{Harlander:2018zpi}. Our $c_2(t)$ corresponds
to~$c_2(t)+(1/4)c_1(t)$ in~Ref.~\cite{Harlander:2018zpi}.}
\begin{align}
   T_{\mu\nu}(x)
   &=c_1(t)\left[
   G_{\mu\rho}^a(t,x)G_{\nu\rho}^a(t,x)
   -\frac{1}{4}\delta_{\mu\nu}G_{\rho\sigma}^a(t,x)G_{\rho\sigma}^a(t,x)
   \right]
\notag\\
   &\qquad{}
   +c_2(t)\delta_{\mu\nu}G_{\rho\sigma}^a(t,x)G_{\rho\sigma}^a(t,x)+O(t),
\label{eq:(1.4)}
\end{align}
where $G_{\mu\nu}^a(t,x)\equiv%
\partial_\mu B_\nu^a(t,x)-\partial_\nu B_\mu^a(t,x)+%
f^{abc}B_\mu^b(t,x)B_\nu^c(t,x)$
and~$D_\nu G_{\nu\mu}^a(t,x)\equiv%
\partial_\nu G_{\nu\mu}^a(t,x)+f^{abc}B_\nu^b(t,x)G_{\nu\mu}^c(t,x)$. In these
expressions, the ``flowed'' gauge field $B_\mu^a(t,x)$ is defined by the
gradient flow~\cite{Narayanan:2006rf,Luscher:2009eq,Luscher:2010iy}, i.e.,
a one-parameter evolution of the gauge field by
\begin{equation}
   \partial_tB_\mu^a(t,x)=D_\nu G_{\nu\mu}^a(t,x),
   \qquad
   B_\mu^a(t=0,x)=A_\mu^a(x).
\label{eq:(1.5)}
\end{equation}
The parameter~$t\geq0$, which possesses the mass dimension~$-2$, is termed the
flow time. Since Eq.~\eqref{eq:(1.4)} is finite~\cite{Luscher:2011bx}, one can
set $D=4$ and the first term on the right-hand side (that is proportional
to~$c_1(t)$) is traceless. The coefficients in this small flow-time expansion,
which are analogous to the Wilson coefficients in OPE, can be calculated by
perturbation theory~\cite{Luscher:2011bx} as
\begin{equation}
   c_1(t)=\frac{1}{g^2}\sum_{\ell=0}^\infty k_1^{(\ell)}
   \left[\frac{g^2}{(4\pi)^2}\right]^\ell,
   \qquad
   c_2(t)=\frac{1}{g^2}\sum_{\ell=1}^\infty k_2^{(\ell)}
   \left[\frac{g^2}{(4\pi)^2}\right]^\ell,
\label{eq:(1.6)}
\end{equation}
where $g$~denotes the renormalized gauge coupling. Throughout this paper, we
assume the $\overline{\text{MS}}$ scheme, in which
\begin{equation}
   g_0^2=\left(\frac{\mu^2e^{\gamma_{\text{E}}}}{4\pi}\right)^\epsilon g^2Z_g.
\label{eq:(1.7)}
\end{equation}
Here, $\mu$ is the renormalization scale, $\gamma_{\text{E}}$ is the Euler
constant, and $Z_g$ is the renormalization factor. In Eq.~\eqref{eq:(1.6)},
\begin{equation}
   k_1^{(0)}=1,
\label{eq:(1.8)}
\end{equation}
because in the tree-level (i.e., LO) approximation
$F_{\mu\rho}^a(x)F_{\nu\rho}^a(x)=G_{\mu\rho}^a(t,x)G_{\nu\rho}^a(t,x)+O(t)$. On the
other hand, there is no ``$k_2^{(0)}$'' in~Eq.~\eqref{eq:(1.6)} because the EMT
is traceless in the tree-level approximation (the trace anomaly emerges from
the one-loop order).

In Eq.~\eqref{eq:(1.6)}, the one-loop order (i.e., NLO) coefficients
$k_i^{(1)}(t)$ ($i=1$, $2$) were computed
in~Refs.~\cite{Suzuki:2013gza,Makino:2014taa} (see
also~Ref.~\cite{Suzuki:2015bqa}). Recently, in~Ref.~\cite{Harlander:2018zpi},
Harlander et al.\ have computed the two-loop order (i.e., NNLO) coefficients
$k_i^{(2)}(t)$ for general vector-like gauge theories; see also
Ref.~\cite{Harlander:2016vzb}. The purpose of the present paper is to study
the effect of the two-loop corrections given in~Ref.~\cite{Harlander:2018zpi}
by performing the lattice computation of thermodynamic quantities in quenched
QCD. For the trace part of the EMT, we also examine the use of the three-loop
order coefficient, $k_2^{(3)}$, which is presented in this paper; this
higher-order coefficient can be obtained for quenched QCD by combining a
two-loop result in~Ref.~\cite{Harlander:2018zpi} and the trace
anomaly~\cite{Adler:1976zt,Nielsen:1977sy,Collins:1976yq}, as we will explain
below. From analyses using lattice data obtained
in~Ref.~\cite{Kitazawa:2016dsl}, we find that the use of the two-loop order
coefficients generally reduces the $t$~dependence of the expectation values of
the EMT in the gradient-flow representation. With the use of the two-loop order
coefficients, therefore, the $t\to0$ extrapolation becomes less sensitive to
the fit function, the fit range, and the choice of the renormalization scale;
the systematic error associated with these factors is considerably reduced. We
expect that this improvement brought about by the two-loop order coefficients
also persists in wider applications of the gradient-flow representation of the
EMT, such as the thermodynamics of full QCD.

This paper is organized as follows. In Sect.~\ref{sec:2}, we explain our
treatment of perturbative coefficients $c_1(t)$ and~$c_2(t)$
of~Eq.~\eqref{eq:(1.6)}. We list the known expansion coefficients
$k_i^{(\ell)}$, and present the three-loop coefficient for $c_2(t)$, $k_2^{(3)}$.
In Sect.~\ref{sec:3}, we perform numerical analyses of the thermodynamic
quantities, which are mainly based on FlowQCD 2016~\cite{Kitazawa:2016dsl}. We
give conclusions in~Sect.~\ref{sec:4}.

\section{Expansion coefficients}
\label{sec:2}
\subsection{$\beta$~function and the running gauge coupling constant}
The $\beta$ function corresponding to~Eq.~\eqref{eq:(1.7)} is given by
\begin{equation}
   \beta(g)\equiv\left.\mu\frac{\partial}{\partial\mu}g\right|_{g_0}
   \stackrel{\epsilon\to0}{\to}
   -g\sum_{\ell=1}^\infty\beta_{\ell-1}
   \left[\frac{g^2}{(4\pi)^2}\right]^\ell,
\label{eq:(2.1)}
\end{equation}
with
coefficients~\cite{Gross:1973id,Politzer:1973fx,Caswell:1974gg,Jones:1974mm,%
Tarasov:1980au,Larin:1993tp,vanRitbergen:1997va}
\begin{align}
   \beta_0&=\frac{11}{3}C_A,
\label{eq:(2.2)}
\\
   \beta_1&=\frac{34}{3}C_A^2,
\label{eq:(2.3)}
\\
   \beta_2&=\frac{2857}{54}C_A^3,
\label{eq:(2.4)}
\\
   \beta_3
   &=\left[\frac{150\,473}{486}+\frac{44}{9}\zeta(3)\right]C_A^4
   +\left[-\frac{40}{3}+352\zeta(3)\right]C_A^2,
\label{eq:(2.5)}
\end{align}
where $C_A$ is the quadratic Casimir for the adjoint representation defined by
\begin{equation}
   f^{acd}f^{bcd}=C_A\delta^{ab}.
\label{eq:(2.6)}
\end{equation}
$C_A=N$ for the gauge group~$G=SU(N)$.

To compute expectation values of the EMT by employing the
representation~\eqref{eq:(1.4)}, we take the
limit~$t\to0$~\cite{Suzuki:2013gza,Makino:2014taa}. First of all this limit
removes the last~$O(t)$ term in~Eq.~\eqref{eq:(1.4)}, the contribution of
operators of higher ($\geq6$) mass dimensions. It also justifies finite-order
truncation of perturbative expansions of the coefficients $c_i(t)$; we treat
$c_i(t)$ in~Eq.~\eqref{eq:(1.4)} as follows. We apply the renormalization group
improvement \cite{Suzuki:2013gza,Makino:2014taa}, i.e., we set
$\mu\propto1/\sqrt{t}$ in~$k_i^{(\ell)}$ and concurrently replace the coupling
constant~$g(\mu)$ with the running gauge coupling satisfying
\begin{equation}
   \mu\frac{dg(\mu)}{d\mu}=\beta(g(\mu)),
\label{eq:(2.7)}
\end{equation}
where the coupling is now a function of $\mu\propto1/\sqrt{t}$. Then the
$t\to0$ limit allows us to neglect the higher-order terms in $g^2$ because the
running gauge coupling~$g(\mu\propto1/\sqrt{t})$ goes to zero due to the
asymptotic freedom. We note that this renormalization group improvement is
legitimate since the coefficients $c_i(t)$ ($i=1$, $2$) are independent of the
renormalization scale~$\mu$ (when the bare coupling~$g_0$ is kept fixed). This
can be seen from the fact that the EMT~\eqref{eq:(1.3)} and the
operator~$G_{\mu\rho}^a(t,x)G_{\nu\rho}^a(t,x)$ are bare quantities.

Although the above argument shows that in principle the coefficients $c_i(t)$
are independent of the choice of the relation between the renormalization
scale~$\mu$ and the flow time~$t$, i.e., of the parameter~$c$
in~$\mu=c/\sqrt{t}$, this independence does not exactly hold in practical
calculations based on fixed-order perturbation theory. In other words, the
difference caused by different choices of~$c$ implies the remaining
perturbative uncertainty. Following Ref.~\cite{Harlander:2018zpi}, we introduce
the combination
\begin{equation}
   L(\mu,t)\equiv\ln(2\mu^2t)+\gamma_{\text{E}}.
\label{eq:(2.8)}
\end{equation}
A conventional choice of~$\mu$ is given by
\begin{equation}
   \mu=\mu_d(t)\equiv\frac{1}{\sqrt{8t}}\Leftrightarrow
   L=-2\ln2+\gamma_{\text{E}}.
\label{eq:(2.9)}
\end{equation}
All the numerical experiments on the basis of the
representation~\eqref{eq:(1.4)} so
far~\cite{Asakawa:2013laa,Taniguchi:2016ofw,Kitazawa:2016dsl,Ejiri:2017wgd,%
Kitazawa:2017qab,Kanaya:2017cpp,Taniguchi:2017ibr,Yanagihara:2018qqg,%
Hirakida:2018uoy,Shirogane:2018zbp} have adopted this choice. On the other
hand, in~Ref.~\cite{Harlander:2018zpi}, it is argued that
\begin{equation}
   \mu=\mu_0(t)\equiv\frac{1}{\sqrt{2e^{\gamma_{\text{E}}}t}}\Leftrightarrow
   L=0,
\label{eq:(2.10)}
\end{equation}
would be an optimal choice on the basis of the two-loop order
coefficients.\footnote{The reduction of the renormalization-scale dependence
from the one-loop order to the two-loop order is studied in detail
in~Ref.~\cite{Harlander:2018zpi}.} In the following numerical analyses, we will
examine both choices $\mu=\mu_0(t)$ and~$\mu=\mu_d(t)$. The difference in the
results with these two choices gives an estimate of higher-order uncertainty,
where $\mu_d(t)\simeq0.667\mu_0(t)$.

Let us now list the known coefficients in~Eq.~\eqref{eq:(1.6)}.

\subsection{One-loop order (NLO) coefficients}
In the one-loop level, we
have~\cite{Suzuki:2013gza,Makino:2014taa,Suzuki:2015bqa}
\begin{align}
   k_1^{(1)}&=-\beta_0L-\frac{7}{3}C_A
\notag\\
   &=C_A\left(-\frac{11}{3}L-\frac{7}{3}\right).
\label{eq:(2.11)}
\\
   k_2^{(1)}&=\frac{1}{8}\beta_0
\notag\\
   &=\frac{11}{24}C_A.
\label{eq:(2.12)}
\end{align}
The number $L$ is defined by~Eq.~\eqref{eq:(2.8)}.

\subsection{Two-loop order (NNLO) coefficients}
The two-loop order coefficients in~Ref.~\cite{Harlander:2018zpi} specialized to
the pure Yang--Mills theory are
\begin{align}
   k_1^{(2)}
   &=
   -\beta_1L
   +C_A^2
   \left(
   -\frac{14\,482}{405}
   -\frac{16\,546}{135}\ln2
   +\frac{1187}{10}\ln3
   \right)
\notag\\
   &=C_A^2\left(
   -\frac{34}{3}L
   -\frac{14\,482}{405}
   -\frac{16\,546}{135}\ln2
   +\frac{1187}{10}\ln3
   \right).
\label{eq:(2.13)}
\\
   k_2^{(2)}
   &=\frac{1}{8}\beta_1-\frac{7}{16}\beta_0C_A
\notag\\
   &=C_A^2
   \left(
   -\frac{3}{16}
   \right).
\label{eq:(2.14)}
\end{align}

\subsection{Three-loop order ($\text{N}^3\text{LO}$) coefficient for~$c_2(t)$,
$k_2^{(3)}$}
In the pure Yang--Mills theory, if one has the small flow-time expansion of the
renormalized operator~$\{F_{\mu\nu}^aF_{\mu\nu}^a\}_R(x)$ in the $\ell$th-loop
order, it is possible to further obtain the coefficient of~$c_2(t)$ one loop
higher, $k_2^{(\ell+1)}$, by using information on the trace
anomaly~\cite{Suzuki:2013gza}. The two-loop order (NNLO)
coefficient~\eqref{eq:(2.14)} can also be obtained in this way from a one-loop
order calculation and has already been used in numerical experiments in
quenched QCD. Repeating this argument, we can now obtain the three-loop order
coefficient, $k_2^{(3)}$.

We recall the trace anomaly~\cite{Adler:1976zt,Nielsen:1977sy,Collins:1976yq}
\begin{equation}
   T_{\mu\mu}(x)=-\frac{\beta(g)}{2g^3}\{F_{\mu\nu}^aF_{\mu\nu}^a\}_R(x),
\label{eq:(2.15)}
\end{equation}
where the $\beta$~function is given by~Eq.~\eqref{eq:(2.1)}. According
to~Eq.~(64) of~Ref.~\cite{Harlander:2018zpi}, we now have the small flow-time
expansion of~$\{F_{\mu\nu}^aF_{\mu\nu}^a\}_R(x)$ to the two-loop order:
\begin{align}
   &\{F_{\mu\nu}^aF_{\mu\nu}^a\}_R(x)
\notag\\
   &=
   \left[1+\frac{g^2}{(4\pi)^2}\left(-\frac{7}{2}C_A\right)
   +\frac{g^4}{(4\pi)^4}C_A^2
   \left(
   -\frac{3}{2}L
   -\frac{1427}{180}+\frac{87}{5}\ln2-\frac{54}{5}\ln3
   \right)
   \right]
\notag\\
   &\qquad{}
   \times G_{\mu\nu}^a(t,x)G_{\mu\nu}^a(t,x)+O(t).
\label{eq:(2.16)}
\end{align}
Plugging this into~Eq.~\eqref{eq:(2.15)} and using Eq.~\eqref{eq:(2.1)}, we
have
\begin{align}
   &T_{\mu\mu}(x)
\notag\\
   &=
   \frac{1}{g^2}
   \bigg\{\frac{g^2}{(4\pi)^2}\frac{1}{2}\beta_0
   +\frac{g^4}{(4\pi)^4}\left(\frac{1}{2}\beta_1-\frac{7}{4}\beta_0C_A\right)
\notag\\
   &\qquad\qquad{}
   +\frac{g^6}{(4\pi)^6}
   \left[
   \frac{1}{2}\beta_2
   -\frac{7}{4}\beta_1C_A
   +\beta_0C_A^2
   \left(
   -\frac{3}{4}L
   -\frac{1427}{360}+\frac{87}{10}\ln2-\frac{27}{5}\ln3
   \right)
   \right]
   \biggr\}
\notag\\
   &\qquad\qquad\qquad{}
   \times G_{\mu\nu}^a(t,x)G_{\mu\nu}^a(t,x)+O(t).
\label{eq:(2.17)}
\end{align}
Comparing this with the trace of~Eq.~\eqref{eq:(1.4)}, we obtain
\begin{align}
   k_2^{(3)}
   &=
   \frac{1}{8}\beta_2
   -\frac{7}{16}\beta_1C_A
   +\beta_0C_A^2
   \left(
   -\frac{3}{16}L
   -\frac{1427}{1440}+\frac{87}{40}\ln2-\frac{27}{20}\ln3
   \right)
\notag\\
   &=C_A^3\left(
   -\frac{11}{16}L
   -\frac{2849}{1440}
   +\frac{319}{40}\ln2
   -\frac{99}{20}\ln3
   \right).
\label{eq:(2.18)}
\end{align}
One can also confirm that Eqs.~\eqref{eq:(2.12)} and~\eqref{eq:(2.14)} are
correctly reproduced. We will examine the use of this $\text{N}^3\text{LO}$
coefficient for the trace anomaly in the numerical analyses below.

\section{Numerical analyses}
\label{sec:3}

In what follows, we use the lattice data obtained
in~Ref.~\cite{Kitazawa:2016dsl} for the $G=SU(3)$ pure Yang--Mills theory and
study the effects of the higher-order coefficients quantitatively. 
We do not repeat the explanation of the lattice setups;
see~Ref.~\cite{Kitazawa:2016dsl} for details. We measure the entropy density
and trace anomaly (which are normalized by the temperature). To obtain these
thermodynamic quantities, the double limit, $a\to0$ and~$t\to0$, is required
because Eq.~\eqref{eq:(1.4)} is the relation in continuum spacetime and in
the small flow-time limit. We first take the continuum limit~$a\to0$ and then
take the small flow-time limit~$t\to0$~\cite{Kitazawa:2016dsl}. In continuum
extrapolation while keeping $t$ in physical units fixed, we adopt only the
range where the flow time satisfies $t\gtrsim a^2$. This is because the flow
time~$t$ is meaningful only for~$t\gtrsim a^2$ with finite lattice spacing~$a$;
the lattice data actually exhibit violently diverging behavior
for~$t\lesssim a^2$ due to finite lattice spacing effects. Hence, at this
stage, we cannot obtain the continuum limit for~$t=0$. Thus, we carry out the
$t\to0$ extrapolation by assuming a certain functional form
of~Eq.~\eqref{eq:(3.1)} with respect to~$t$.

\begin{figure}[b!]
\centering
\begin{subfigure}{0.45\columnwidth}
\centering
\includegraphics[width=\columnwidth]{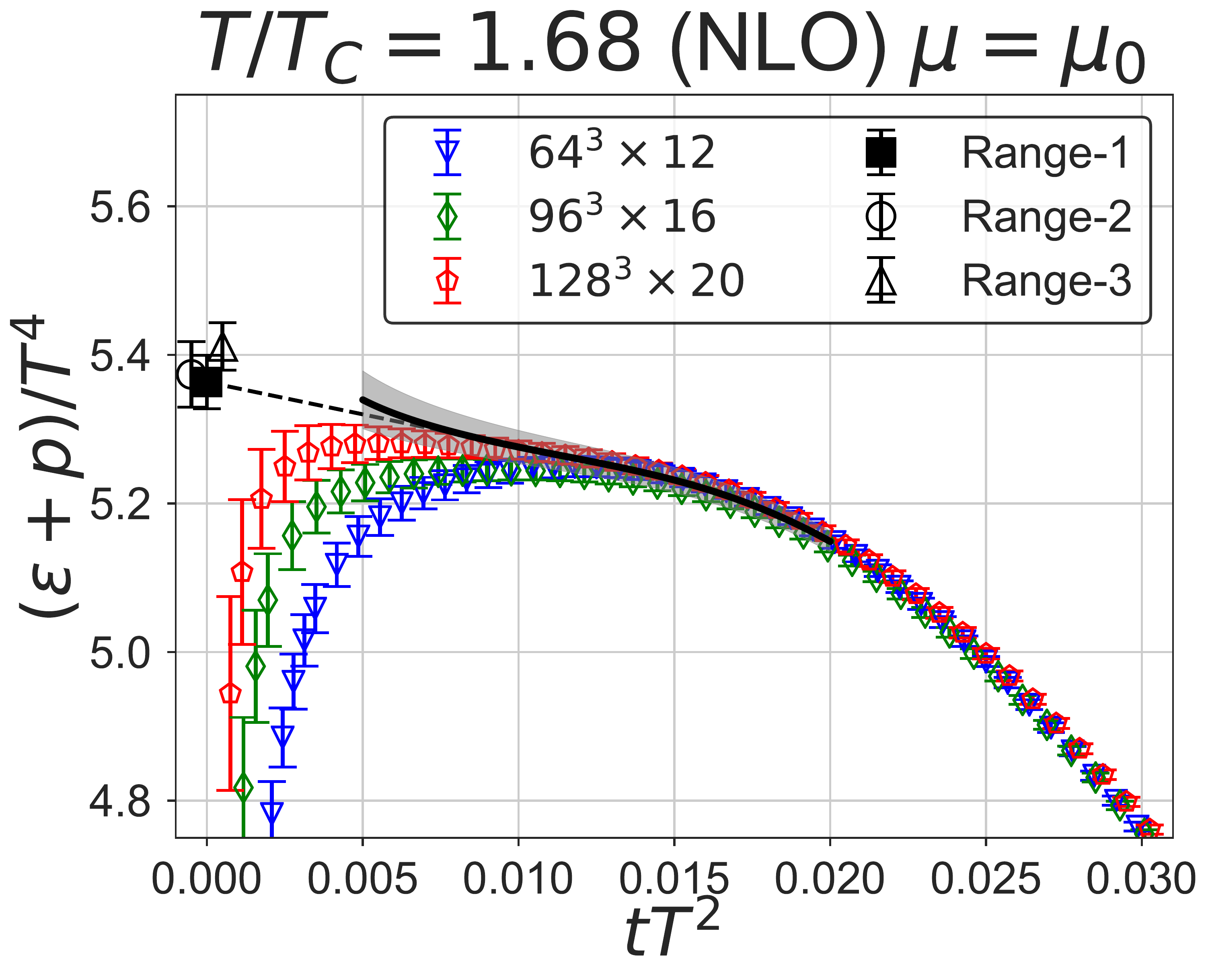}
\caption{}
\label{}
\end{subfigure}
\hspace{10mm}
\begin{subfigure}{0.45\columnwidth}
\centering
\includegraphics[width=\columnwidth]{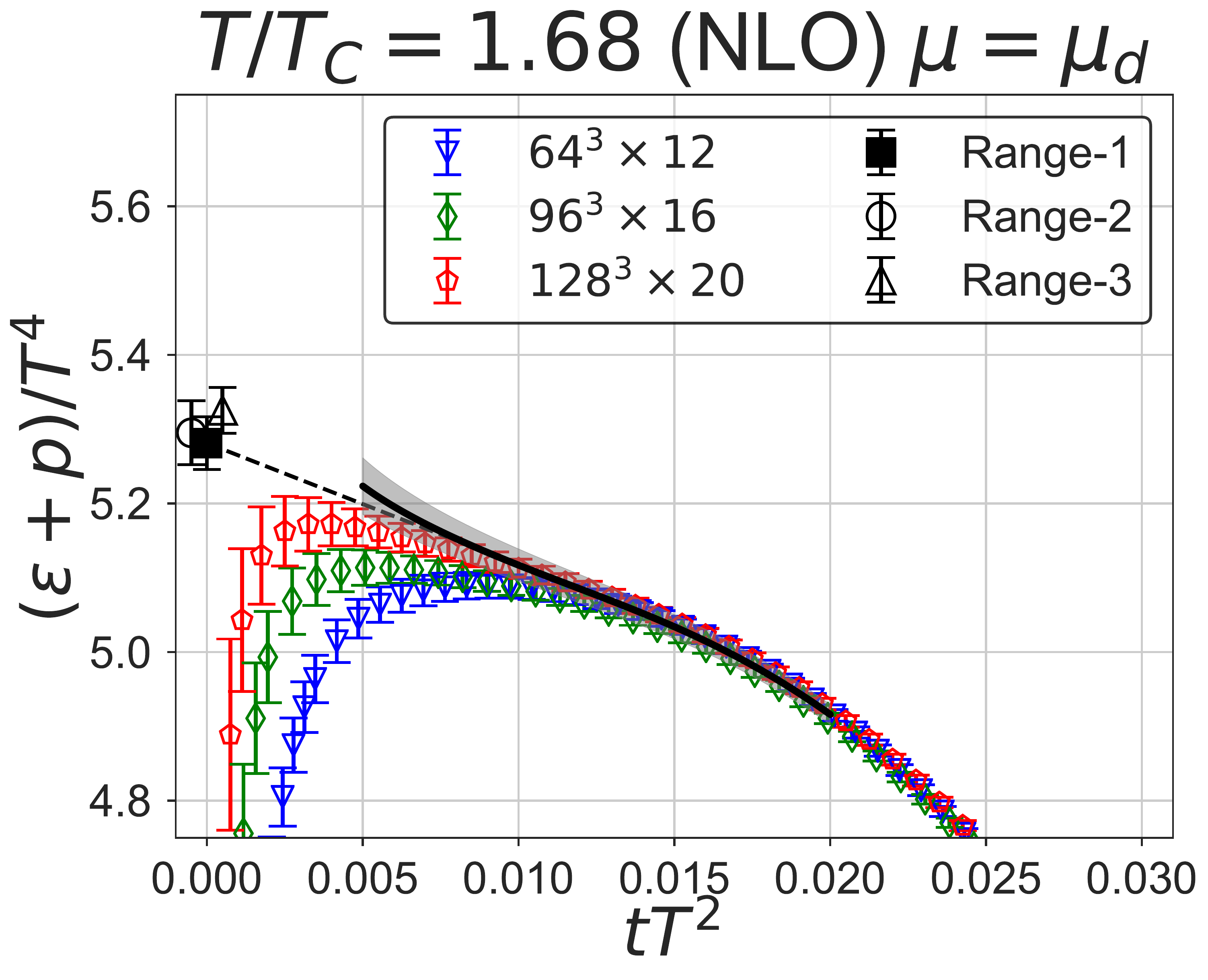}
\caption{}
\label{}
\end{subfigure}
\begin{subfigure}{0.45\columnwidth}
\centering
\includegraphics[width=\columnwidth]{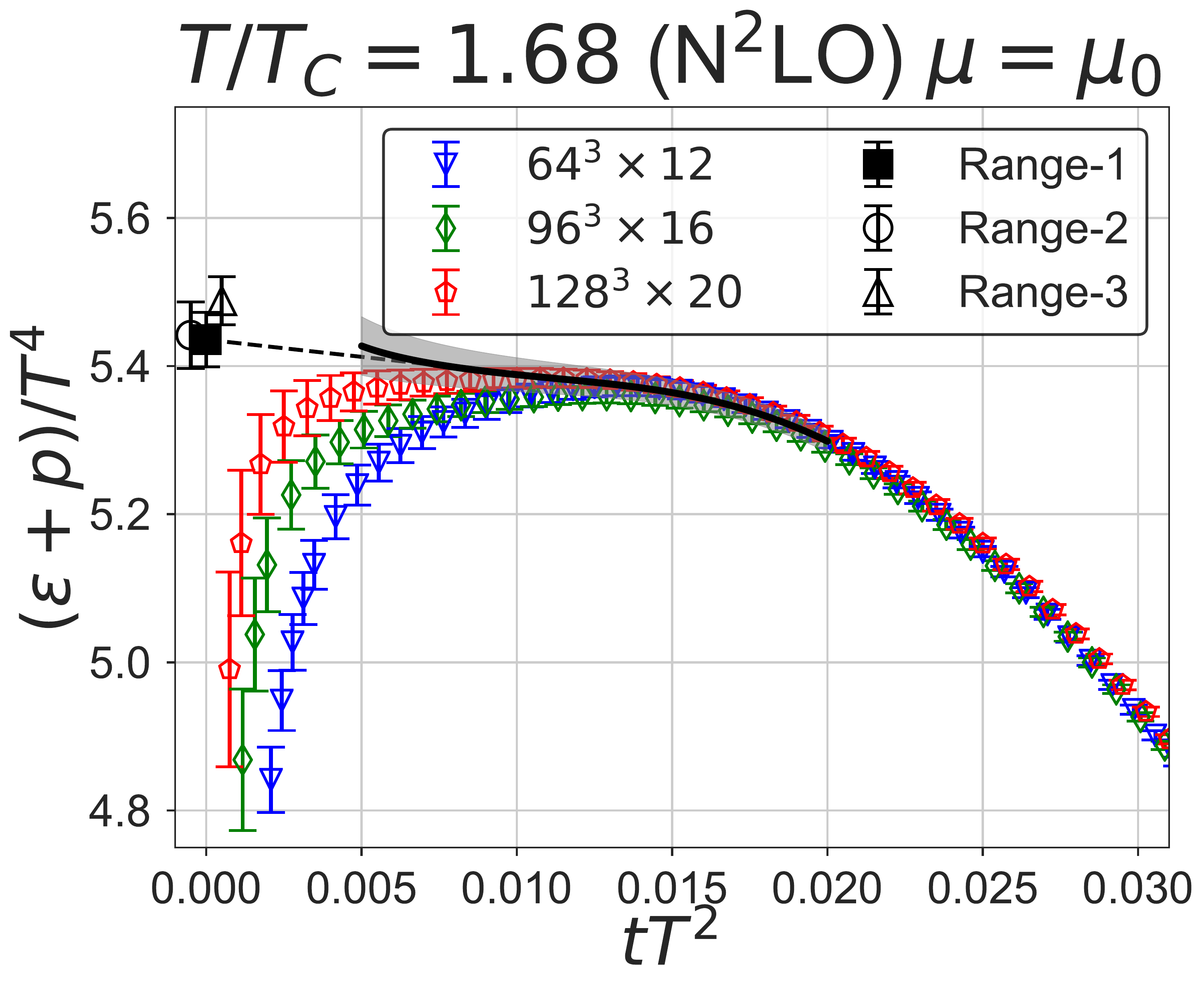}
\caption{}
\label{}
\end{subfigure}
\hspace{10mm}
\begin{subfigure}{0.45\columnwidth}
\centering
\includegraphics[width=\columnwidth]{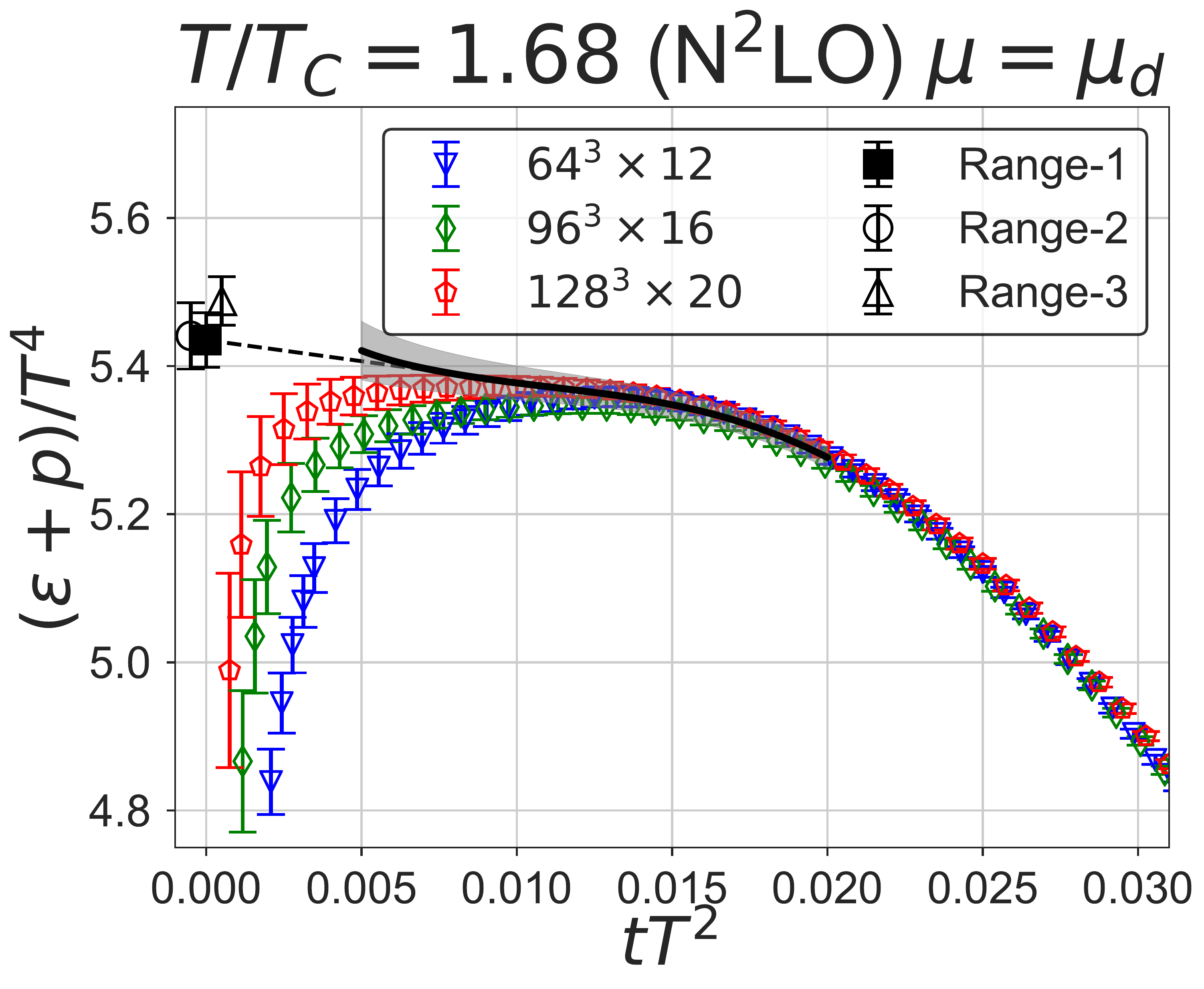}
\caption{}
\label{}
\end{subfigure}
\caption{Equation~\eqref{eq:(3.1)} as a function of~$tT^2$ for~$T/T_c=1.68$. In
each panel, the order of perturbation theory and the choice of the
renormalization scale are indicated. The errors are statistical only. The
extrapolation of the continuum limit (the gray band) to~$t=0$ is plotted by the
black circle (obtained by the fit range~\eqref{eq:(3.2)}), the white circle
(obtained by the fit range~\eqref{eq:(3.3)}), and the white triangle (obtained
by the fit range~\eqref{eq:(3.4)}).}
\label{fig:1}
\end{figure}

Let us start with the entropy density, $\varepsilon+p$.
From~Eq.~\eqref{eq:(1.1)}, $(\varepsilon+p)/T^4$ is obtained by taking the
$t\to0$ limit of the thermal expectation value:
\begin{align}
   -\frac{4}{3T^4}c_1(t)\left\langle
   G_{0\rho}^a(t,x)G_{0\rho}^a(t,x)
   -\frac{1}{4}G_{\rho\sigma}^a(t,x)G_{\rho\sigma}^a(t,x)
   \right\rangle.
\label{eq:(3.1)}
\end{align}
In~Fig.~\ref{fig:1}, we plot Eq.~\eqref{eq:(3.1)} as a function of~$tT^2$; the
temperature is~$T/T_c=1.68$. The plots for the other temperatures listed in the
left-most column of~Table~\ref{table:2} are deferred to~Appendix~\ref{app:A}.
In each panel of~Fig.~\ref{fig:1}, we show lattice results
for~Eq.~\eqref{eq:(3.1)} with three different lattice spacings. The
coefficient~$c_1(t)$ used in each panel is (a)~the one-loop order (i.e., NLO)
with the choice of the renormalization scale~$\mu_0(t)$~\eqref{eq:(2.10)},
(b)~the NLO with~$\mu_d(t)$~\eqref{eq:(2.9)}, (c)~the two-loop order (i.e.,
NNLO or~$\text{N}^2\text{LO}$) with~$\mu_0(t)$, and (d)~the
$\text{N}^2\text{LO}$ with~$\mu_d(t)$, respectively. In calculations
of~$c_1(t)$, the running coupling as a function of $tT^2$ is required, which is
originally a function of $\mu(t)/\Lambda_{\overline{\text{MS}}}$. For this
conversion, we use the central values of Eqs.~(23) and~(A2)
in~Ref.~\cite{Kitazawa:2016dsl}. We use the four-loop running gauge coupling in
the $\overline{\text{MS}}$ scheme, adopting the approximate formula~(9.5)
in~Ref.~\cite{Tanabashi:2018oca}.\footnote{In~Ref.~\cite{Kitazawa:2016dsl}, the
$\Lambda$ parameter is obtained  at the three-loop level, while we use the
four-loop running coupling. The effect of the discrepancy in the perturbation
order would be able to be taken into account by varying
$\Lambda_{\overline{\text{MS}}}$ in our error analysis below.}

We then take the continuum limit $a\to0$ at each fixed value of~$tT^2$.
(See~Ref.~\cite{Kitazawa:2016dsl} for details of this procedure.) The continuum
limit results are shown by gray bands. In~Fig.~\ref{fig:1} and in the
corresponding figures in~Appendix~\ref{app:A},
Figs.~\ref{fig:A1}--\ref{fig:A7}, we observe that the use of the two-loop order
coefficient generally reduces the $t$~dependence of the continuum limit (it
becomes flatter in~$t$).\footnote{An explanation for this flatter behavior will
be provided in another paper (H.~Suzuki and H.~Takaura, manuscript in
preparation).} This behavior generally allows us to perform stable $t\to0$
extrapolation.

We carry out the $t\to0$ extrapolation with a linear function in~$t$
as~Ref.~\cite{Kitazawa:2016dsl}. The following three reasonable fit ranges are
adopted for the $t\to0$ extrapolation~\cite{Kitazawa:2016dsl}:\footnote{The
finite lattice spacing and volume effects are controlled
by~$tT^2$~\cite{Kitazawa:2016dsl}.}
\begin{align}
   &\text{Range~1: $0.01\leq tT^2\leq0.015$},
\label{eq:(3.2)}
\\
   &\text{Range~2: $0.005\leq tT^2\leq0.015$},
\label{eq:(3.3)}
\\
   &\text{Range~3: $0.01\leq tT^2\leq0.02$}.
\label{eq:(3.4)}
\end{align}
In~Table~\ref{table:1}, the coefficients of the linear term in~$t$, determined
in~$t\to0$ extrapolation using Range~1, are shown. The tendency for the slopes
to get smaller at~$\text{N}^2\text{LO}$ than at NLO is quantitatively observed.
In addition to the linear function in~$t$, we also use the linear function
of~$[g(\mu)^2/(4\pi)^2]^{\ell+1}$, where $\ell=1$ for the NLO approximation and
$\ell=2$ for the $\text{N}^2\text{LO}$ approximation.\footnote{Recall that the
renormalization scale~$\mu$ is a function of~$t$ through~Eq.~\eqref{eq:(2.8)}.}
This functional form is suggested from a detailed study of the asymptotic
behavior of~Eq.~\eqref{eq:(3.1)} for~$t\to0$ (H.~Suzuki and H.~Takaura,
manuscript in preparation). We estimate the systematic error associated with
the $t\to0$ extrapolation by examining the variation obtained from the
different extrapolation function. As mentioned, the use of the two-loop
coefficient leads to a flatter behavior with respect to~$t$. Hence, $t\to0$
extrapolation becomes less sensitive to the fit function, the fit range, and
the choice of the renormalization scale.
\begin{table}[t!]
\begin{center}
\begin{tabular}{c|c|c}
\hline\hline
   $(\varepsilon+p)/T^4$ \\
\hline
$T/T_c$ & NLO & $\text{N}^2\text{LO}$ \\
\hline
0.93 & $0.6(1.5)$ & $0.8(1.6)$ \\
1.02 & $-5.6(1.7)$ & $-2.0(1.7)$ \\
1.12 & $-5.0(1.9)$ & $0.3(1.9)$ \\
1.40 & $-4.7(1.6)$ & $0.2(1.7)$ \\
1.68 & $-8.7(1.6)$ & $-4.7(1.7)$ \\
2.10 & $-5.5(1.5)$ & $-2.2(1.6)$ \\
2.31 & $-2.4(2.4)$ & $0.7(2.4)$ \\
2.69 & $-7.0(1.5)$ & $-4.4(1.5)$ \\
\hline\hline
$(\varepsilon-3p)/T^4$ \\
\hline
$T/T_c$ & $\text{N}^2\text{LO}$ & $\text{N}^3\text{LO}$ \\
\hline
0.93 & $1.3(1.1)$ & $1.2(1.1)$ \\
1.02 & $1.9(1.0)$ & $0.6(1.0)$ \\
1.12 & $2.5(1.0)$ & $1.0(1.0)$ \\
1.40 & $0.8(1.0)$ & $0.1(1.0)$ \\
1.68 & $-0.9(0.7)$ & $-1.3(0.7)$ \\
\hline\hline
\end{tabular}
\caption{Linear coefficients in the $t\to0$ extrapolation. The renormalization
scale $\mu_0(t)$ and Range~1 are used.}
\label{table:1}
\end{center}
\end{table}

\begin{table}[t!]
\begin{center}
\begin{tabular}{l|l|l|l} 
\hline\hline
   $(\varepsilon+p)/T^4$ \\
\hline
   $T/T_c$ & NLO & $\text{N}^2\text{LO}$ & FlowQCD 2016 \\
\hline
0.93 & $0.082(34)(02)(01)(00)(05)$ & $0.083(35)(02)(01)(00)(01)$ & 0.082(33)($^{+3}_{-6}$)(0) \\
1.02 & $2.128(64)(09)(25)(25)(57)$ & $2.163(66)(09)(25)(04)(06)$ & 2.104(63)($^{+16}_{-2}$)(8) \\
1.12 & $3.651(47)(32)(41)(50)(60)$ & $3.709(49)(34)(41)(05)(01)$ & 3.603(46)($^{+39}_{-0}$)(13) \\
1.40 & $4.777(36)(49)(51)(74)(72)$ & $4.847(37)(53)(51)(01)(01)$ & 4.706(35)($^{+49}_{-0}$)(17) \\
1.68 & $5.363(36)(48)(54)(82)(158)$ & $5.436(37)(52)(54)(01)(37)$ & 5.285(35)($^{+44}_{-0}$)(18) \\
2.10 & $5.694(35)(71)(54)(80)(117)$ & $5.762(35)(75)(53)(01)(22)$ & 5.617(34)($^{+66}_{-0}$)(18) \\
2.31 & $5.731(56)(87)(53)(77)(55)$ & $5.797(56)(92)(52)(01)(08)$ & 5.657(55)($^{+82}_{-15}$)(18) \\
2.69 & $5.986(33)(75)(52)(75)(175)$ & $6.050(33)(80)(52)(02)(55)$ & 5.914(32)($^{+70}_{-0}$)(18) \\
\hline\hline
   $(\varepsilon-3p)/T^4$ \\
\hline
   $T/T_c$ & $\text{N}^2\text{LO}$ & $\text{N}^3\text{LO}$ & FlowQCD 2016 \\
\hline
0.93 & $0.066(32)(03)(00)(00)(12)$ & $0.066(32)(03)(00)(00)(02)$ & 0.066(32)($^{+3}_{-2}$)(0) \\
1.02 & $1.947(58)(07)(00)(03)(20)$ & $1.934(57)(07)(00)(03)(02)$ & 1.945(57)($^{+8}_{-7}$)(0) \\
1.12 & $2.564(33)(11)(00)(04)(29)$ & $2.548(33)(11)(01)(01)(04)$ & 2.560(33)($^{+12}_{-8}$)(0) \\
1.40 & $1.779(24)(14)(00)(03)(12)$ & $1.769(24)(14)(00)(01)(01)$ & 1.777(24)($^{+14}_{-3}$)(0) \\
1.68 & $1.203(19)(10)(00)(02)(17)$ & $1.196(18)(10)(00)(01)(10)$ & 1.201(19)($^{+10}_{-0}$)(0) \\
\hline\hline
\end{tabular}
\caption{Summary of the entropy density and the trace anomaly obtained by
using coefficients in different orders of perturbation theory. The statistical
errors are shown in the first parentheses. The numbers in the other parentheses
show systematic errors: the error associated with the fit range, the $3\%$
uncertainty of~$\Lambda_{\overline{\text{MS}}}$, the renormalization scale, and
the $t\to0$ extrapolation function are shown in the second, third, fourth, and
fifth parentheses, respectively. The results of~Ref.~\cite{Kitazawa:2016dsl}
(FlowQCD 2016) are also tabulated in the last column: the numbers in the first,
second, and third parentheses are the statistical error, the systematic error
associated with the choice of the fit range (only the linear~$t$ fit is adopted
in~Ref.~\cite{Kitazawa:2016dsl}), and the systematic error associated
with~$\Lambda_{\overline{\text{MS}}}$ (varying it by $1\%$
in~Ref.~\cite{Kitazawa:2016dsl}). (The difference in the central values between
FlowQCD 2016 and our analyses at the lower orders stems from the choice of the
renormalization scale; $\mu_0(t)$ is adopted in the present work rather
than~$\mu_d(t)$).}
\label{table:2}
\end{center}
\end{table}

In~Table~\ref{table:2}, the result of~$(\epsilon+p)/T^4$ is summarized. The
central values and the statistical errors are obtained by the linear $t$ fit in
Range~1~\eqref{eq:(3.2)} with the choice of the renormalization
scale~$\mu=\mu_0(t)$~\eqref{eq:(2.10)}. The systematic errors associated with
(i)~the fit range (estimated by other choices, Range~2~\eqref{eq:(3.3)} and
Range~3~\eqref{eq:(3.4)}), (ii)~the uncertainty
of~$\Lambda_{\overline{\text{MS}}}$ of $3\%$~\cite{Aoki:2016frl}, (iii)~the
renormalization scale (estimated from another
choice~$\mu=\mu_d(t)$~\eqref{eq:(2.9)}), and (iv)~the $t\to0$ extrapolation
(estimated by adopting a different extrapolation function) are shown. Because
of the reduction of the $t$~dependence with the two-loop coefficient, one can
see that the systematic errors associated with the choice of the
renormalization scale and the fit function are greatly reduced in the
$\text{N}^2\text{LO}$ approximation. This clearly shows an advantage of the
two-loop order coefficient.

\begin{figure}[b!]
\centering
\begin{subfigure}{0.45\columnwidth}
\centering
\includegraphics[width=\columnwidth]{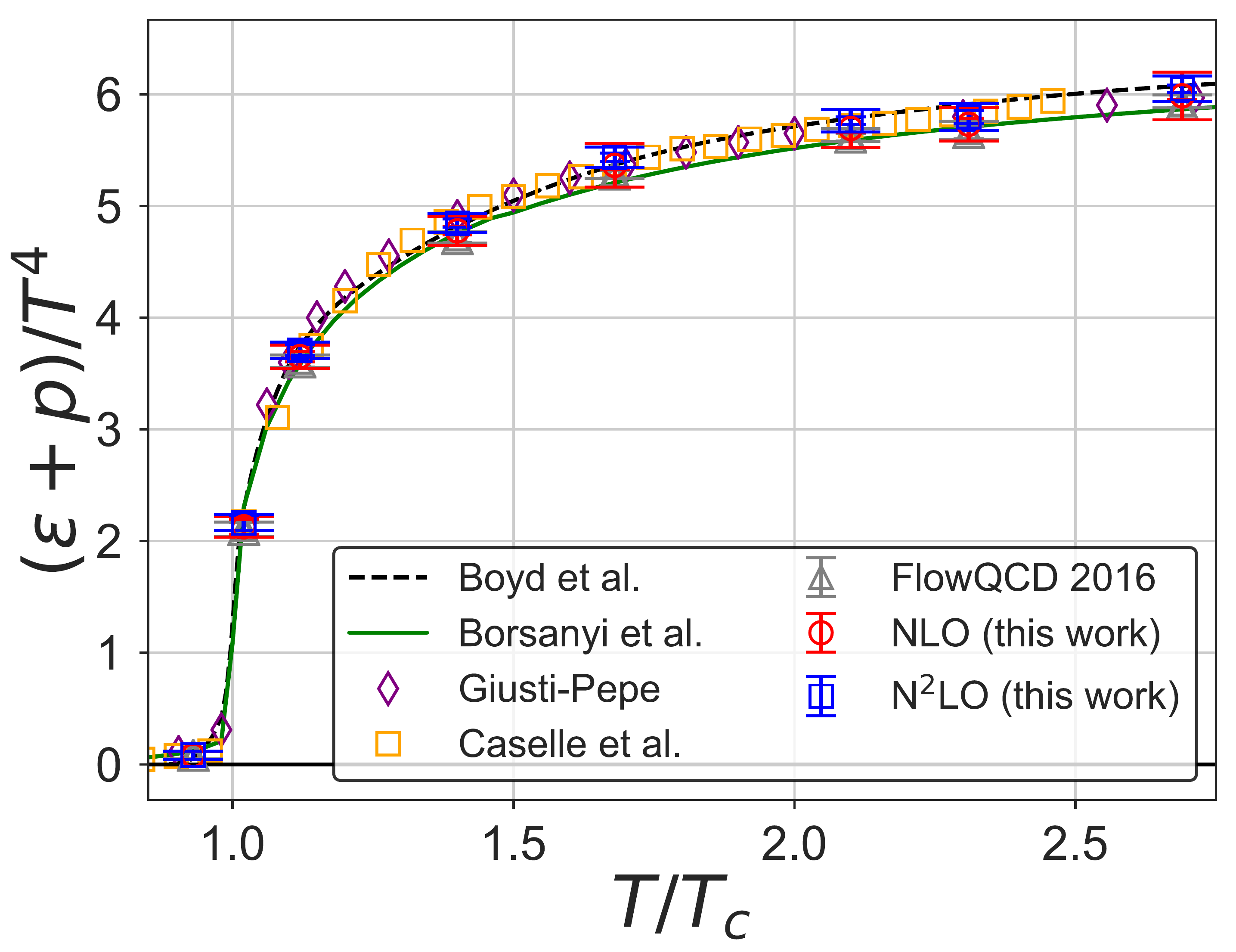}
\end{subfigure}
\hspace{10mm}
\begin{subfigure}{0.45\columnwidth}
\centering
\includegraphics[width=\columnwidth]{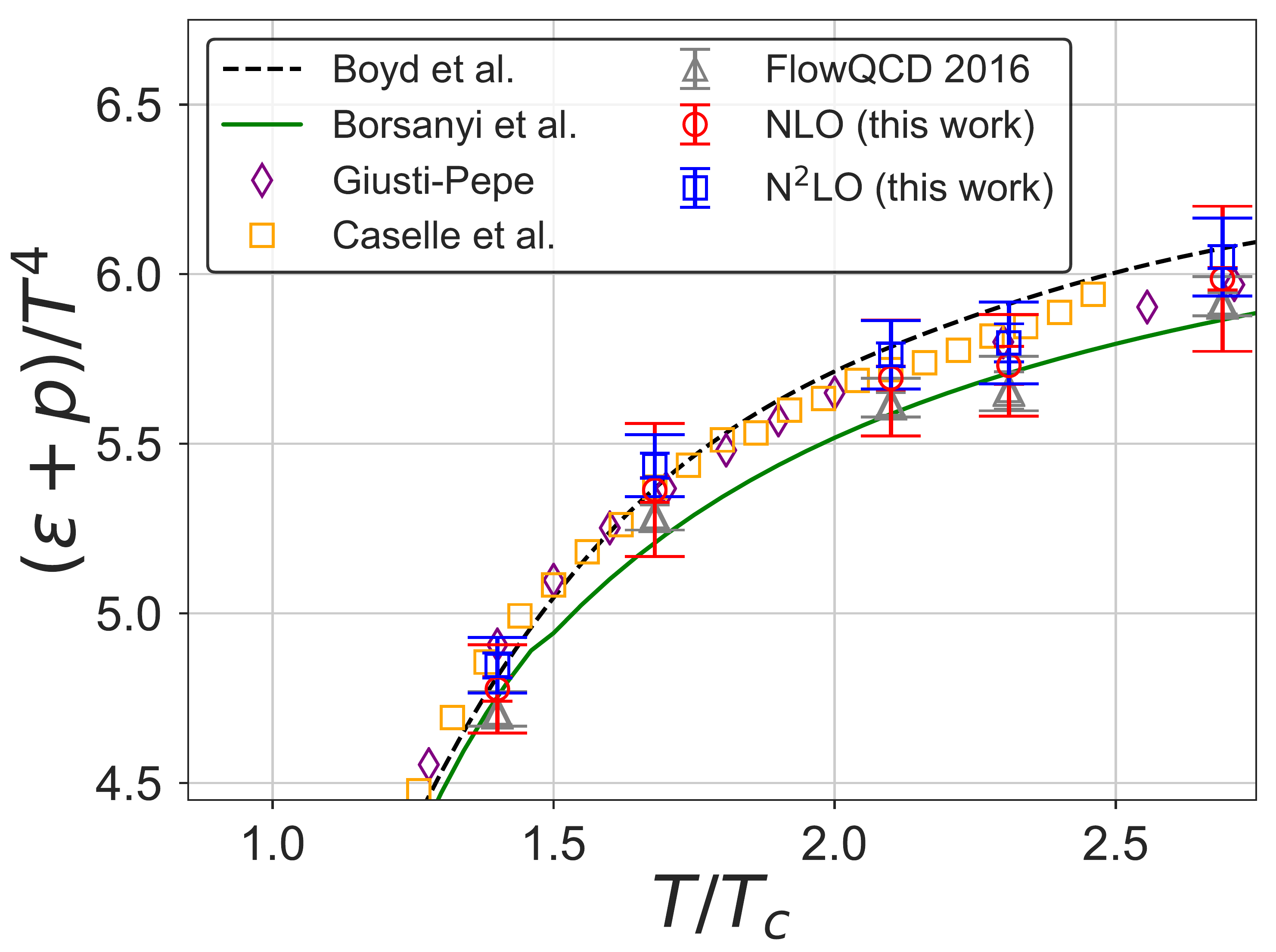}
\end{subfigure}
\caption{Summary of the entropy density~$(\varepsilon+p)/T^4$ as a function
of~$T/T_c$. In the right-hand panel, the region
$4.5\lesssim(\varepsilon+p)/T^4\lesssim6.5$ is magnified. The results from the
present paper are the red circles~(NLO) and the blue squares
($\text{N}^2\text{LO}$). The error bars include the systematic error as well as
the statistical error; see~Table~\ref{table:2} and the main text for details.
For comparison, we also show the results
of~Refs.~\cite{Boyd:1996bx,Borsanyi:2012ve,Giusti:2016iqr,Caselle:2018kap,%
Kitazawa:2016dsl}.
}
\label{fig:2}
\end{figure}

In~Fig.~\ref{fig:2}, $(\epsilon+p)/T^4$ is plotted as a function of~$T/T_c$.
The error bar represents the total error, obtained by combining all the errors
in quadrature. Our $\text{N}^2\text{LO}$ results are consistent with the
results
of~Refs.~\cite{Boyd:1996bx,Borsanyi:2012ve,Giusti:2016iqr,Caselle:2018kap,%
Kitazawa:2016dsl}, especially with Refs.~\cite{Boyd:1996bx,Caselle:2018kap}.

Reference~\cite{Kitazawa:2016dsl} is the preceding analysis performed with the
same method as this work but with the NLO coefficient. In our NLO analysis, the
systematic errors are investigated in more detail, including some additional
error sources. This leads to larger errors than
in~Ref.~\cite{Kitazawa:2016dsl}, while the central values
of~Ref.~\cite{Kitazawa:2016dsl} are consistent with the present work.
Figure~\ref{fig:2} clearly shows that the use of the two-loop coefficient
generally reduces the systematic errors.

\begin{figure}[tbhp]
\centering
\begin{subfigure}{0.45\columnwidth}
\centering
\includegraphics[width=\columnwidth]{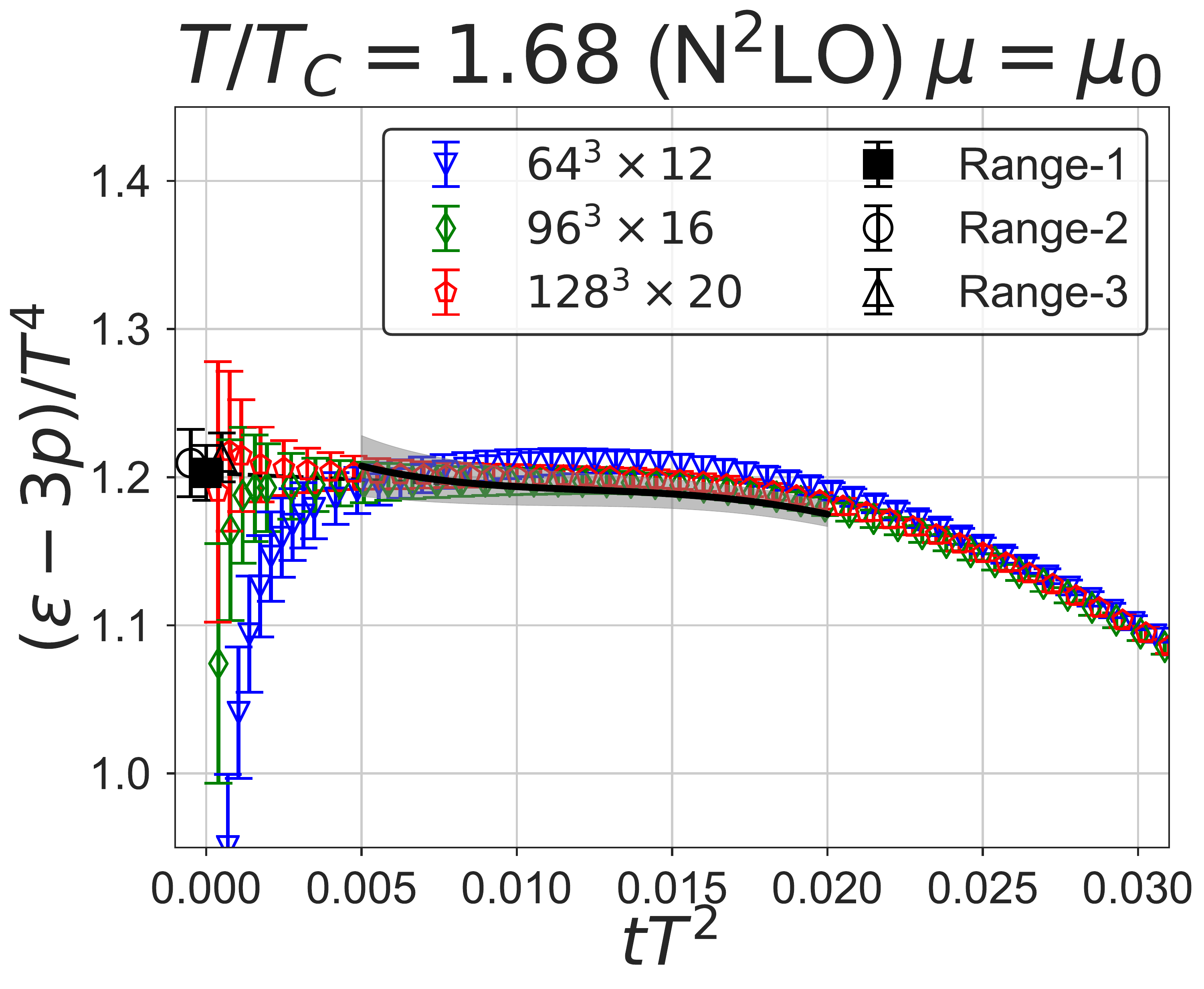}
\caption{}
\label{}
\end{subfigure}
\hspace{10mm}
\begin{subfigure}{0.45\columnwidth}
\centering
\includegraphics[width=\columnwidth]{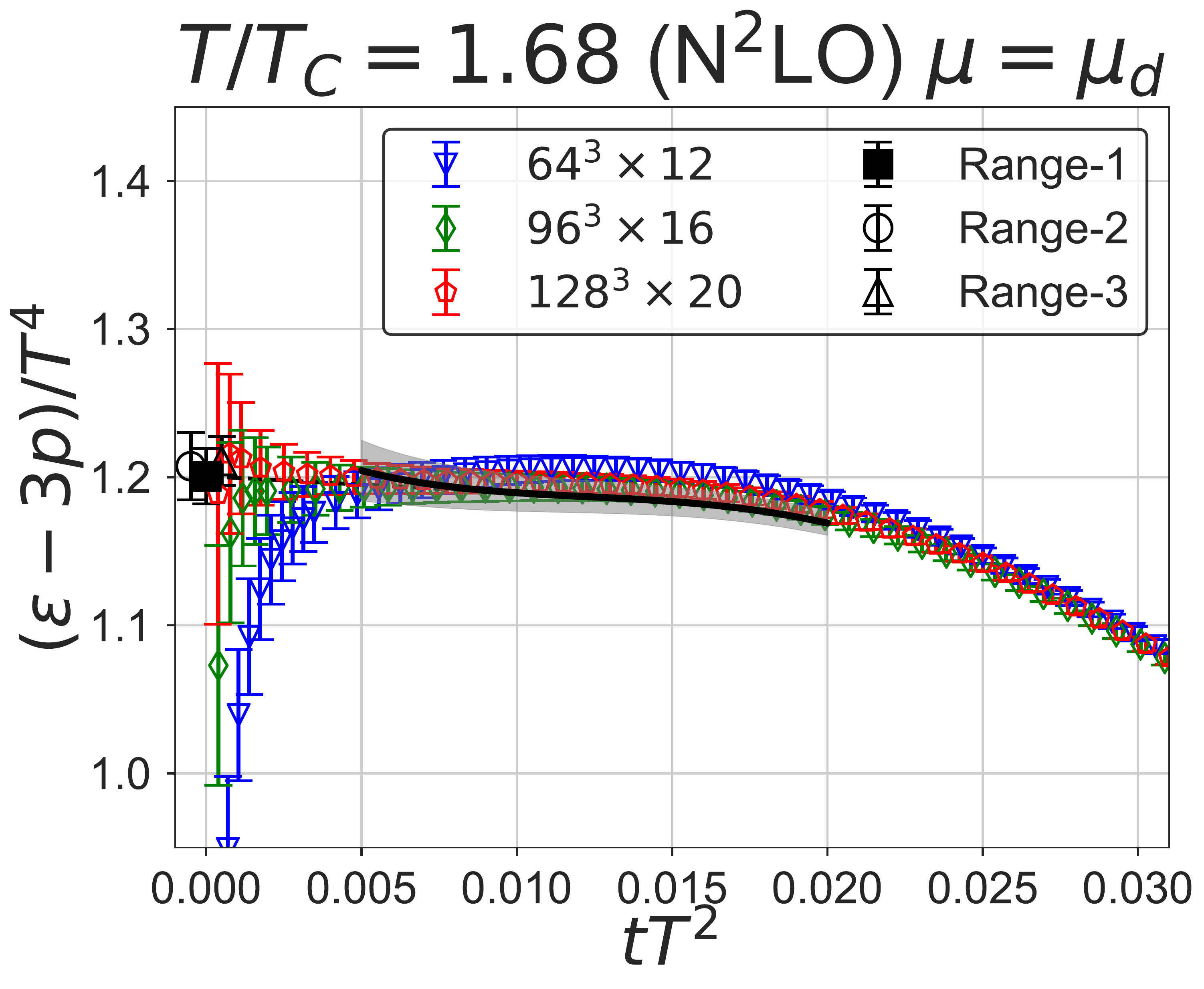}
\caption{}
\label{}
\end{subfigure}
\begin{subfigure}{0.45\columnwidth}
\centering
\includegraphics[width=\columnwidth]{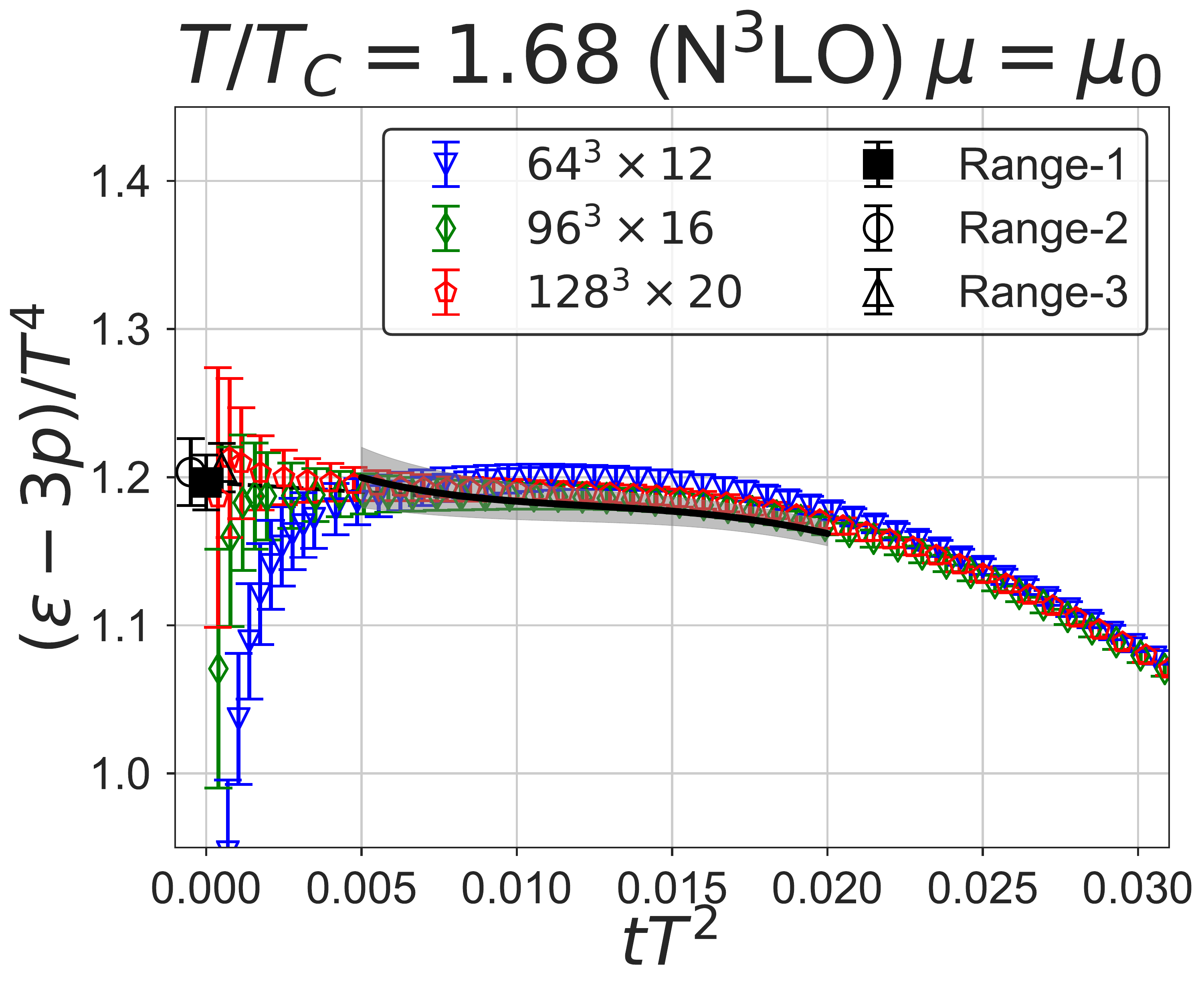}
\caption{}
\label{}
\end{subfigure}
\hspace{10mm}
\begin{subfigure}{0.45\columnwidth}
\centering
\includegraphics[width=\columnwidth]{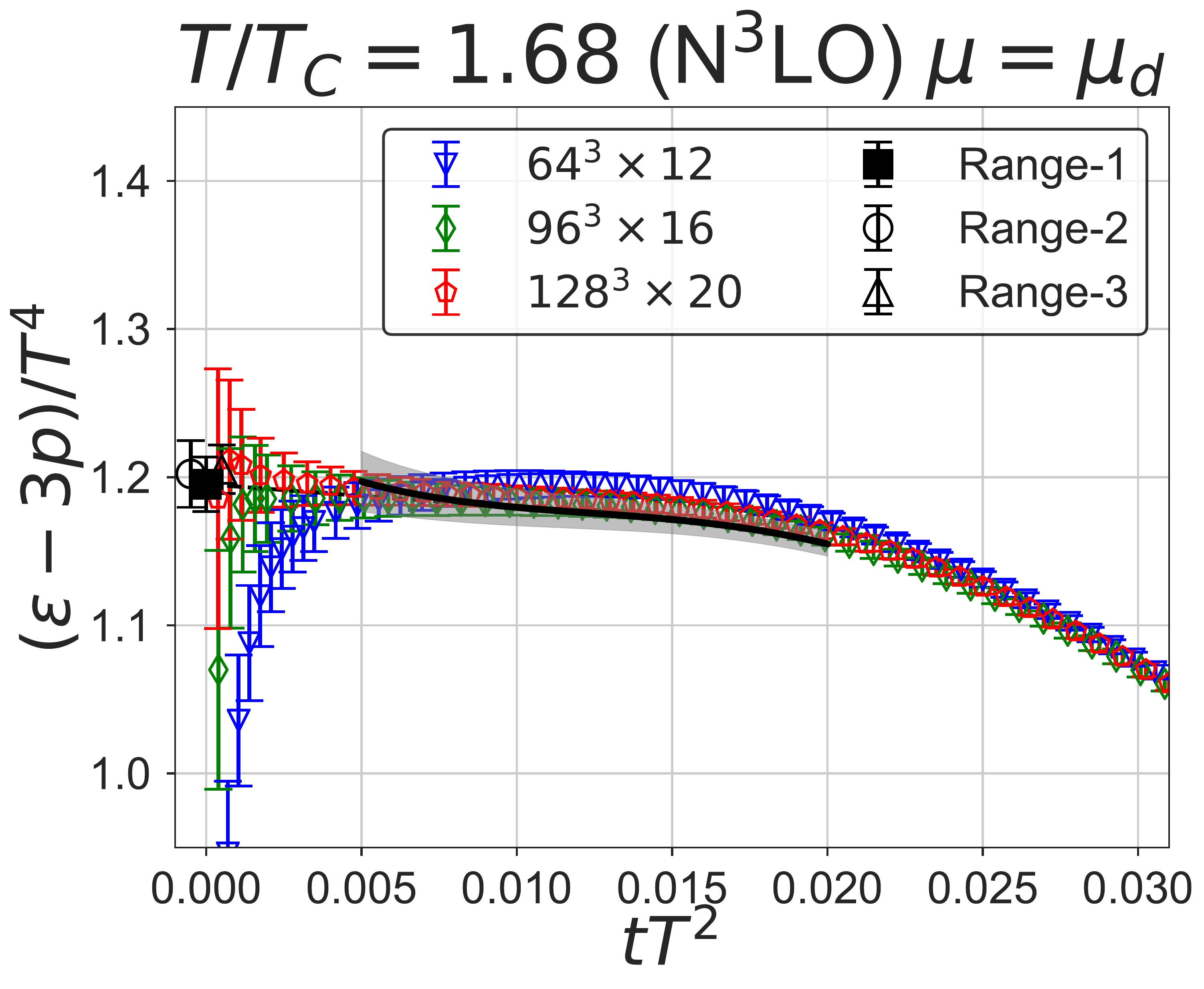}
\caption{}
\label{}
\end{subfigure}
\caption{Equation~\eqref{eq:(3.2)} as a function of~$tT^2$ for~$T/T_c=1.68$.
In each panel, the order of perturbation theory and the choice of the
renormalization scale are indicated. The errors are statistical only. The
extrapolation of the continuum limit (the gray band) to~$t=0$ is plotted by the
black circle (obtained by the fit range~\eqref{eq:(3.2)}), the white circle
(obtained by the fit range~\eqref{eq:(3.3)}), and the white triangle (obtained
by the fit range~\eqref{eq:(3.4)}).}
\label{fig:3}
\end{figure}

\begin{figure}[bthp]
\centering
\begin{subfigure}{0.45\columnwidth}
\centering
\includegraphics[width=\columnwidth]{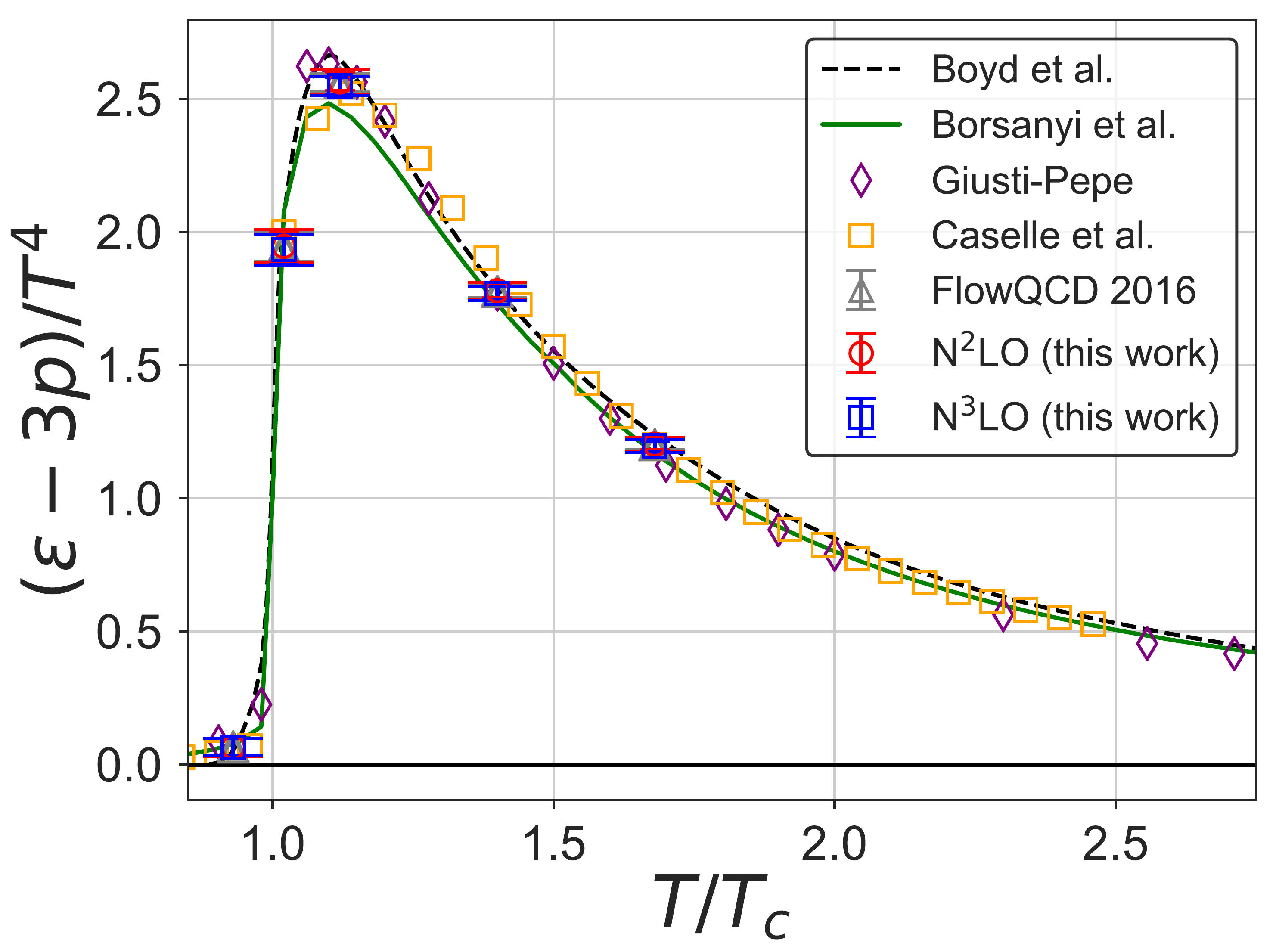}
\end{subfigure}
\hspace{10mm}
\begin{subfigure}{0.45\columnwidth}
\centering
\includegraphics[width=\columnwidth]{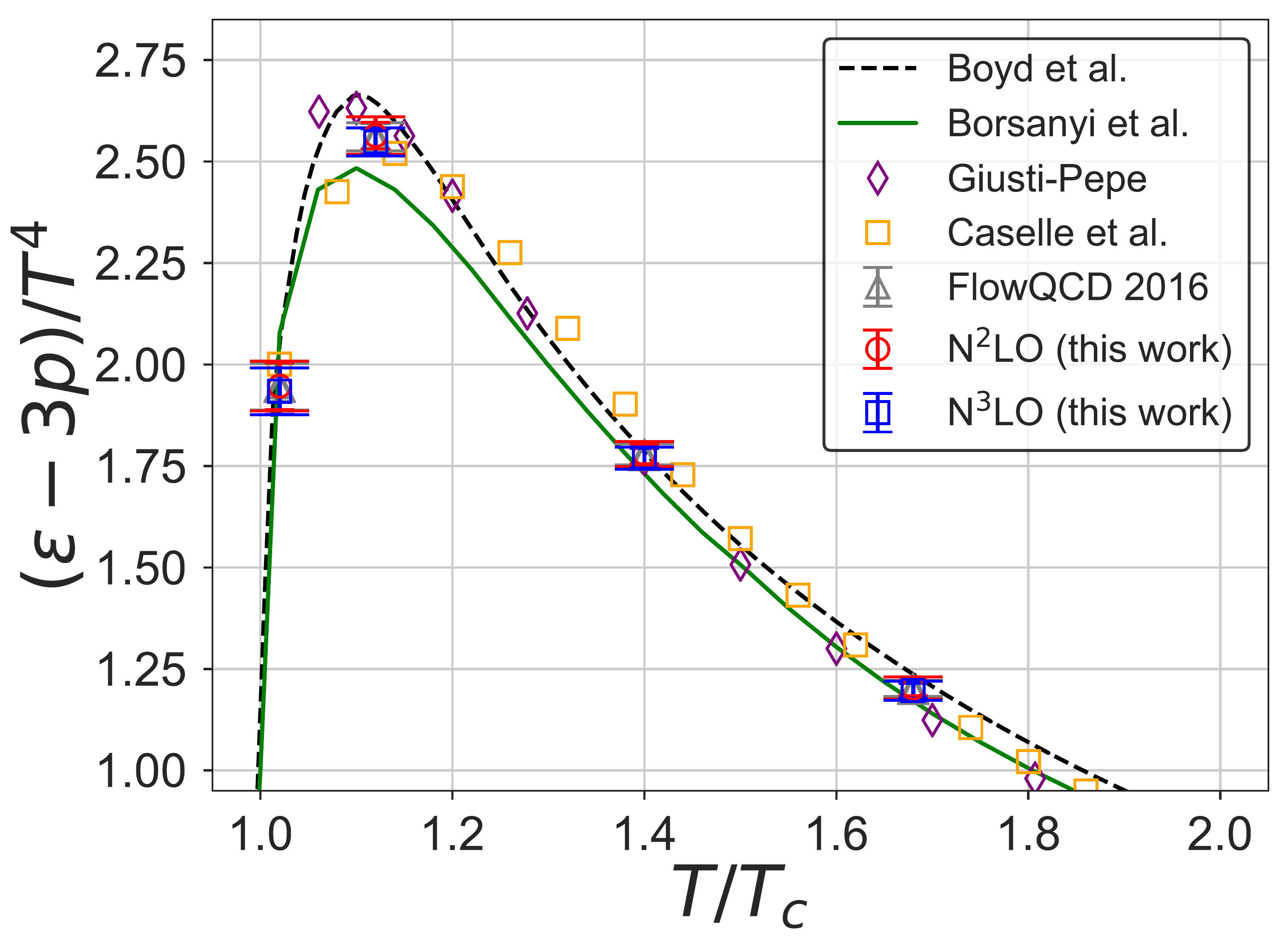}
\end{subfigure}
\caption{Summary of the trace anomaly~$(\varepsilon-3p)/T^4$ as a function
of~$T/T_c$. In the right-hand panel, the region
$1.00\lesssim(\varepsilon-3p)/T^4\lesssim2.75$ is magnified. The results from
the present paper are the red circles~($\text{N}^2\text{LO}$) and the blue
squares ($\text{N}^3\text{LO}$). The error bars include the systematic error as
well as the statistical error; see Table~\ref{table:2} for details. For
comparison, we also show the results
of~Refs.~\cite{Boyd:1996bx,Borsanyi:2012ve,Giusti:2016iqr,Caselle:2018kap,%
Kitazawa:2016dsl}.
}
\label{fig:4}
\end{figure}

We now turn to the trace anomaly, $\varepsilon-3p$, which is investigated in a
parallel manner to the entropy density. The expectation value
\begin{equation}
   -\frac{4}{T^4}c_2(t)
   \left\langle G_{\mu\nu}^a(t,x)G_{\mu\nu}^a(t,x)\right\rangle,
\label{eq:(3.5)}
\end{equation}
as a function of~$tT^2$ is plotted in Fig.~\ref{fig:3} for~$T/T_c=1.68$.
(Results for other temperatures are deferred to~Appendix~\ref{app:A}.) As we
noted, the two-loop order ($\text{N}^2\text{LO}$) and the three-loop order
($\text{N}^3\text{LO}$) coefficients are available for the trace anomaly. We
observe that already with the $\text{N}^2\text{LO}$ coefficient the continuum
limit (the gray band) is almost constant in~$t$ within our fit ranges. Thus,
naturally, the extrapolation of the continuum limit to~$t=0$ is quite
insensitive to the choices $\text{N}^2\text{LO}$ or~$\text{N}^3\text{LO}$,
and~$\mu=\mu_0(t)$ or~$\mu=\mu_d(t)$. Similarly to the entropy density above,
we use the linear function of~$t$ for the $t\to0$ extrapolation. We also
use the linear function of~$[g(\mu)^2/(4\pi)^2]^\ell$, where $\ell=2$ for the
$\text{N}^2\text{LO}$ approximation and $\ell=3$ for the $\text{N}^3\text{LO}$
approximation, as suggested from the study of the asymptotic $t\to0$ behavior
of~Eq.~\eqref{eq:(3.5)} (H.~Suzuki and H.~Takaura, manuscript in preparation).
The difference caused by this is treated as a systematic error. The results are
summarized in~Table~\ref{table:2} and~Fig.~\ref{fig:4}. All the results are
almost degenerate as seen from~Fig.~\ref{fig:4}. However, it is worth noting
that the use of the $\text{N}^3\text{LO}$ coefficient certainly reduces the
dependences on the choice of the renormalization scale and the fit function, as
seen from the fourth and the fifth parentheses in~Table~\ref{table:2}.

\section{Conclusions}
\label{sec:4}
We investigated the thermodynamics in quenched QCD using the gradient-flow
representation of the EMT. In particular, we studied the effect of the
$\text{N}^2\text{LO}$ coefficients in the gradient-flow formalism, which have
become available recently. For the trace anomaly, we used the
$\text{N}^3\text{LO}$ coefficient, which was obtained in this paper for
quenched QCD. It turned out that the use of the $\text{N}^2\text{LO}$
(or~$\text{N}^3\text{LO}$) coefficients considerably reduces the systematic
errors, especially concerning the choice of the renormalization scale and the
$t\to0$ extrapolation. We expect that the use of the $\text{N}^2\text{LO}$
coefficients will also make precise studies possible in full QCD, which has
been investigated with the NLO coefficients so far.

\section*{Acknowledgements}
We would like to thank Robert V. Harlander, Kazuyuki Kanaya, Yusuke Taniguchi,
and Ryosuke Yanagihara for helpful discussions.
This work was supported by JSPS Grants-in-Aid for Scientific Research, Grant
Numbers JP17K05442 (M.K.) and JP16H03982 (H.S.).
Numerical simulation was carried out on the IBM System Blue Gene Solution at
KEK under its Large-Scale Simulation Program (No.~16/17-07).

\appendix

\section{Numerical results (continued)}
\label{app:A}
\begin{figure}[tbhp]
\centering
\begin{subfigure}{0.35\columnwidth}
\centering
\includegraphics[width=\columnwidth]{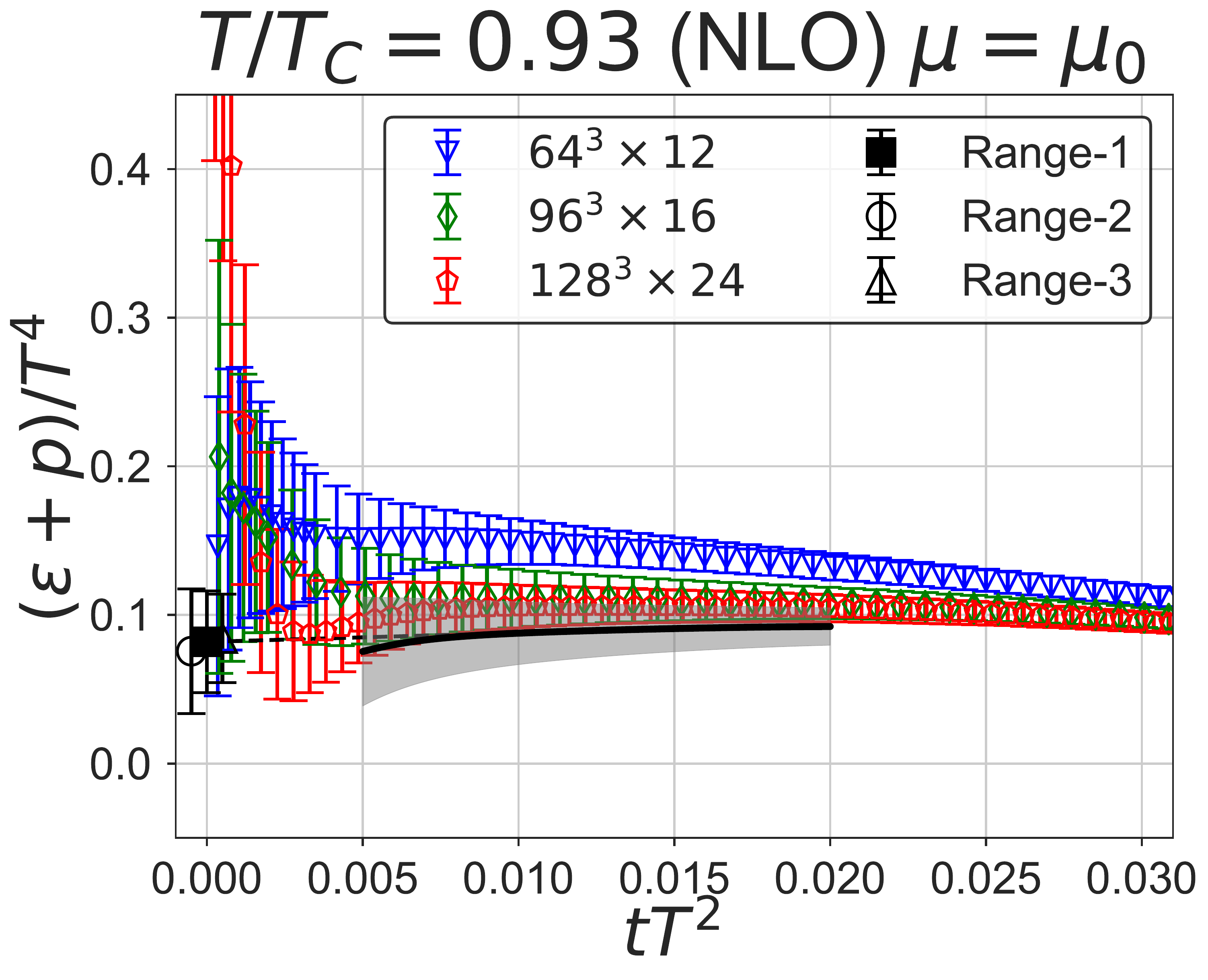}
\caption{}
\label{}
\end{subfigure}
\hspace{10mm}
\begin{subfigure}{0.35\columnwidth}
\centering
\includegraphics[width=\columnwidth]{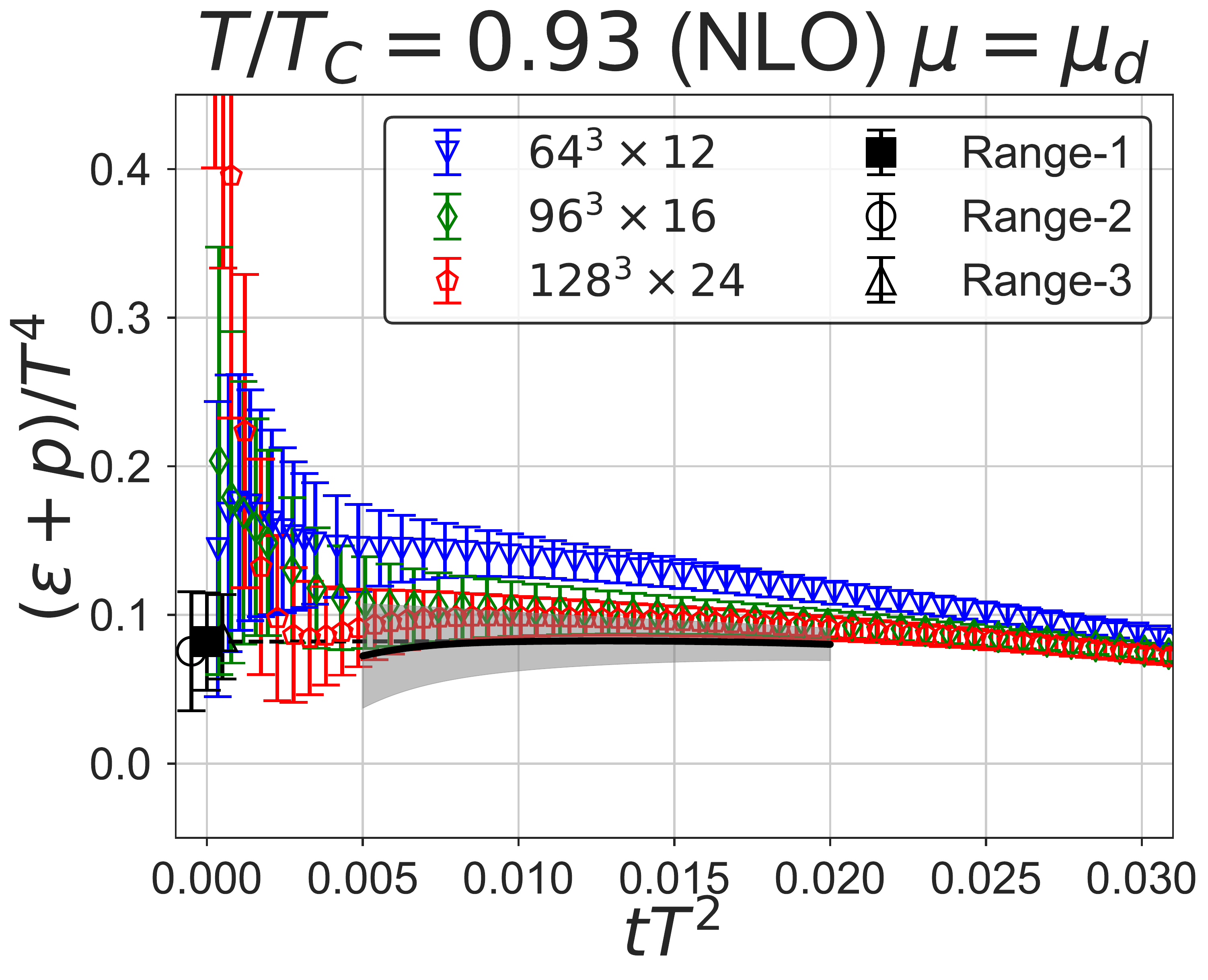}
\caption{}
\label{}
\end{subfigure}
\begin{subfigure}{0.35\columnwidth}
\centering
\includegraphics[width=\columnwidth]{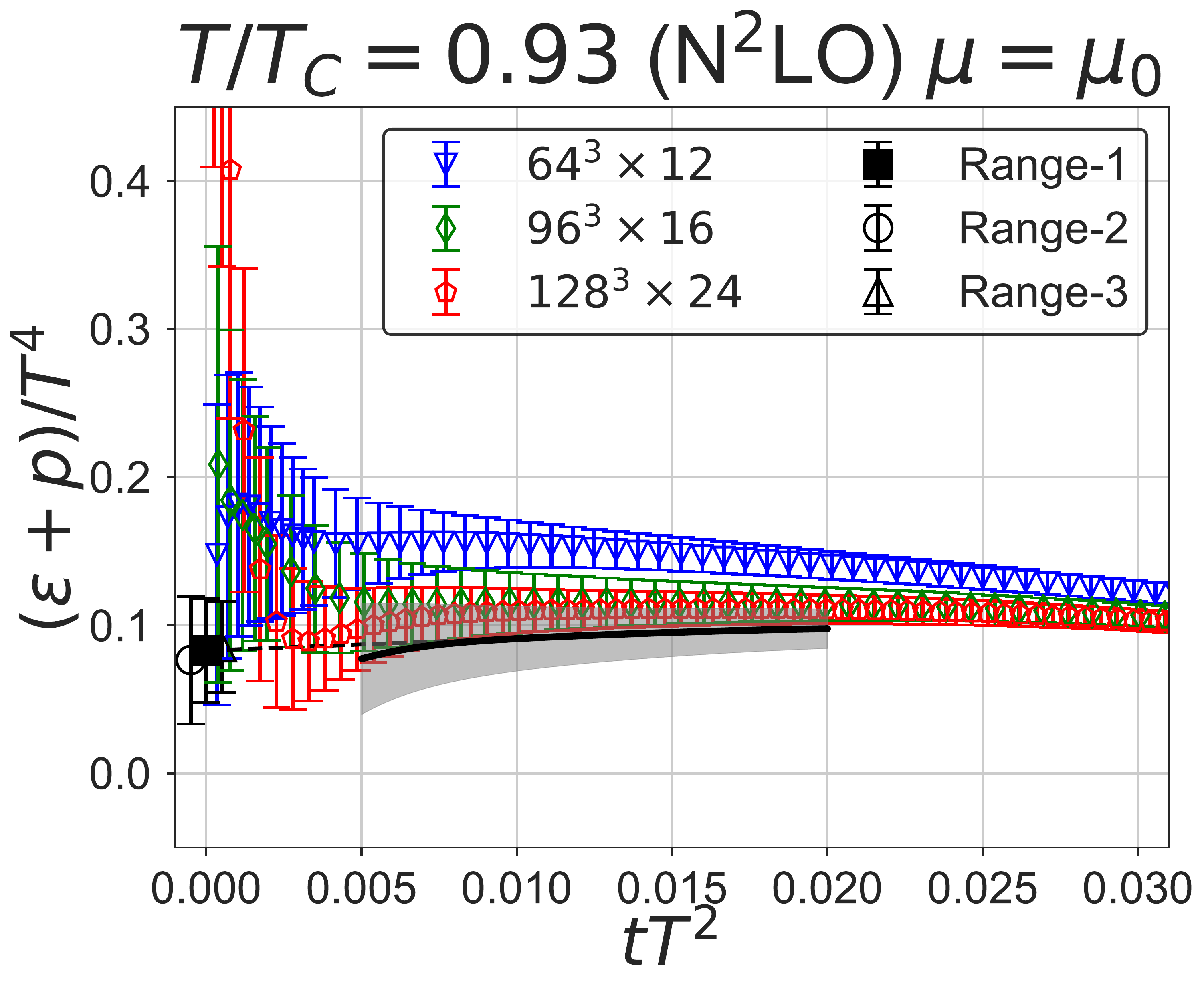}
\caption{}
\label{}
\end{subfigure}
\hspace{10mm}
\begin{subfigure}{0.35\columnwidth}
\centering
\includegraphics[width=\columnwidth]{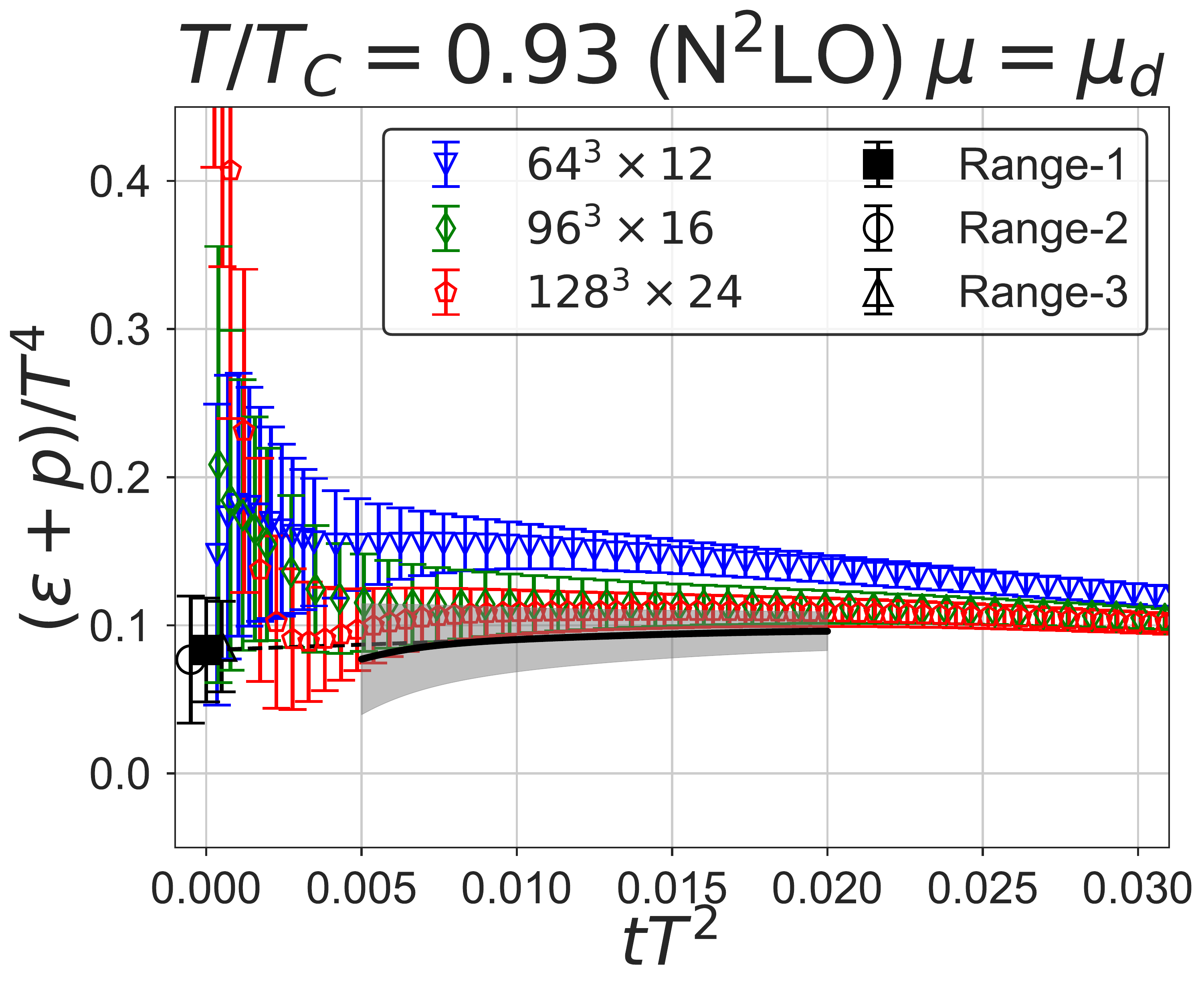}
\caption{}
\label{}
\end{subfigure}
\caption{Same as~Fig.~\ref{fig:1}. $T/T_c=0.93$.}
\label{fig:A1}
\end{figure}

\begin{figure}[]
\centering
\begin{subfigure}{0.35\columnwidth}
\centering
\includegraphics[width=\columnwidth]{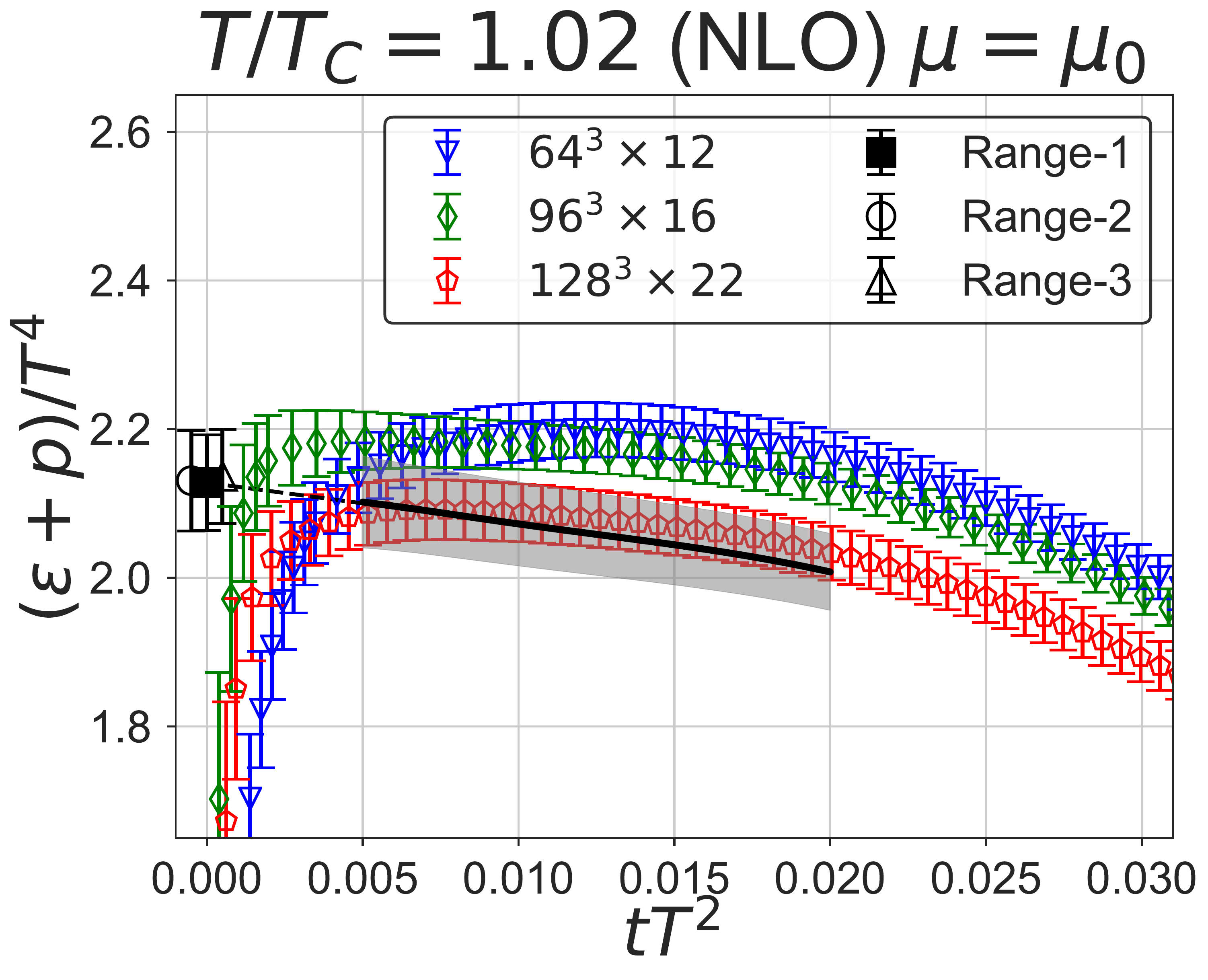}
\caption{}
\label{}
\end{subfigure}
\hspace{10mm}
\begin{subfigure}{0.35\columnwidth}
\centering
\includegraphics[width=\columnwidth]{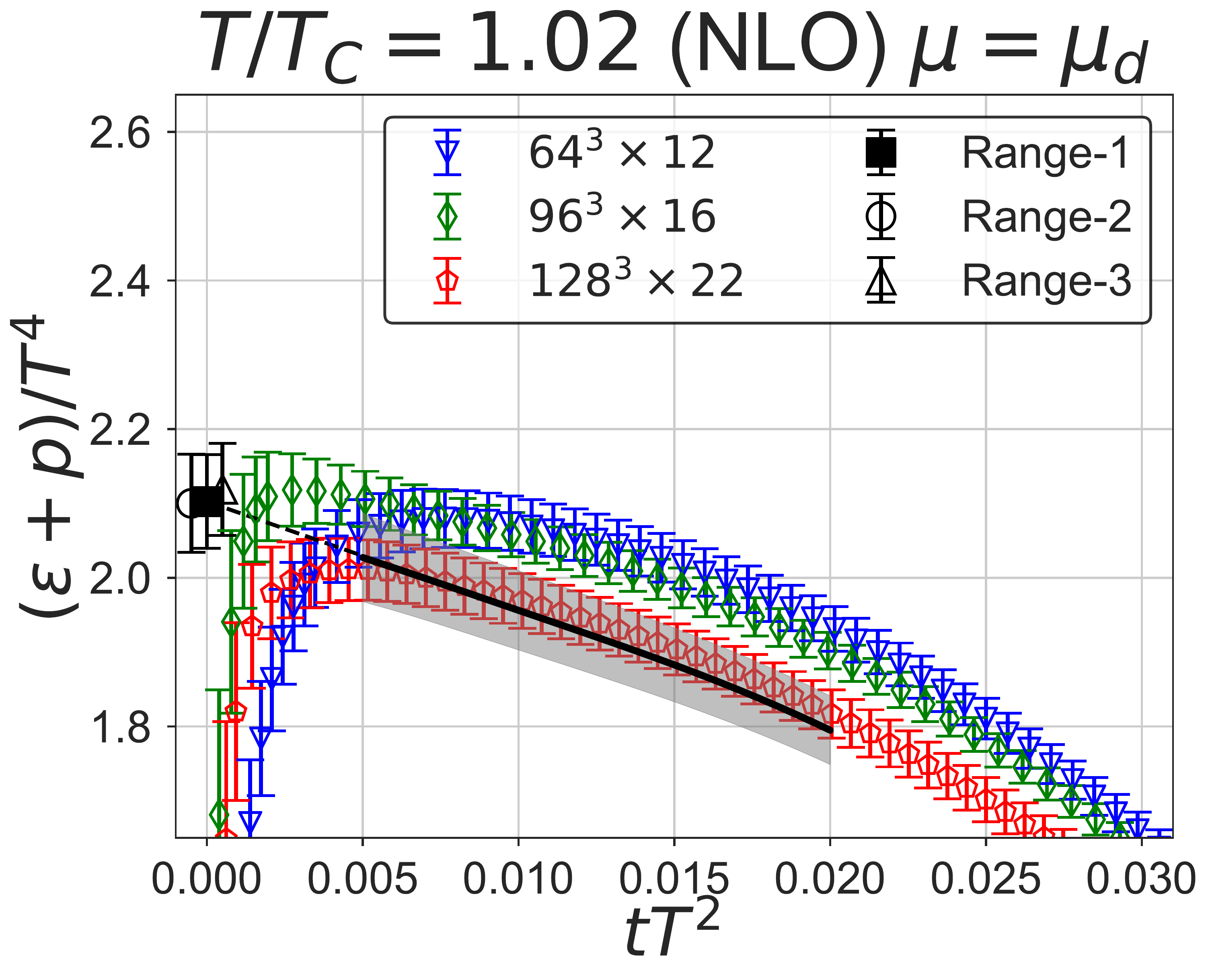}
\caption{}
\label{}
\end{subfigure}
\begin{subfigure}{0.35\columnwidth}
\centering
\includegraphics[width=\columnwidth]{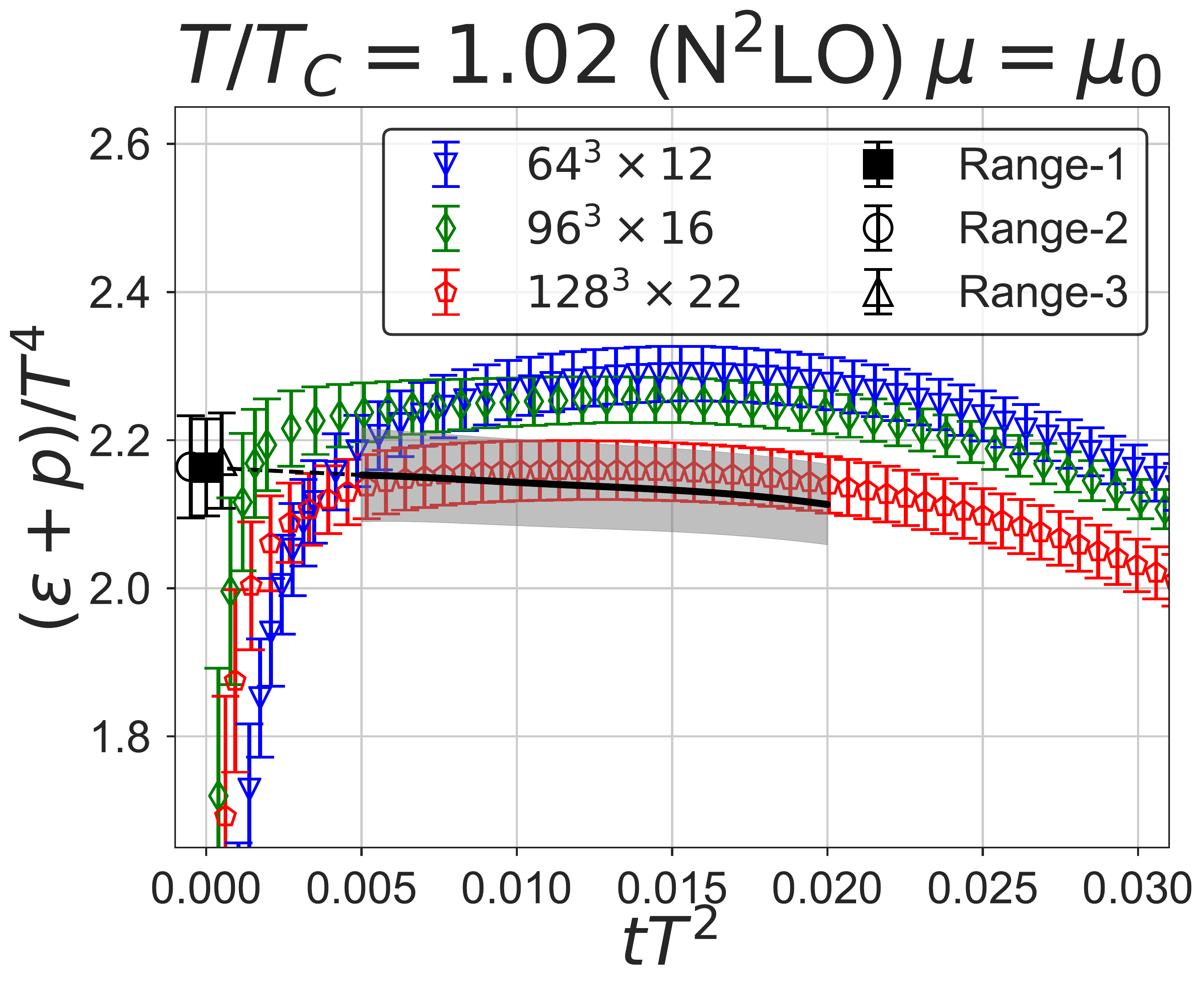}
\caption{}
\label{}
\end{subfigure}
\hspace{10mm}
\begin{subfigure}{0.35\columnwidth}
\centering
\includegraphics[width=\columnwidth]{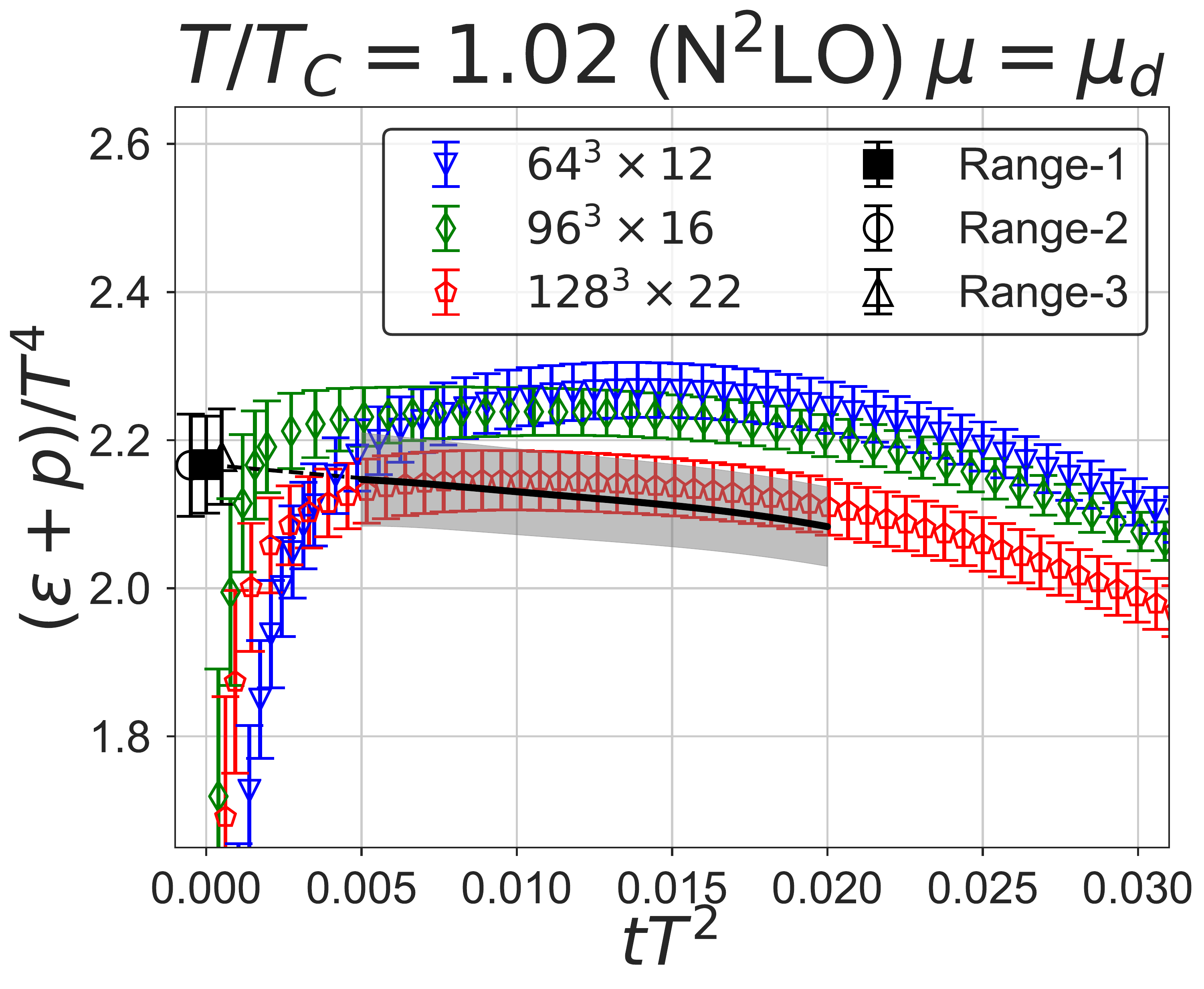}
\caption{}
\label{}
\end{subfigure}
\caption{Same as~Fig.~\ref{fig:1}. $T/T_c=1.02$.}
\label{fig:A2}
\end{figure}

\begin{figure}[htbp]
\centering
\begin{subfigure}{0.35\columnwidth}
\centering
\includegraphics[width=\columnwidth]{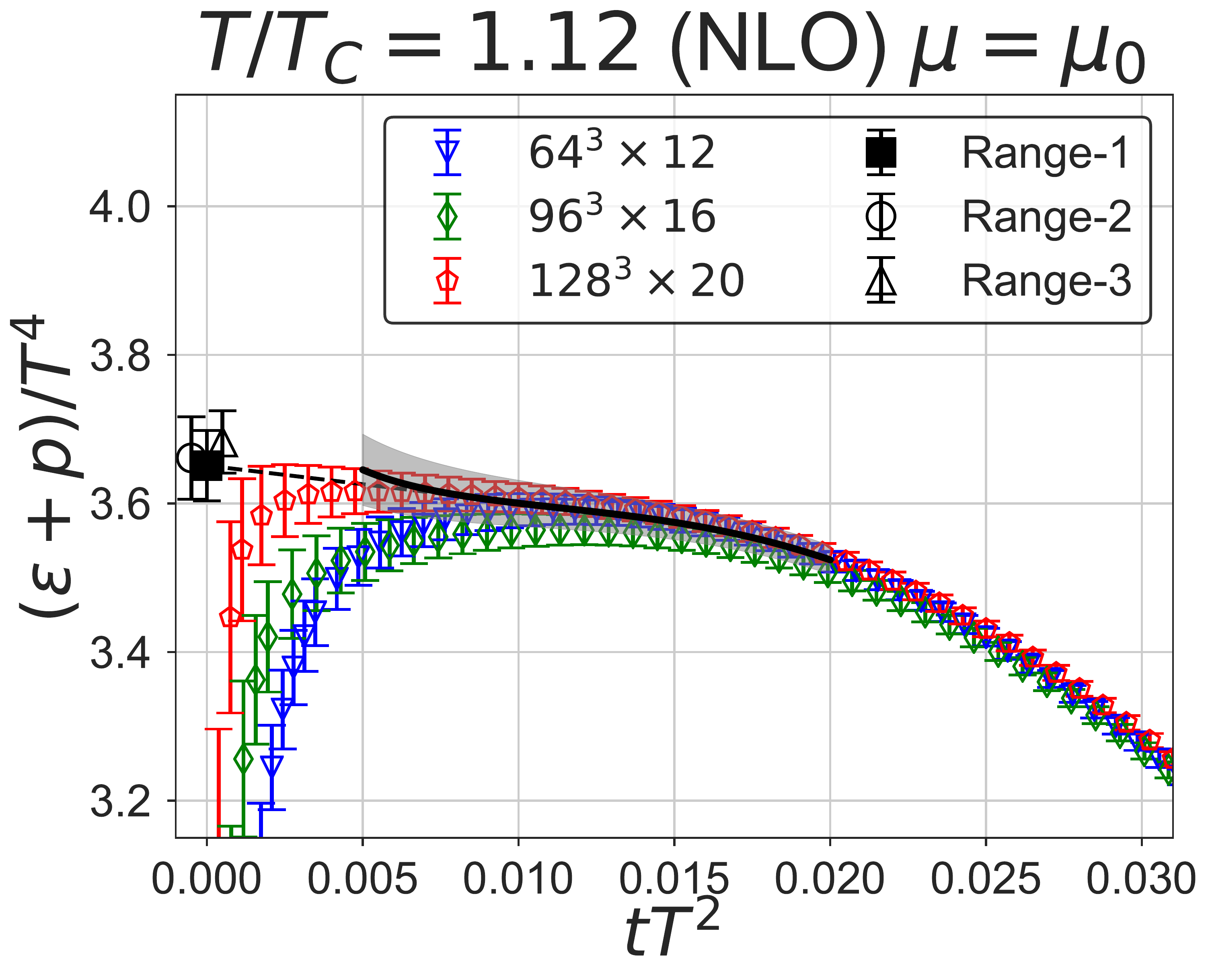}
\caption{}
\label{}
\end{subfigure}
\hspace{10mm}
\begin{subfigure}{0.35\columnwidth}
\centering
\includegraphics[width=\columnwidth]{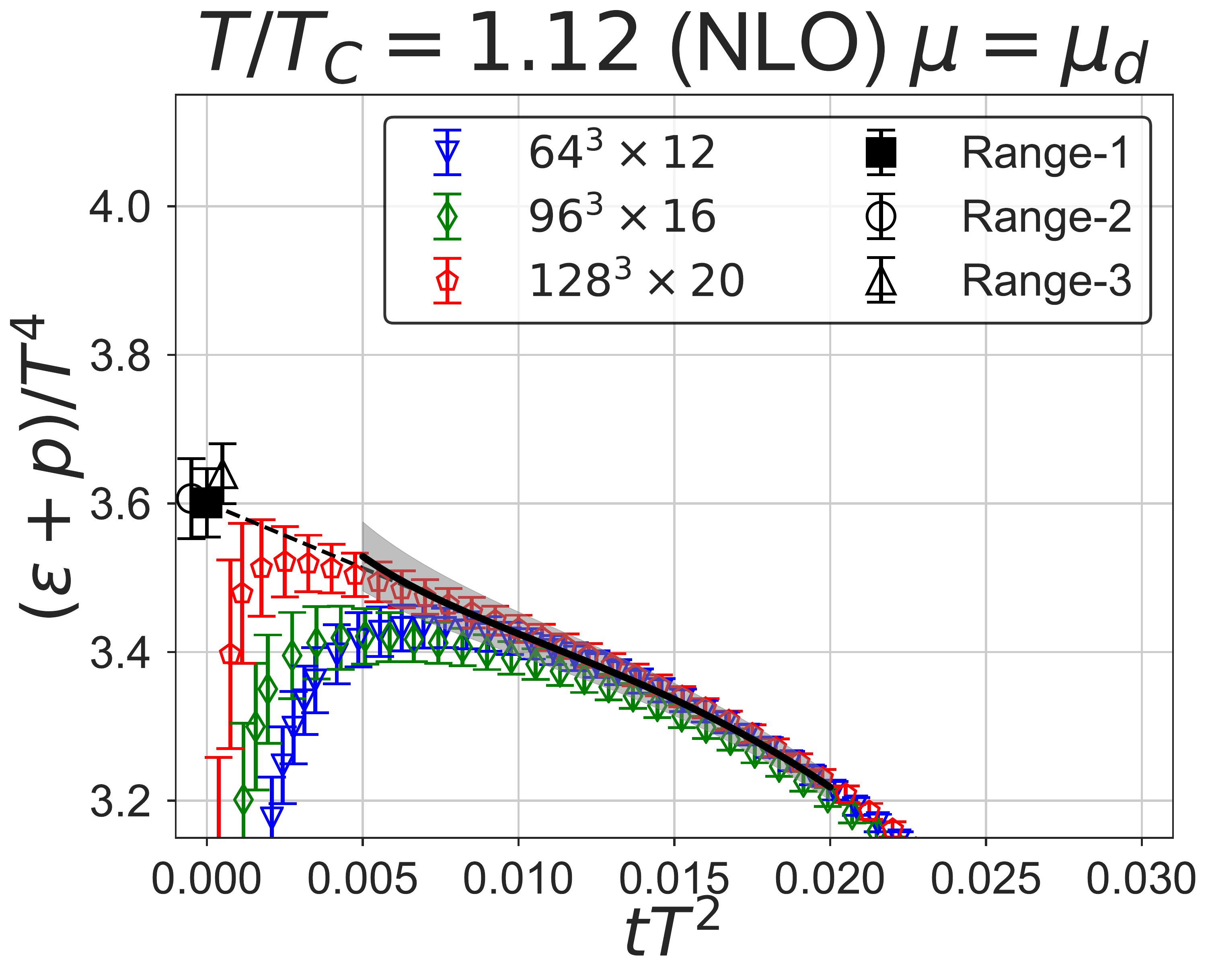}
\caption{}
\label{}
\end{subfigure}
\begin{subfigure}{0.35\columnwidth}
\centering
\includegraphics[width=\columnwidth]{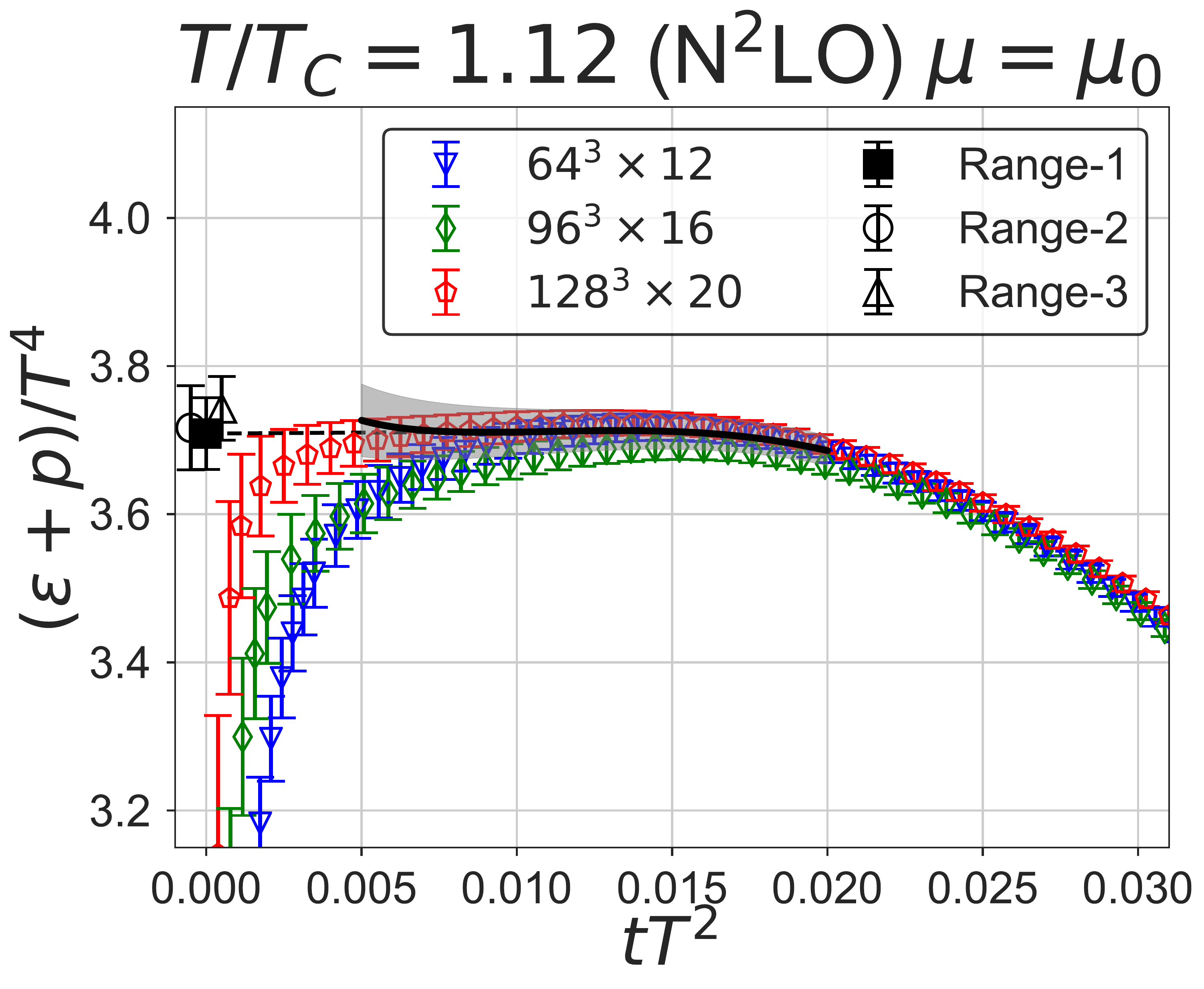}
\caption{}
\label{}
\end{subfigure}
\hspace{10mm}
\begin{subfigure}{0.35\columnwidth}
\centering
\includegraphics[width=\columnwidth]{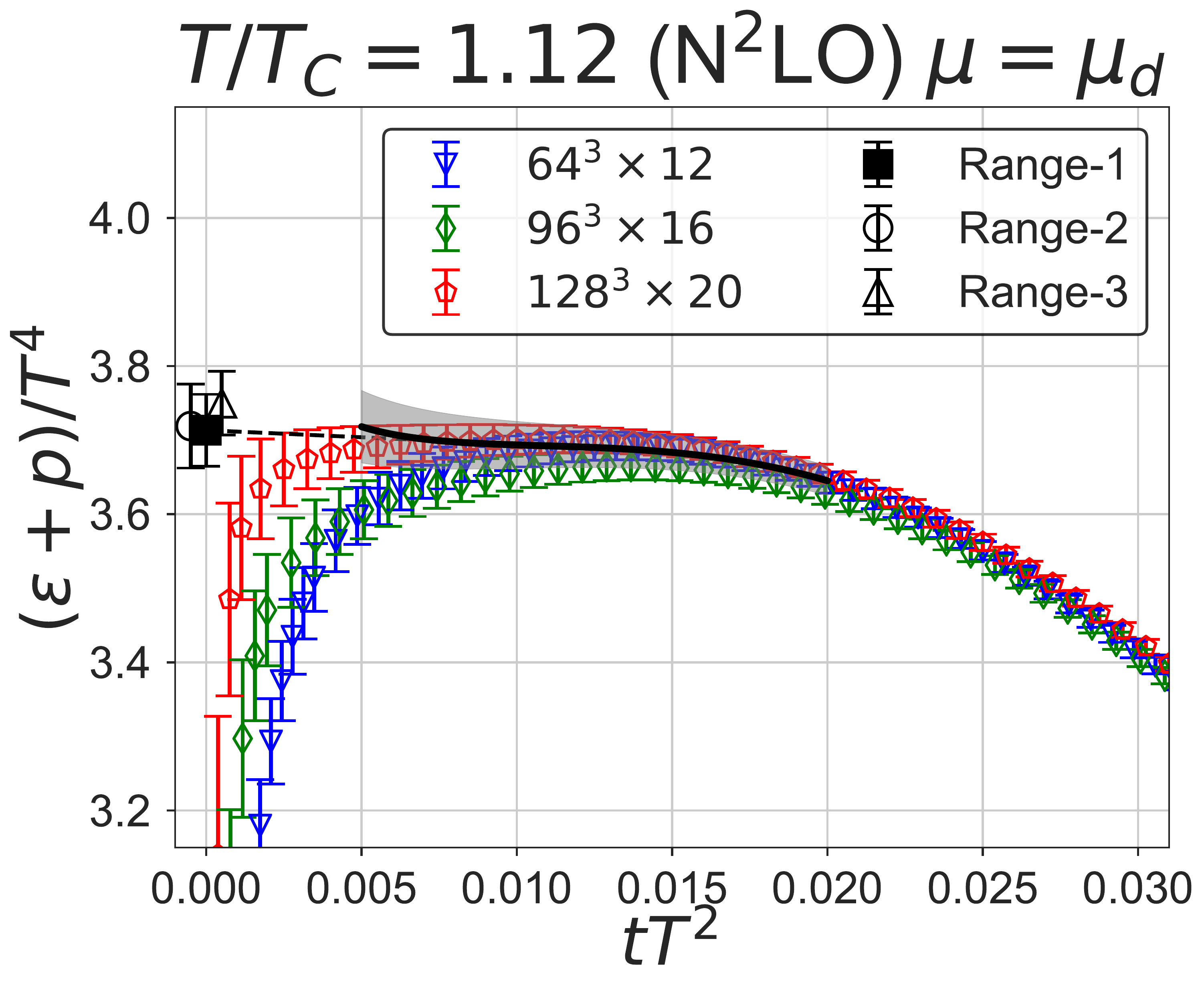}
\caption{}
\label{}
\end{subfigure}
\caption{Same as~Fig.~\ref{fig:1}. $T/T_c=1.12$.}
\label{fig:A3}
\end{figure}

\begin{figure}[htbp]
\centering
\begin{subfigure}{0.35\columnwidth}
\centering
\includegraphics[width=\columnwidth]{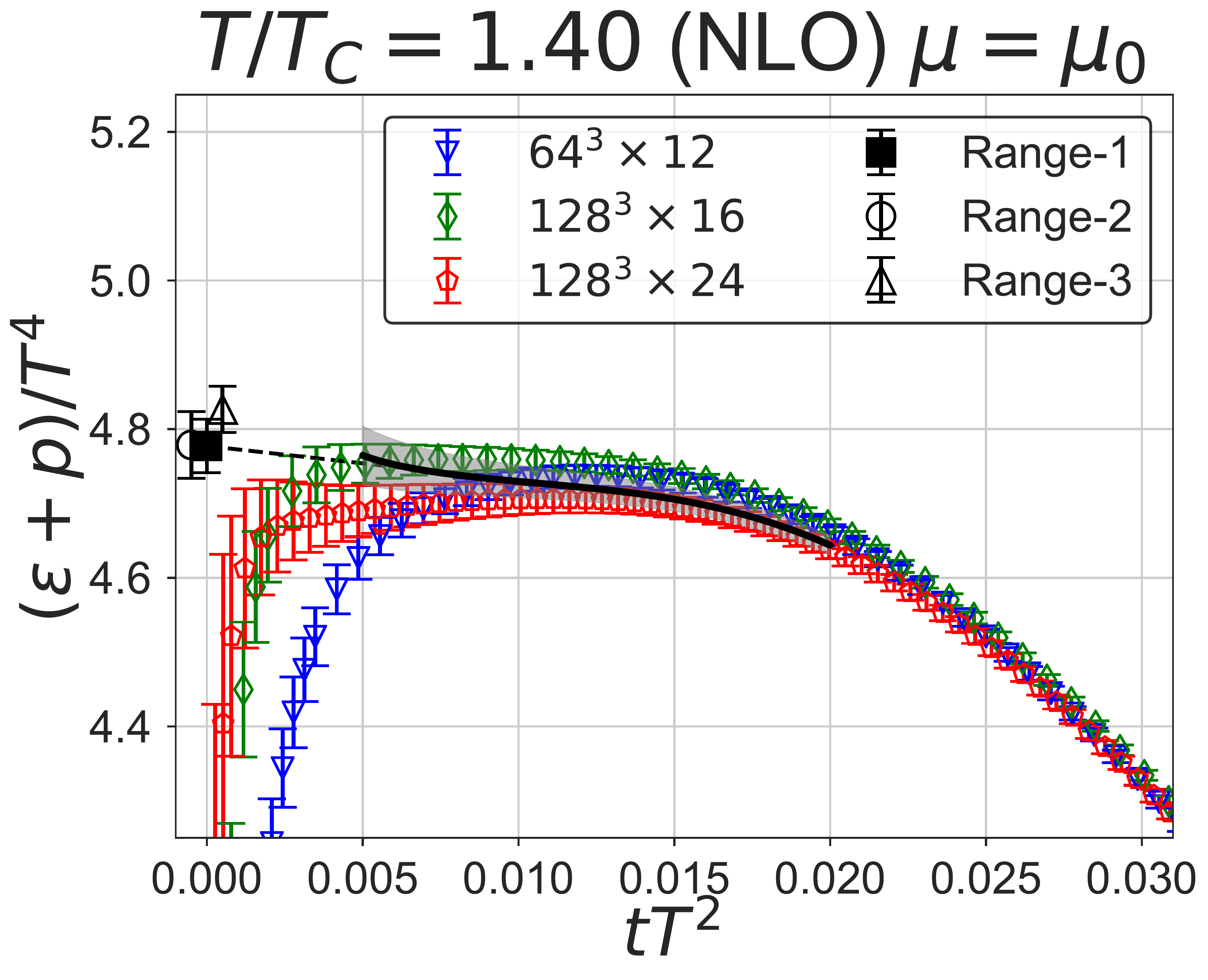}
\caption{}
\label{}
\end{subfigure}
\hspace{10mm}
\begin{subfigure}{0.35\columnwidth}
\centering
\includegraphics[width=\columnwidth]{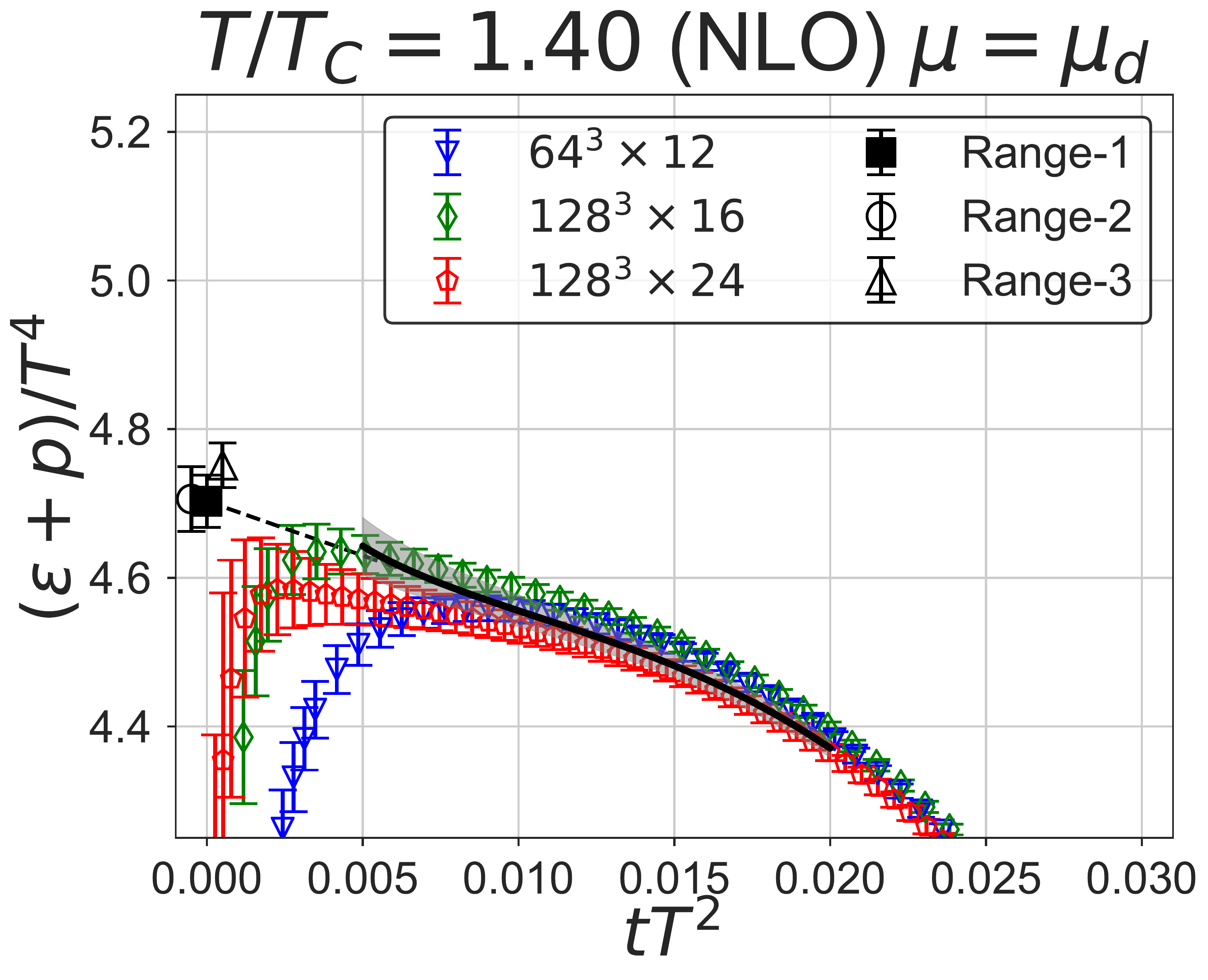}
\caption{}
\label{}
\end{subfigure}
\begin{subfigure}{0.35\columnwidth}
\centering
\includegraphics[width=\columnwidth]{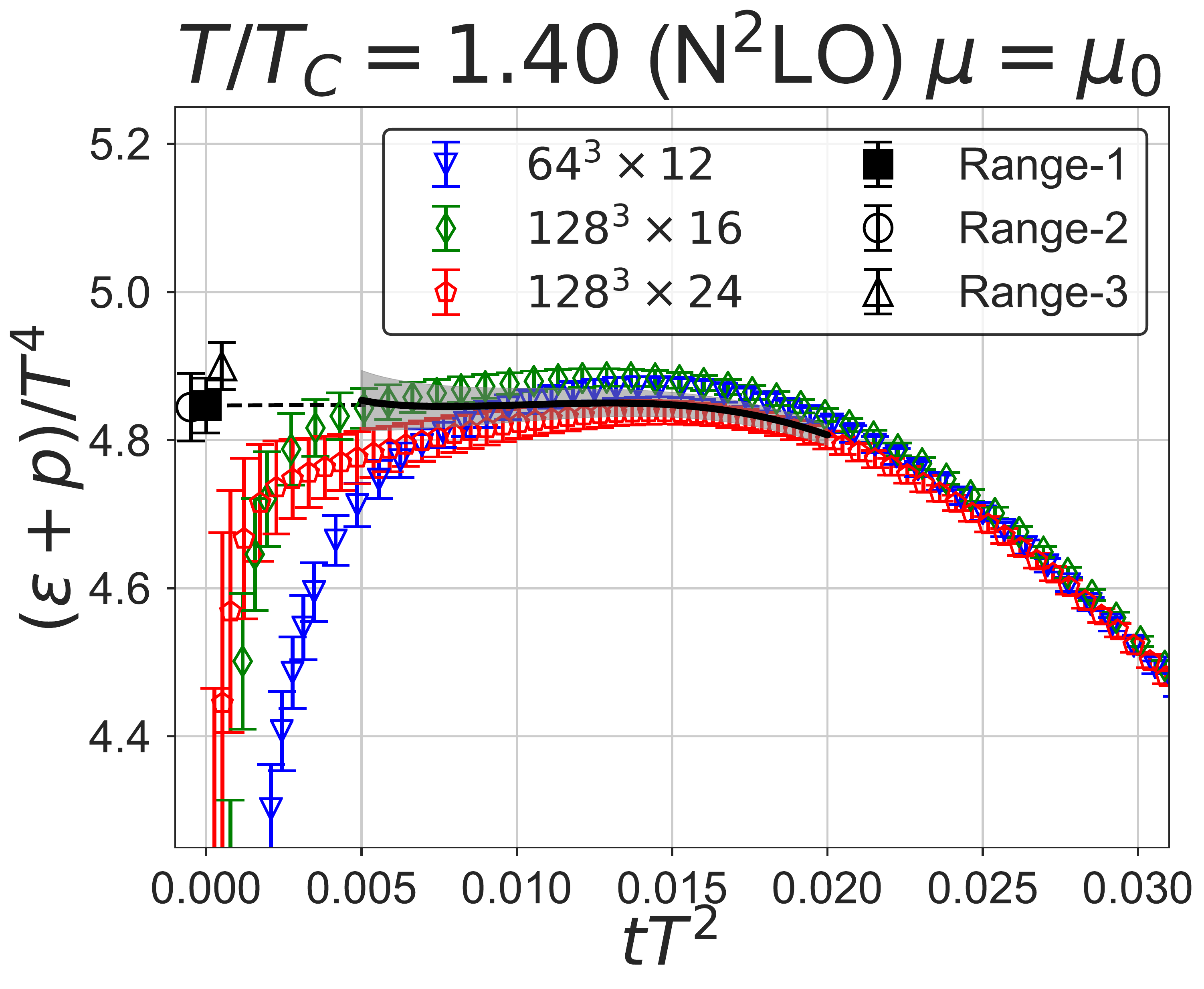}
\caption{}
\label{}
\end{subfigure}
\hspace{10mm}
\begin{subfigure}{0.35\columnwidth}
\centering
\includegraphics[width=\columnwidth]{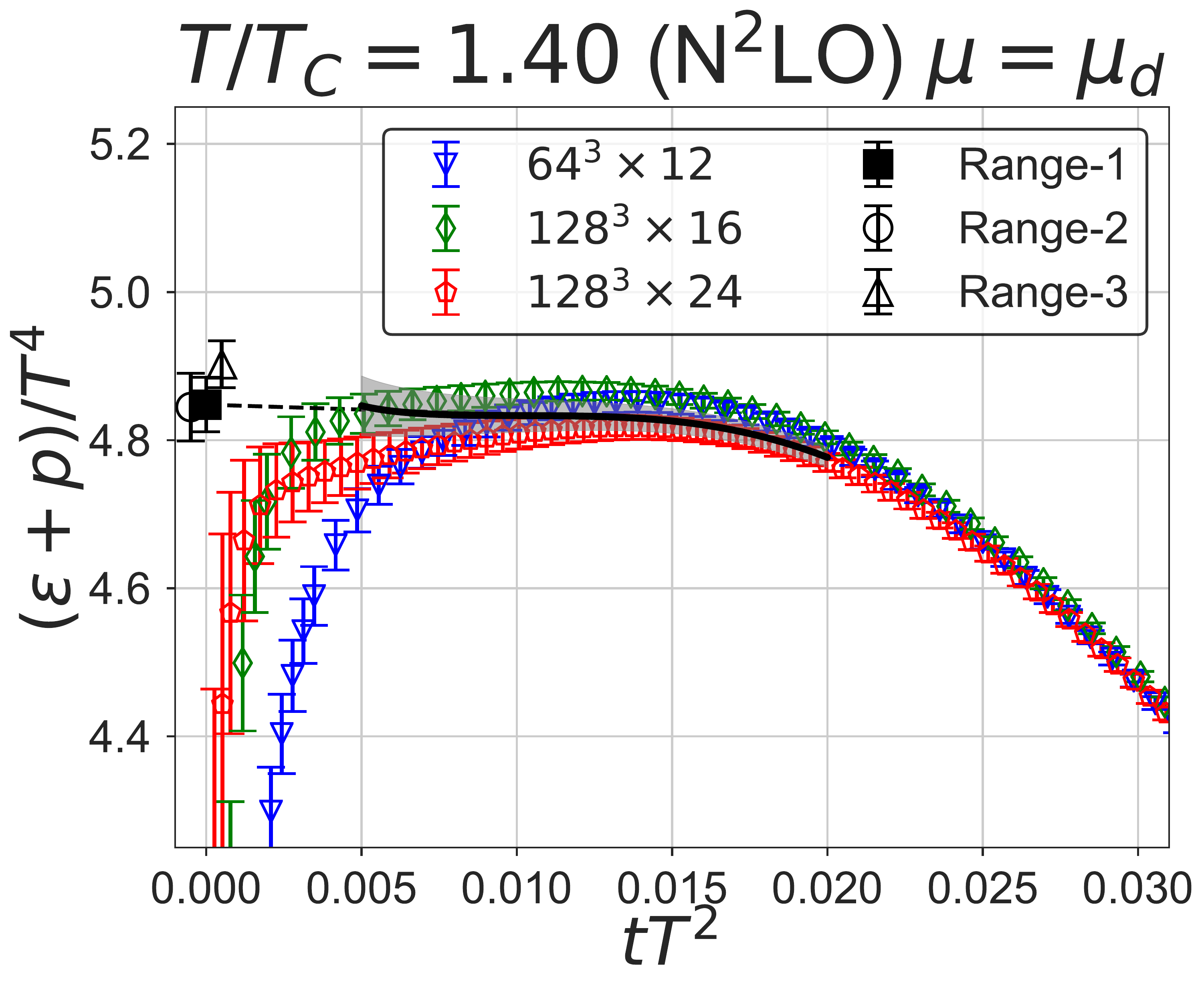}
\caption{}
\label{}
\end{subfigure}
\caption{Same as~Fig.~\ref{fig:1}. $T/T_c=1.40$.}
\label{fig:A4}
\end{figure}

\begin{figure}[htbp]
\centering
\begin{subfigure}{0.35\columnwidth}
\centering
\includegraphics[width=\columnwidth]{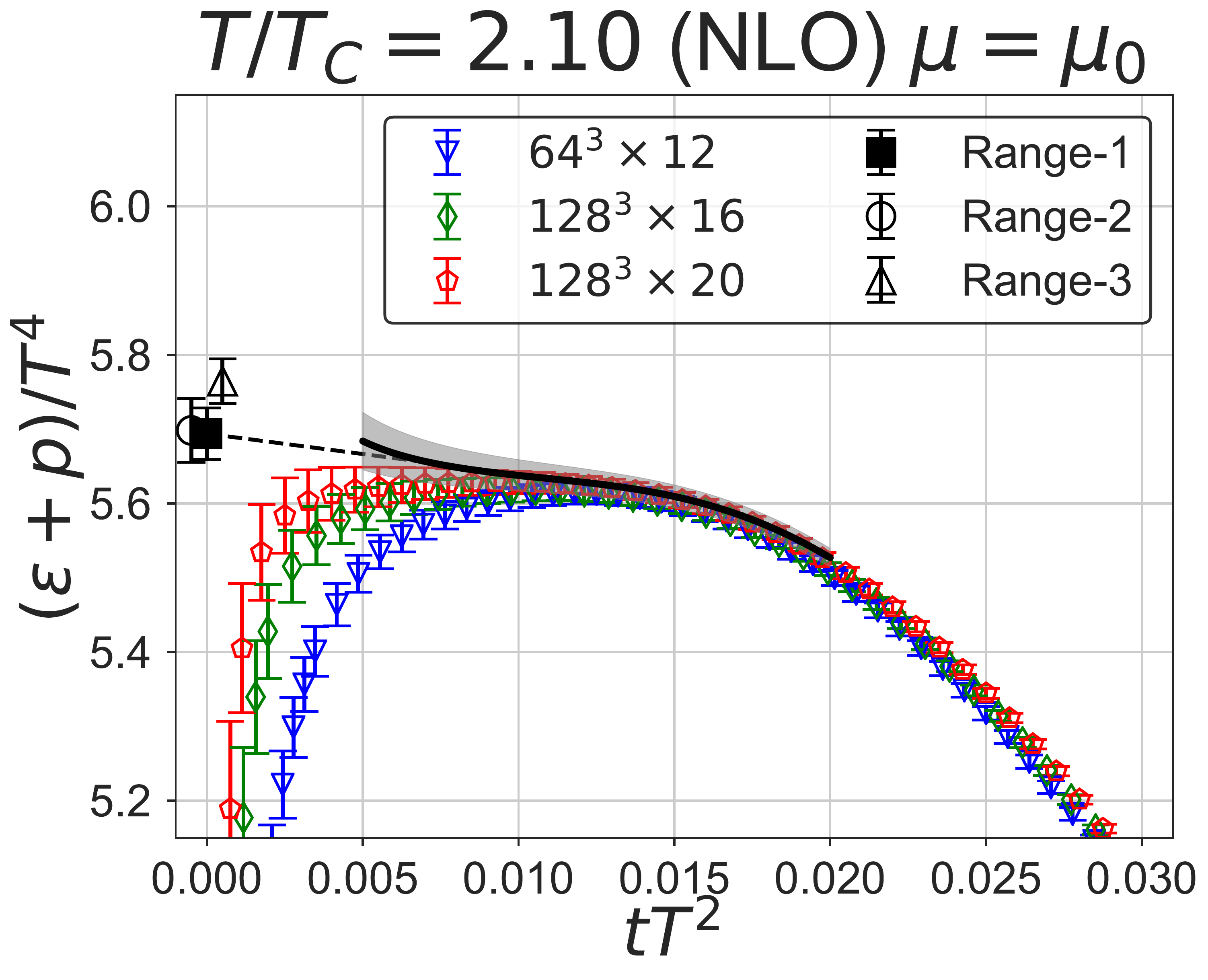}
\caption{}
\label{}
\end{subfigure}
\hspace{10mm}
\begin{subfigure}{0.35\columnwidth}
\centering
\includegraphics[width=\columnwidth]{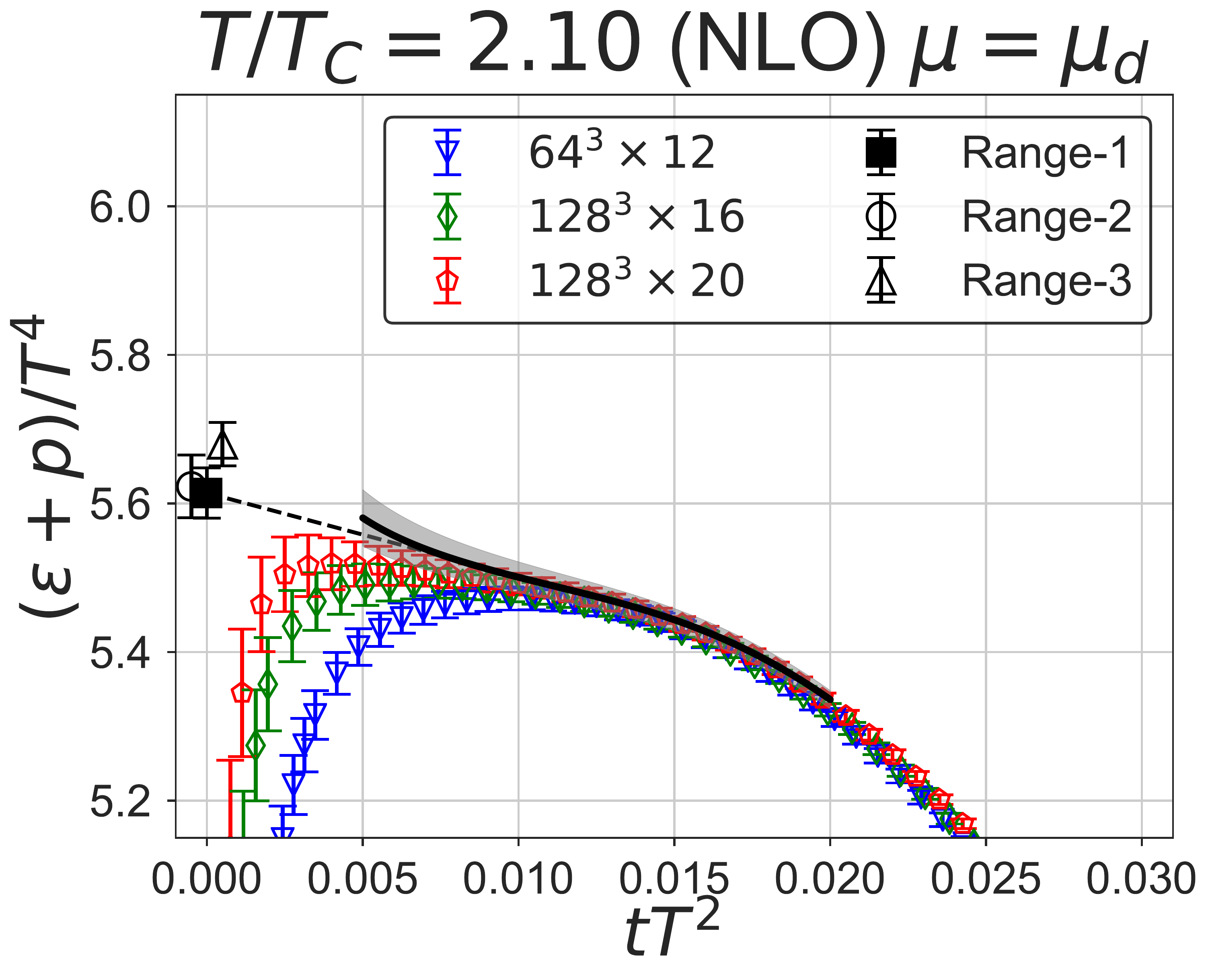}
\caption{}
\label{}
\end{subfigure}
\begin{subfigure}{0.35\columnwidth}
\centering
\includegraphics[width=\columnwidth]{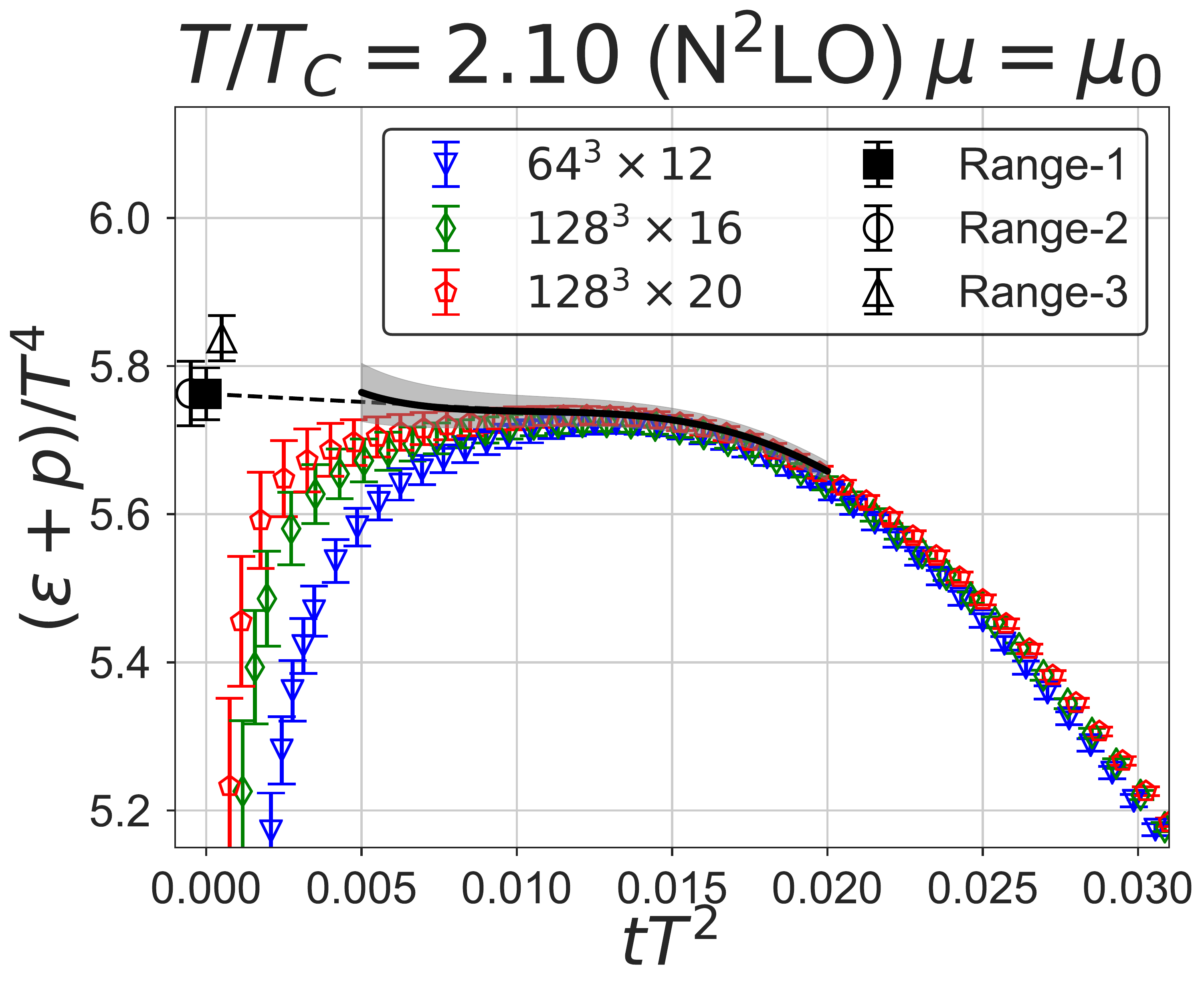}
\caption{}
\label{}
\end{subfigure}
\hspace{10mm}
\begin{subfigure}{0.35\columnwidth}
\centering
\includegraphics[width=\columnwidth]{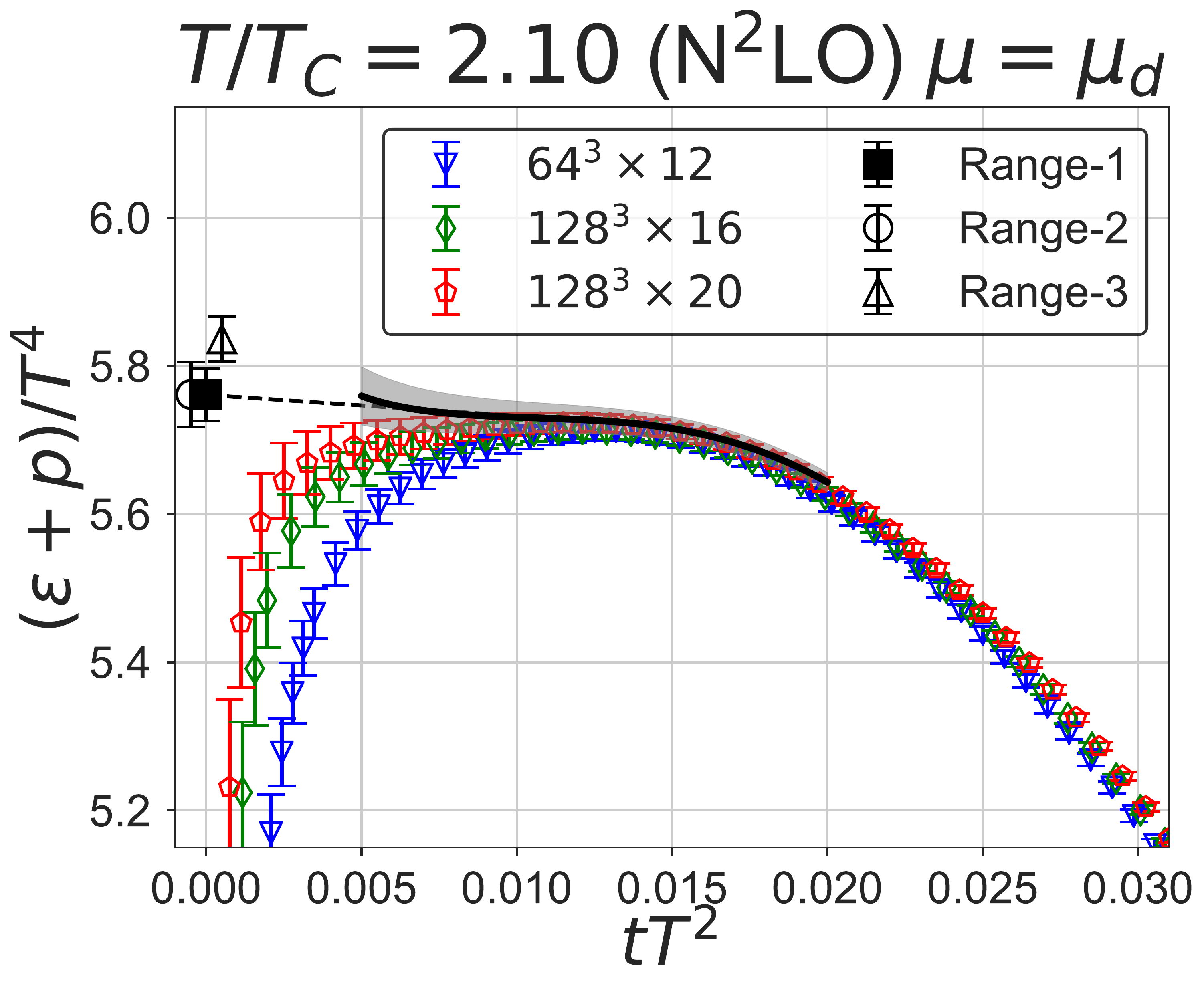}
\caption{}
\label{}
\end{subfigure}
\caption{Same as~Fig.~\ref{fig:1}. $T/T_c=2.10$.}
\label{fig:A5}
\end{figure}

\begin{figure}[htbp]
\centering
\begin{subfigure}{0.35\columnwidth}
\centering
\includegraphics[width=\columnwidth]{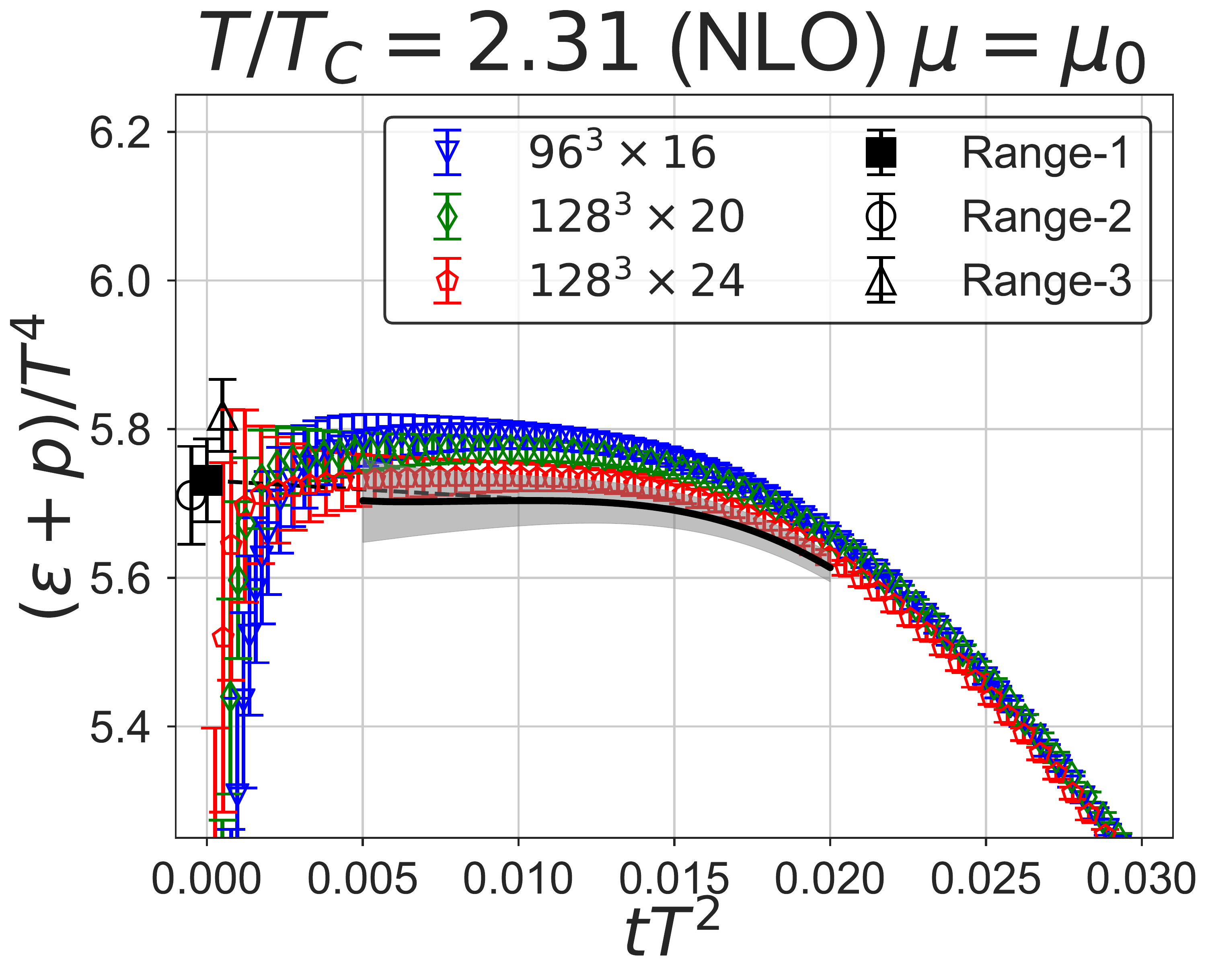}
\caption{}
\label{}
\end{subfigure}
\hspace{10mm}
\begin{subfigure}{0.35\columnwidth}
\centering
\includegraphics[width=\columnwidth]{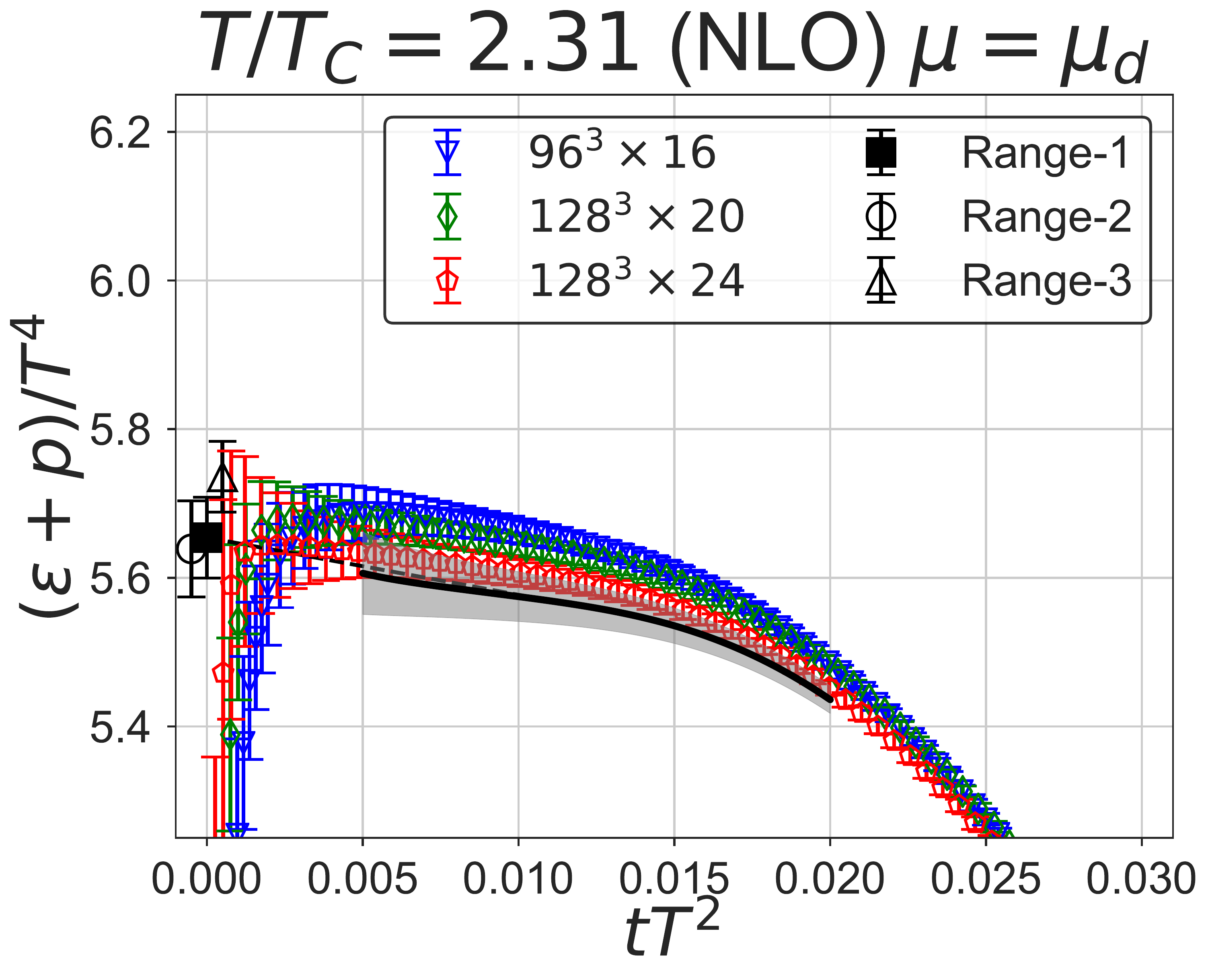}
\caption{}
\label{}
\end{subfigure}
\begin{subfigure}{0.35\columnwidth}
\centering
\includegraphics[width=\columnwidth]{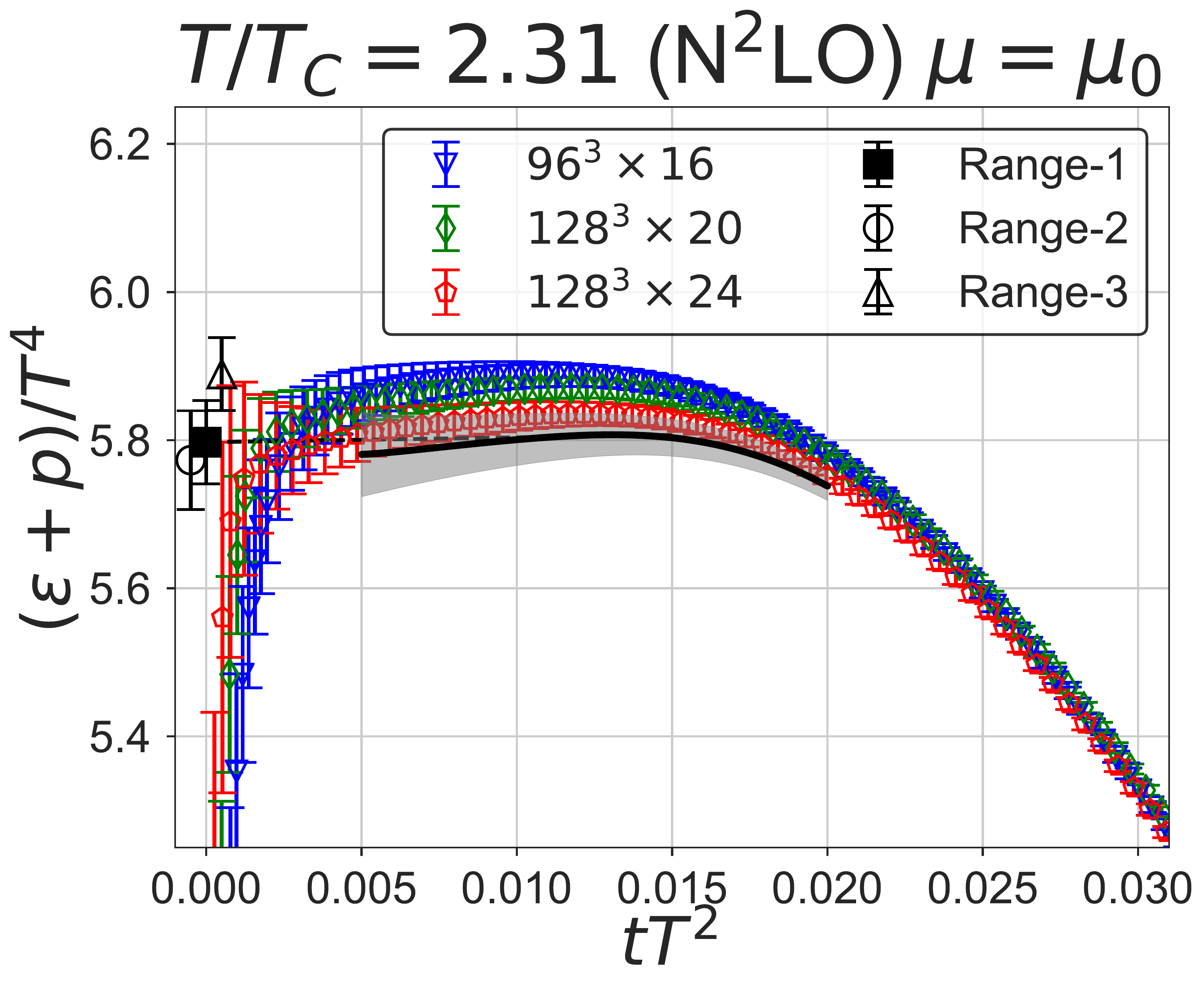}
\caption{}
\label{}
\end{subfigure}
\hspace{10mm}
\begin{subfigure}{0.35\columnwidth}
\centering
\includegraphics[width=\columnwidth]{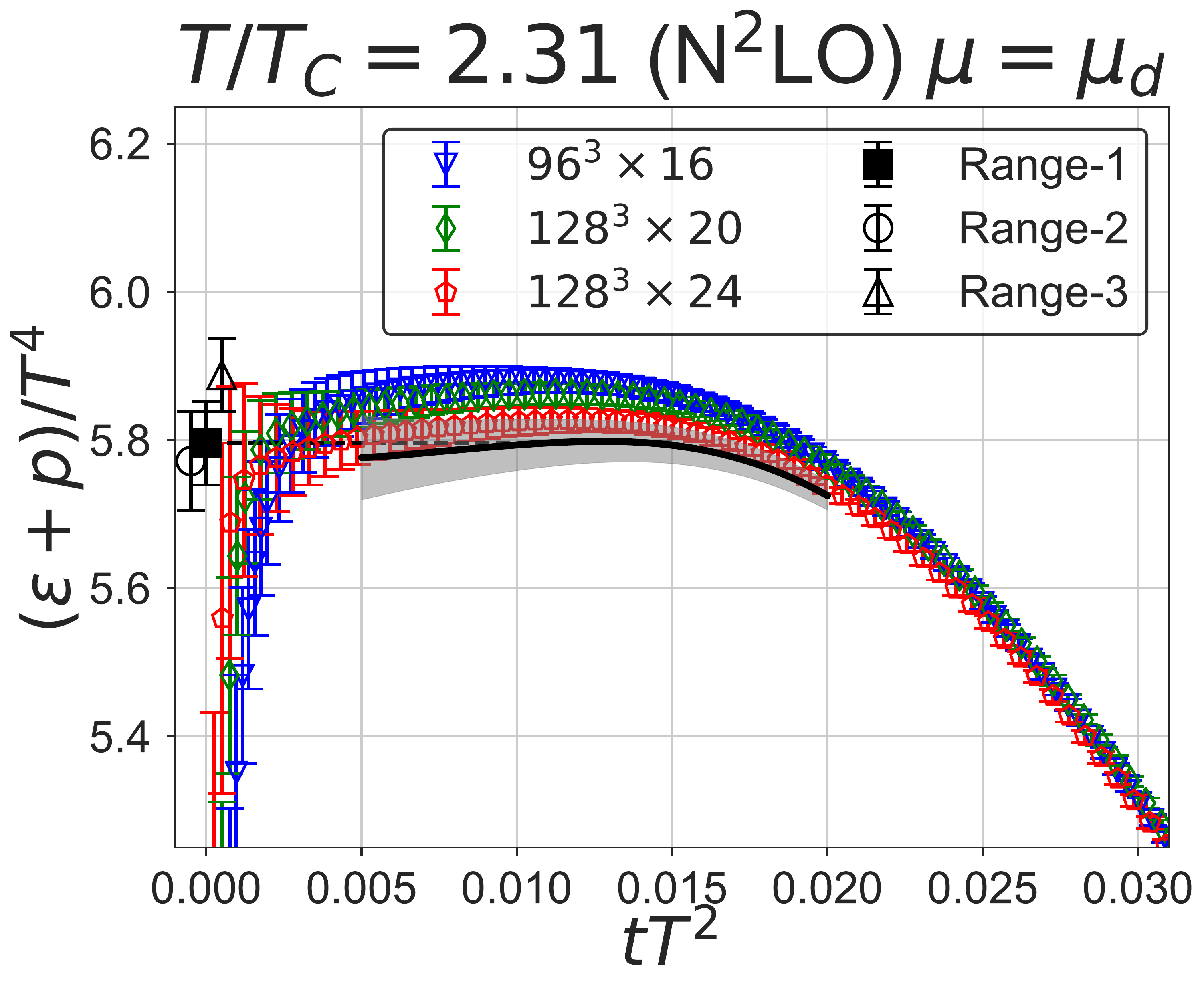}
\caption{}
\label{}
\end{subfigure}
\caption{Same as~Fig.~\ref{fig:1}. $T/T_c=2.31$.}
\label{fig:A6}
\end{figure}

\begin{figure}[htbp]
\centering
\begin{subfigure}{0.35\columnwidth}
\centering
\includegraphics[width=\columnwidth]{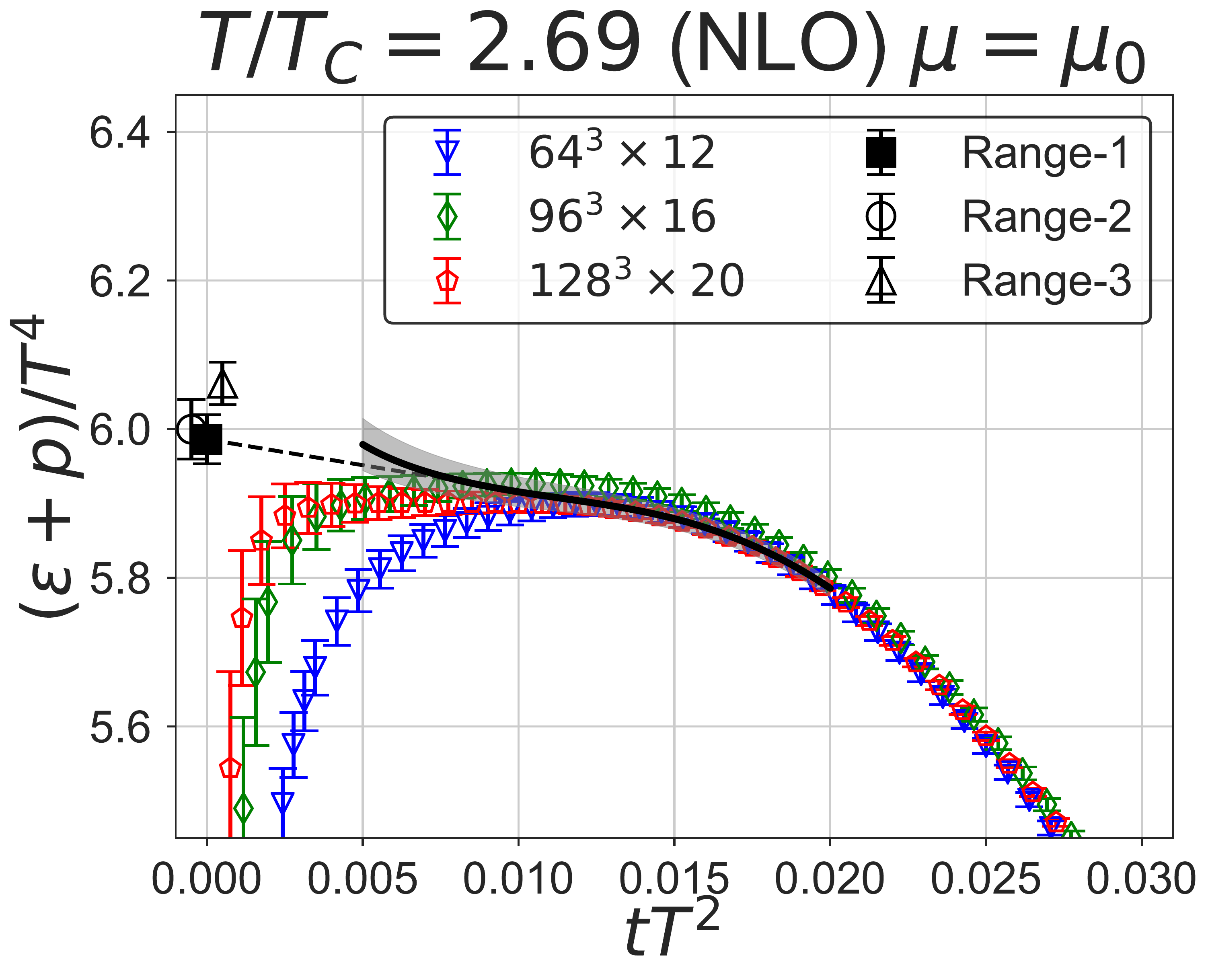}
\caption{}
\label{}
\end{subfigure}
\hspace{10mm}
\begin{subfigure}{0.35\columnwidth}
\centering
\includegraphics[width=\columnwidth]{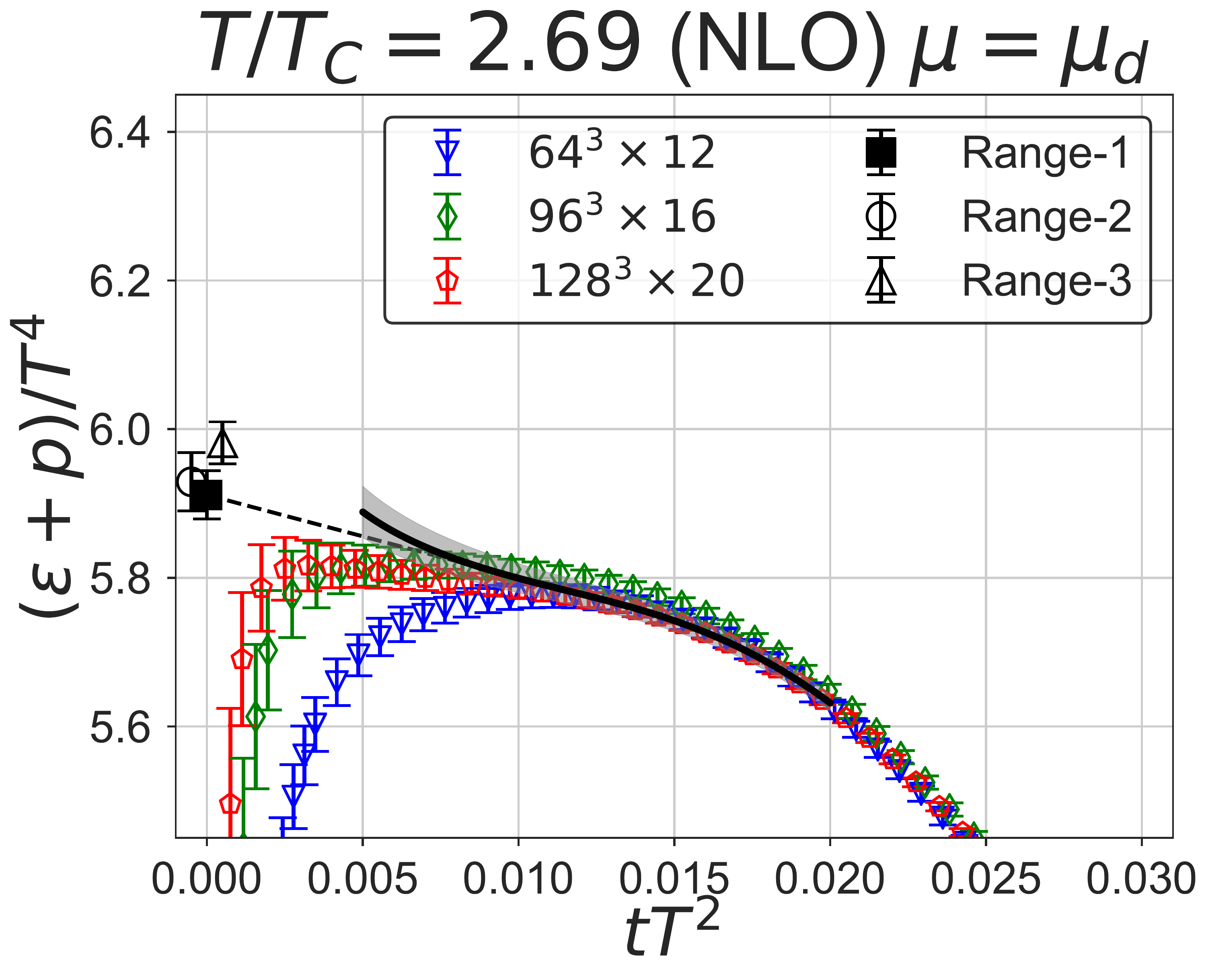}
\caption{}
\label{}
\end{subfigure}
\begin{subfigure}{0.35\columnwidth}
\centering
\includegraphics[width=\columnwidth]{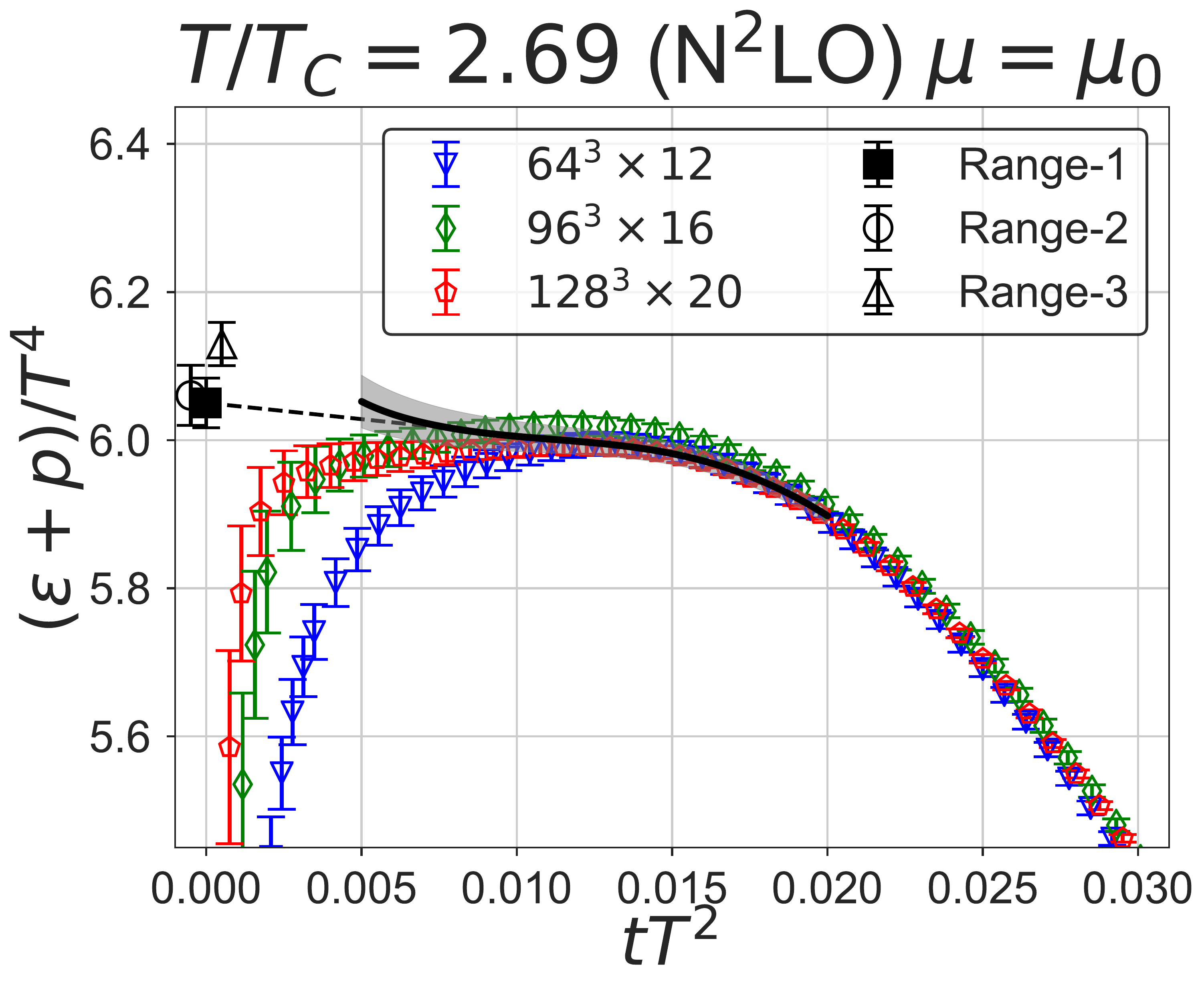}
\caption{}
\label{}
\end{subfigure}
\hspace{10mm}
\begin{subfigure}{0.35\columnwidth}
\centering
\includegraphics[width=\columnwidth]{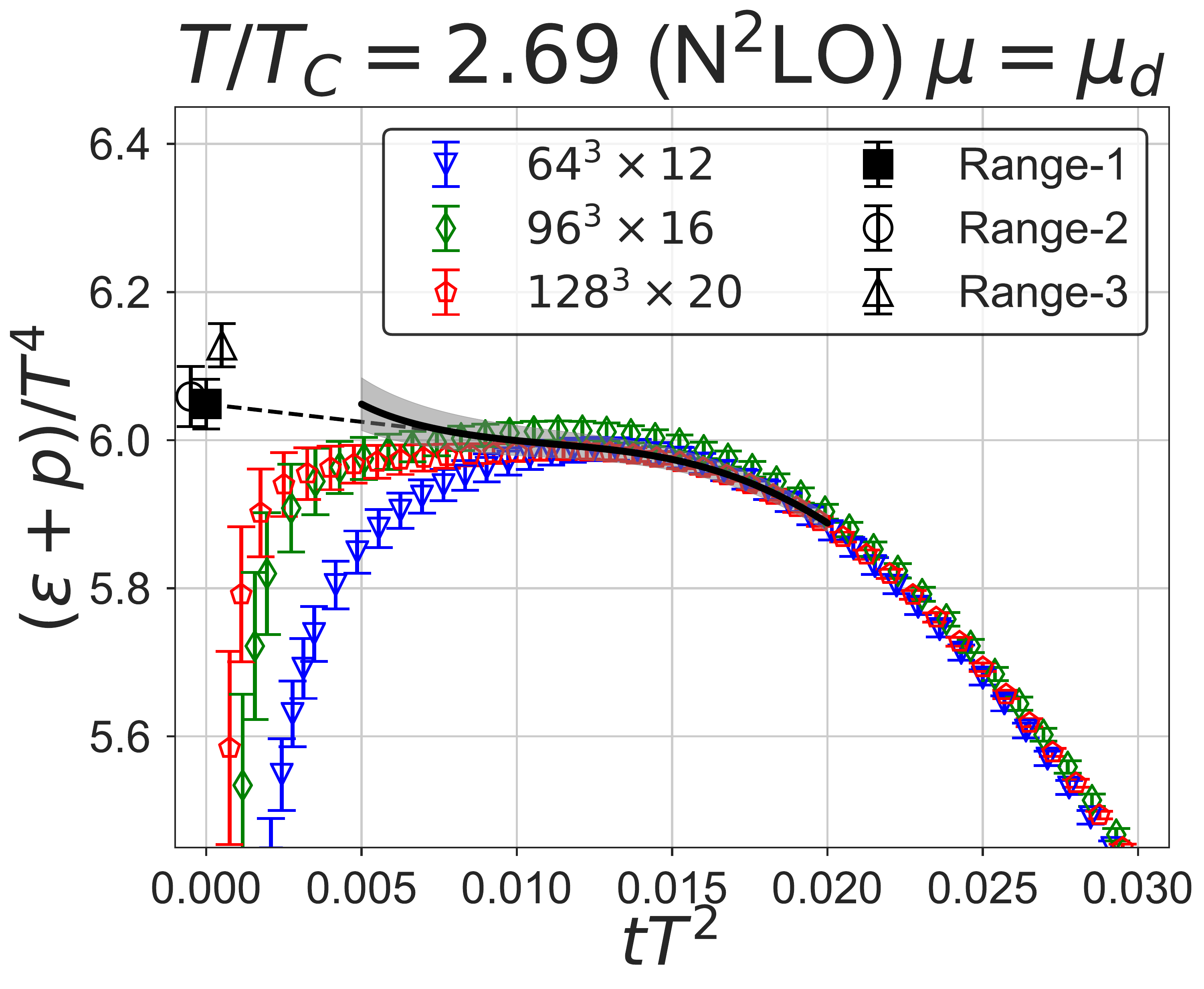}
\caption{}
\label{}
\end{subfigure}
\caption{Same as~Fig.~\ref{fig:1}. $T/T_c=2.69$.}
\label{fig:A7}
\end{figure}

\begin{figure}[htbp]
\centering
\begin{subfigure}{0.35\columnwidth}
\centering
\includegraphics[width=\columnwidth]{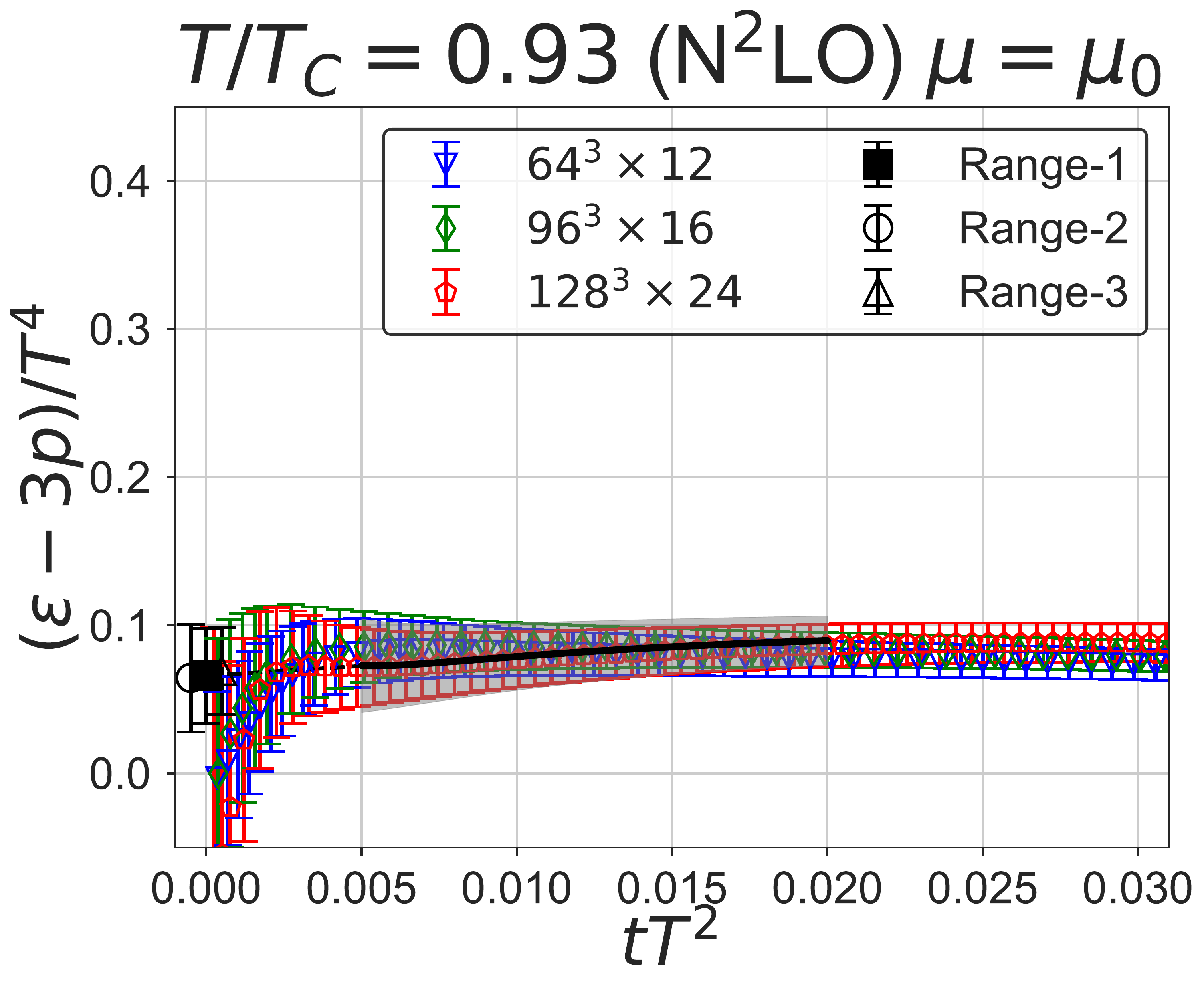}
\caption{}
\label{}
\end{subfigure}
\hspace{10mm}
\begin{subfigure}{0.35\columnwidth}
\centering
\includegraphics[width=\columnwidth]{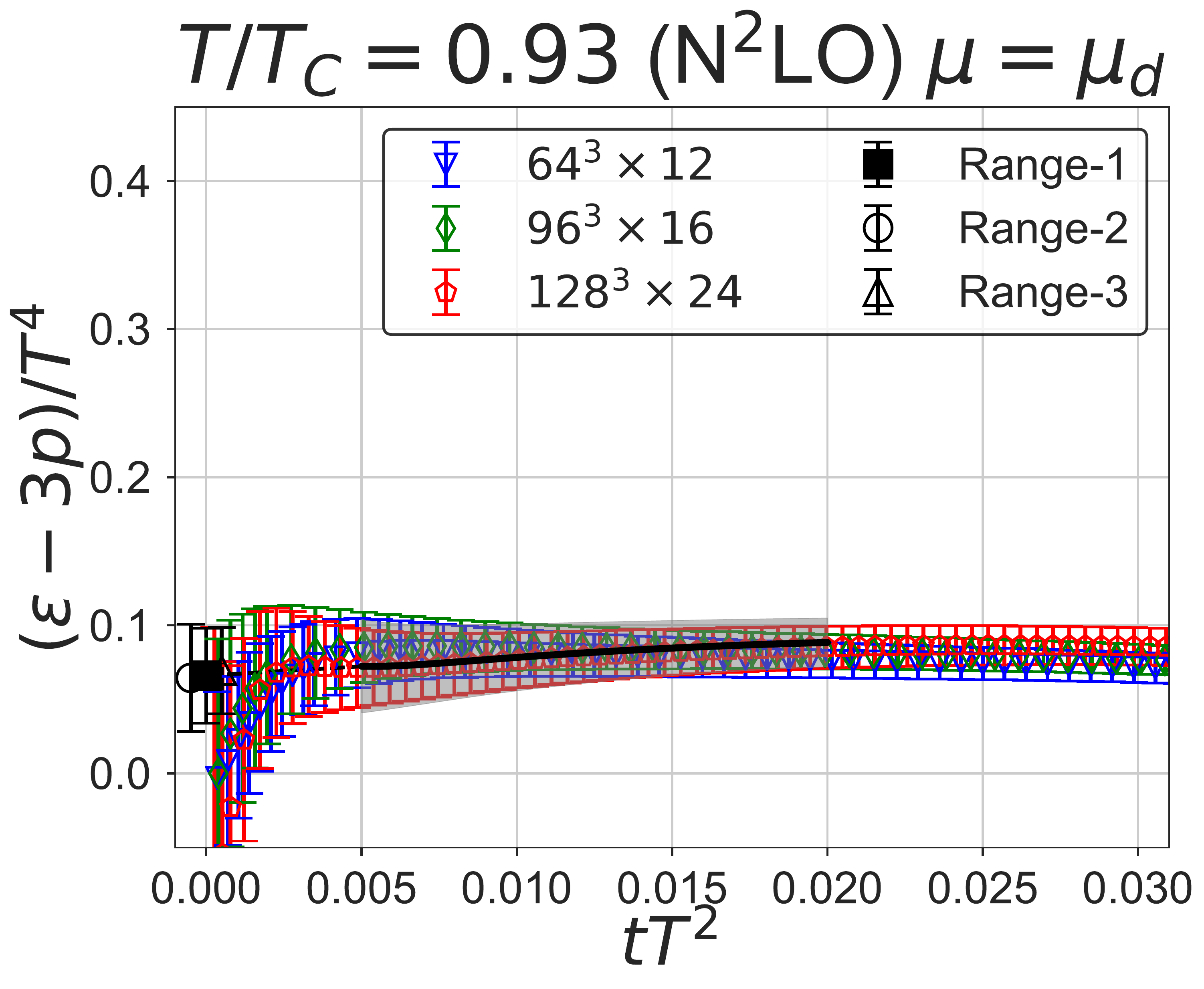}
\caption{}
\label{}
\end{subfigure}
\begin{subfigure}{0.35\columnwidth}
\centering
\includegraphics[width=\columnwidth]{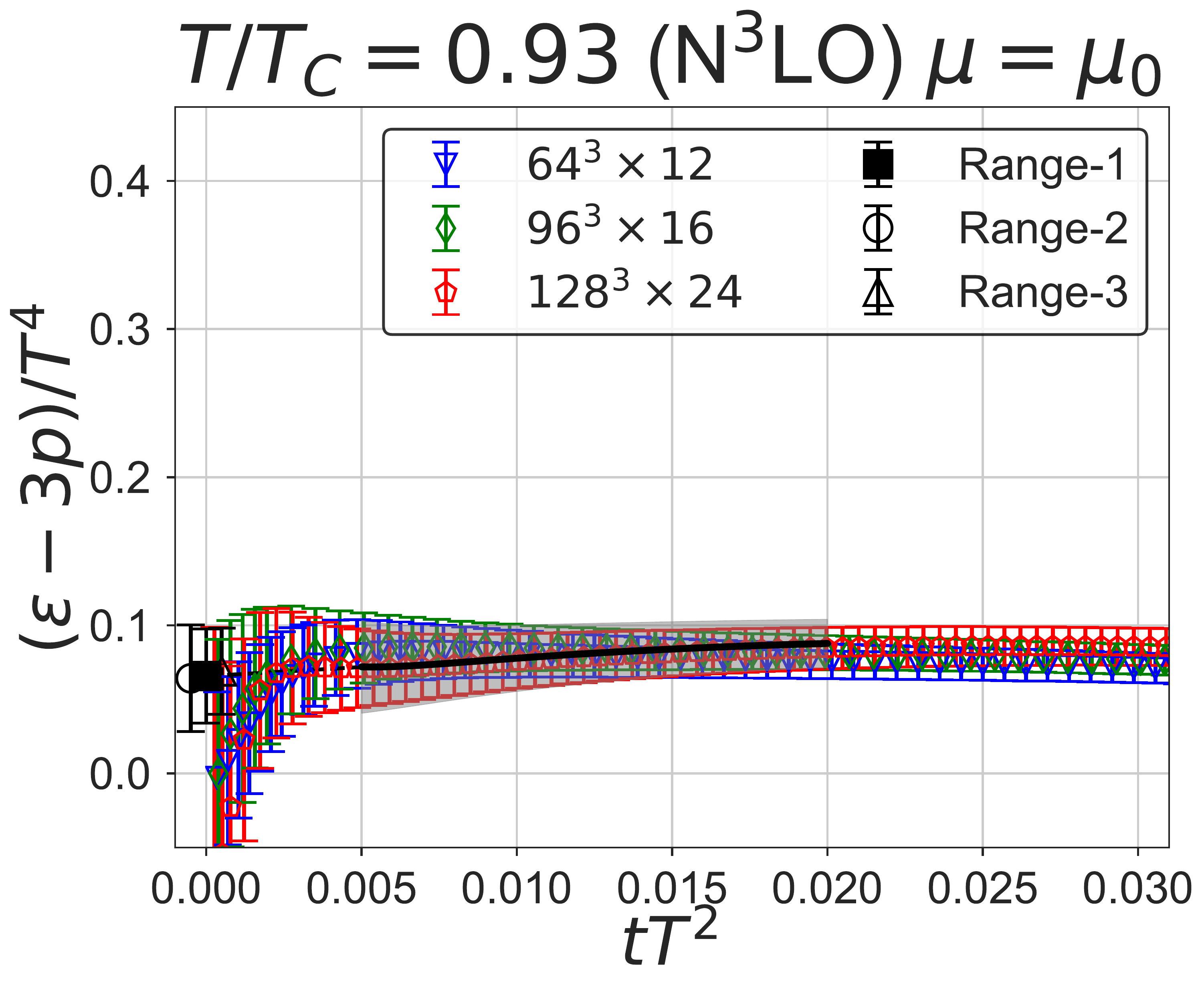}
\caption{}
\label{}
\end{subfigure}
\hspace{10mm}
\begin{subfigure}{0.35\columnwidth}
\centering
\includegraphics[width=\columnwidth]{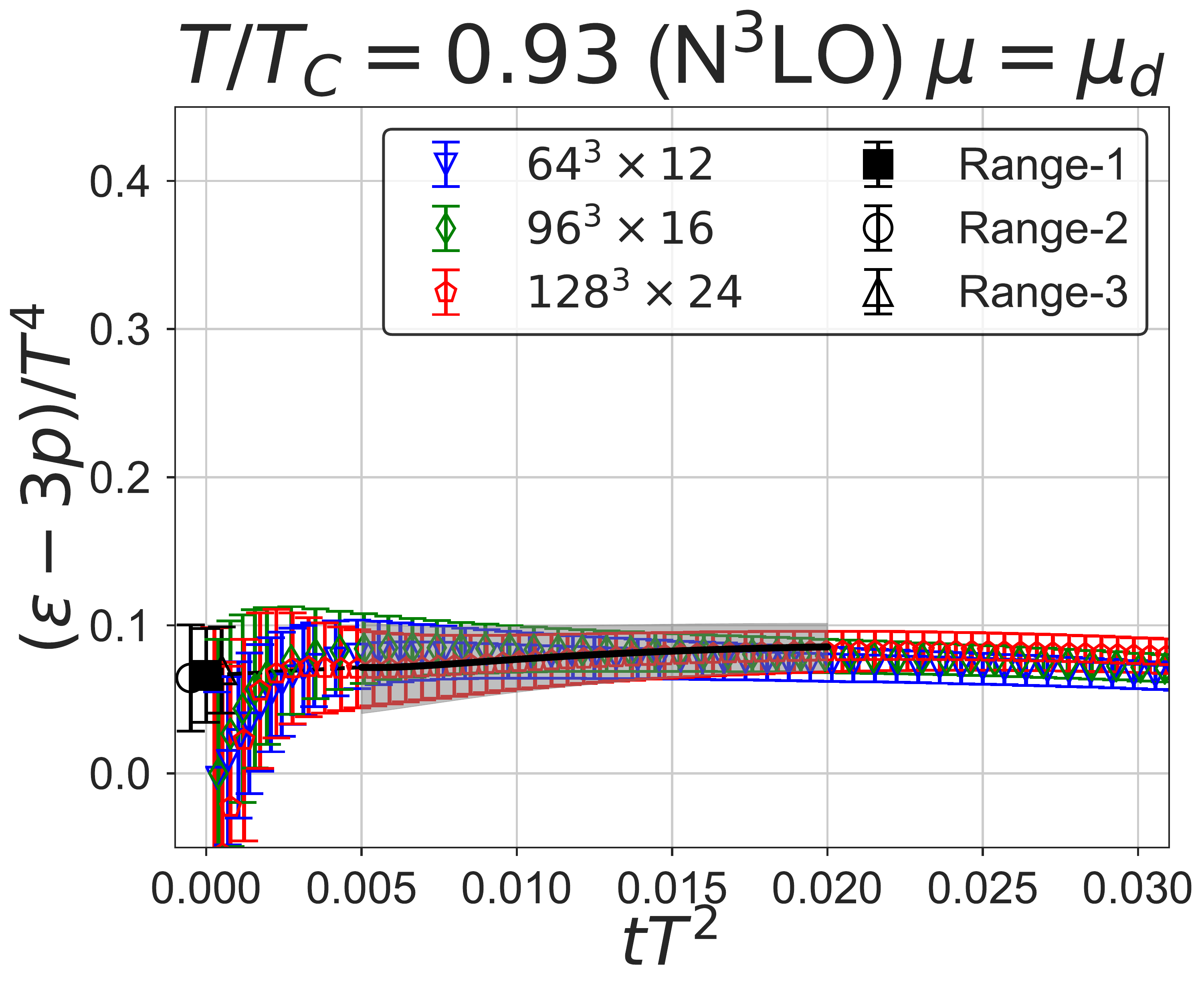}
\caption{}
\label{}
\end{subfigure}
\caption{Same as~Fig.~\ref{fig:3}. $T/T_c=0.93$.}
\label{fig:A8}
\end{figure}

\begin{figure}[htbp]
\centering
\begin{subfigure}{0.35\columnwidth}
\centering
\includegraphics[width=\columnwidth]{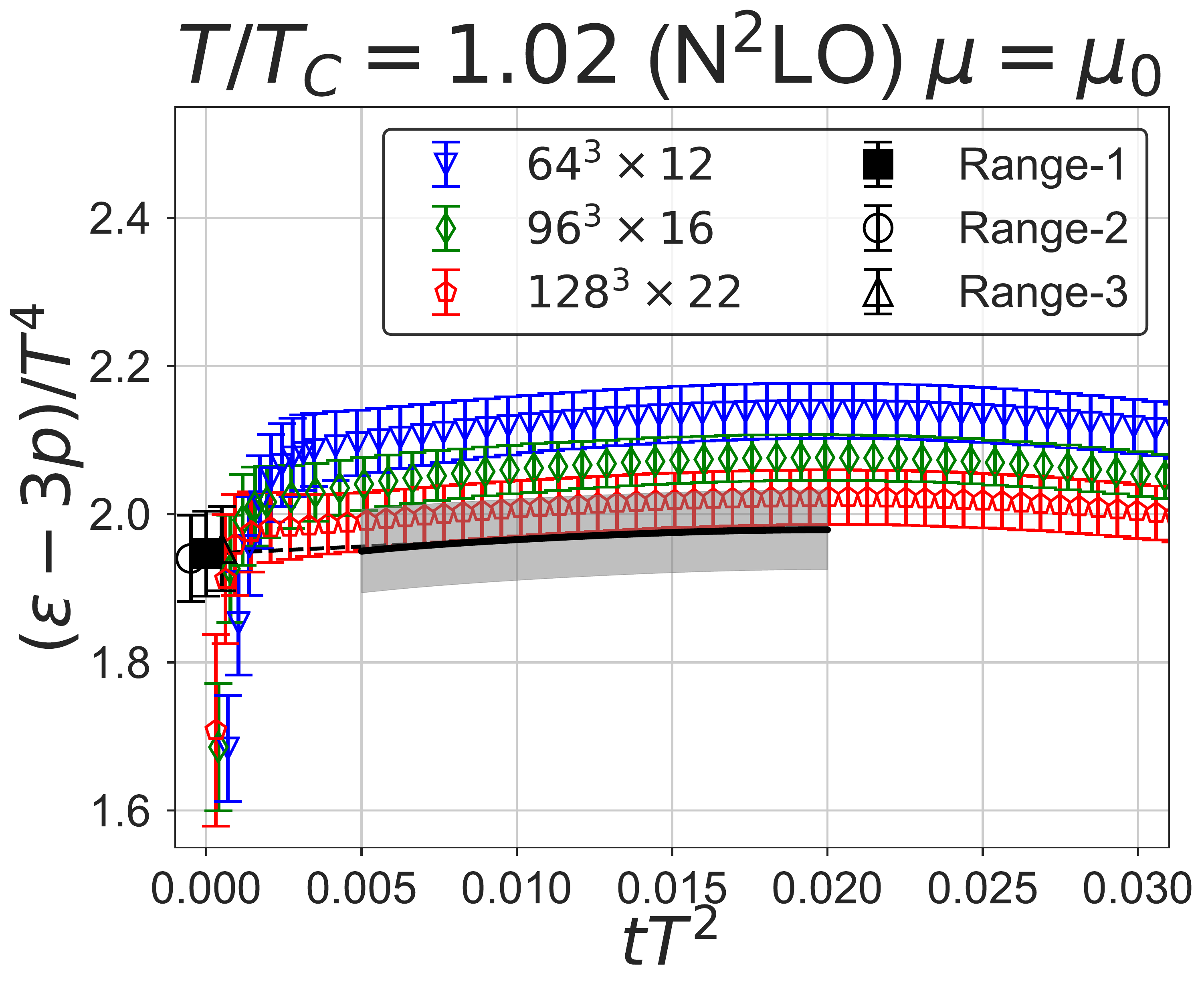}
\caption{}
\label{}
\end{subfigure}
\hspace{10mm}
\begin{subfigure}{0.35\columnwidth}
\centering
\includegraphics[width=\columnwidth]{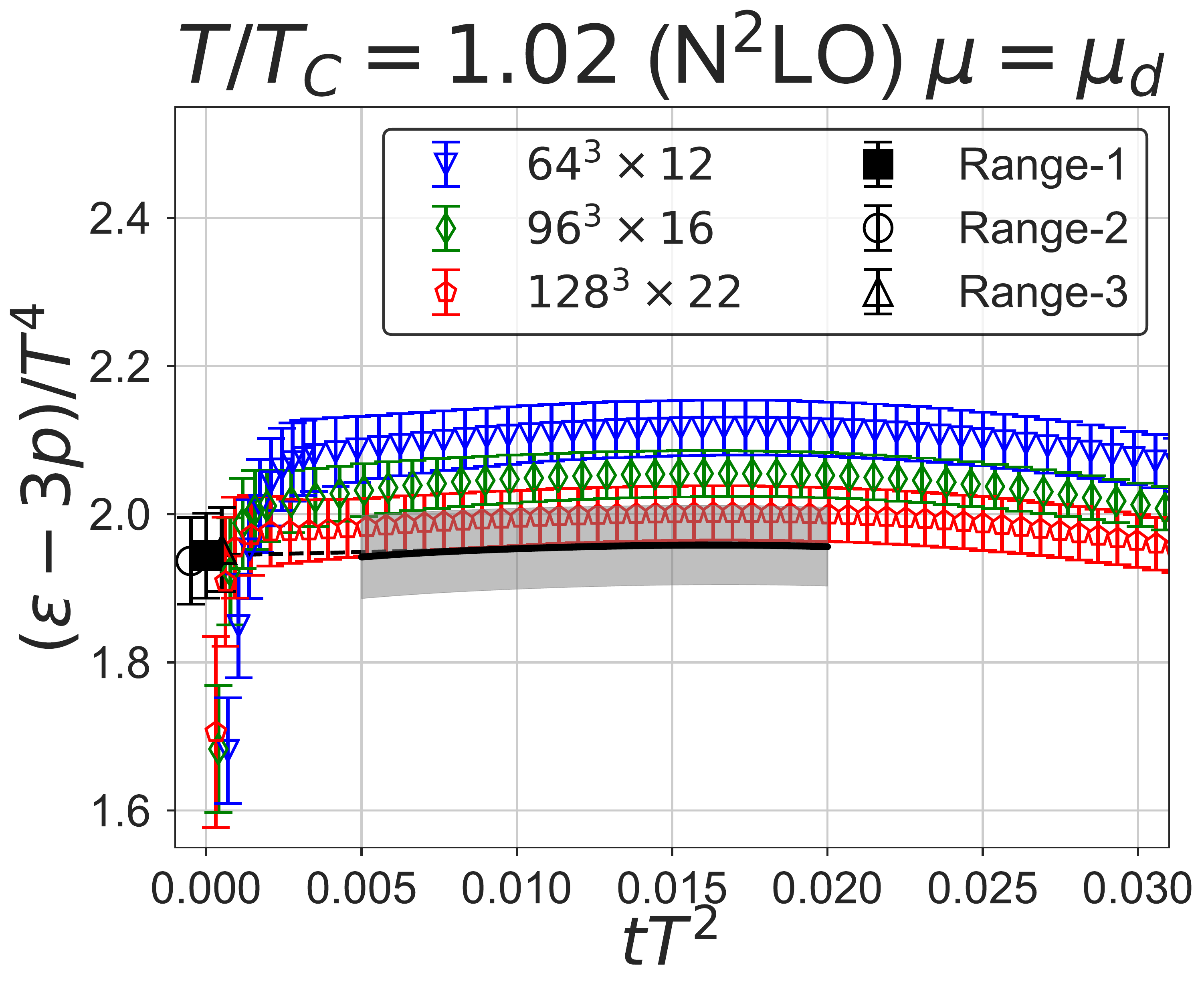}
\caption{}
\label{}
\end{subfigure}
\begin{subfigure}{0.35\columnwidth}
\centering
\includegraphics[width=\columnwidth]{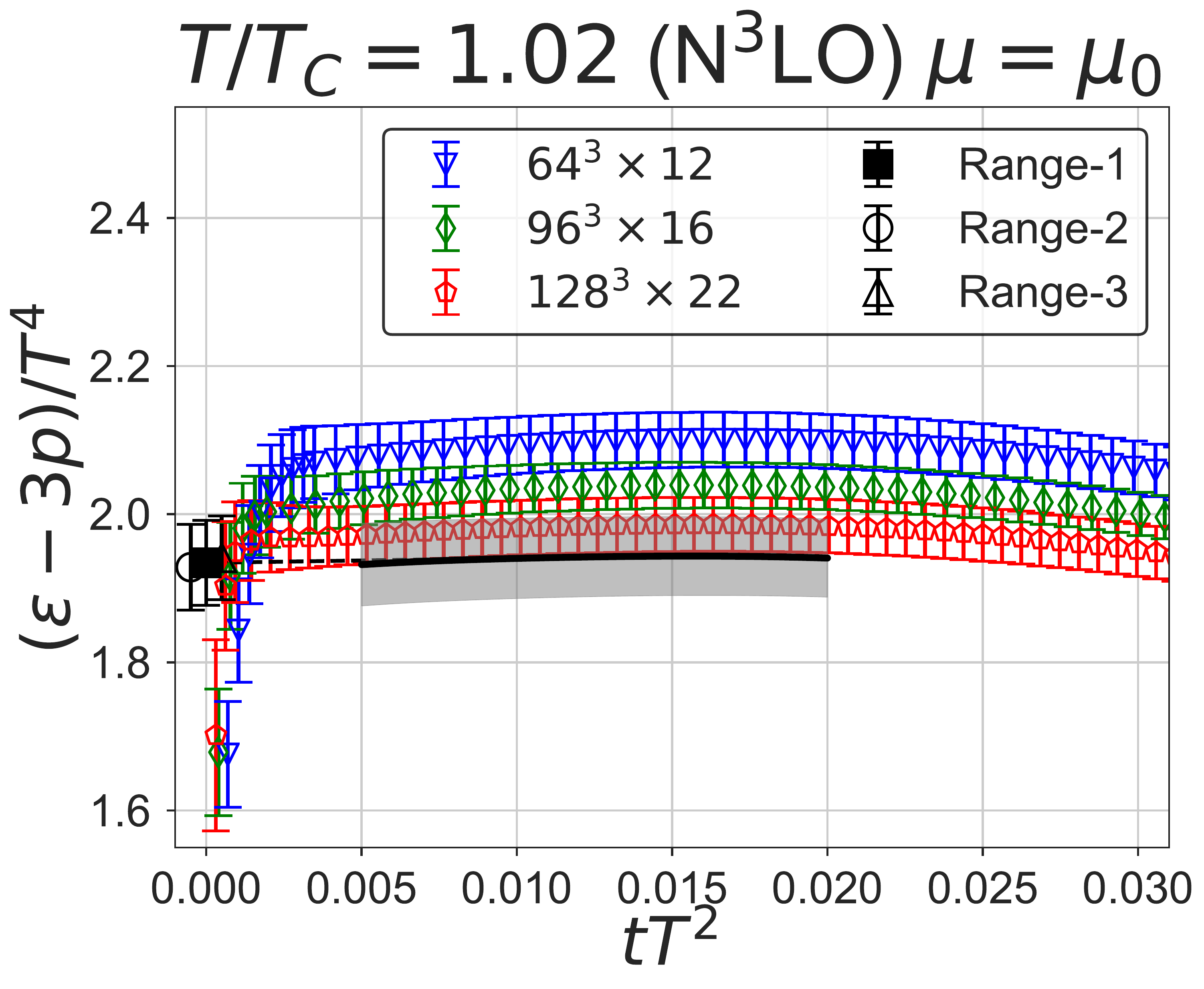}
\caption{}
\label{}
\end{subfigure}
\hspace{10mm}
\begin{subfigure}{0.35\columnwidth}
\centering
\includegraphics[width=\columnwidth]{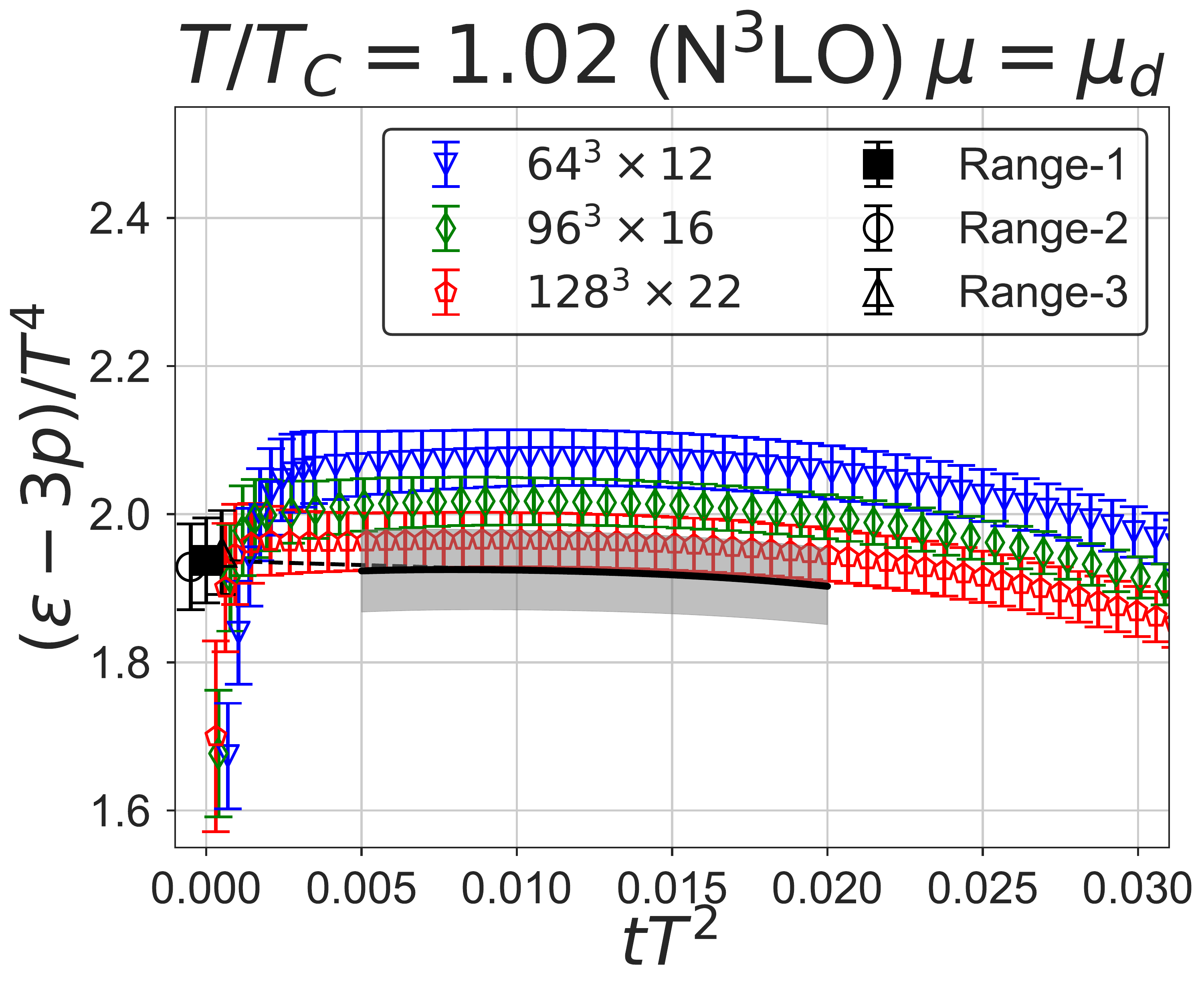}
\caption{}
\label{}
\end{subfigure}
\caption{Same as~Fig.~\ref{fig:3}. $T/T_c=1.02$.}
\label{fig:A9}
\end{figure}

\begin{figure}[htbp]
\centering
\begin{subfigure}{0.35\columnwidth}
\centering
\includegraphics[width=\columnwidth]{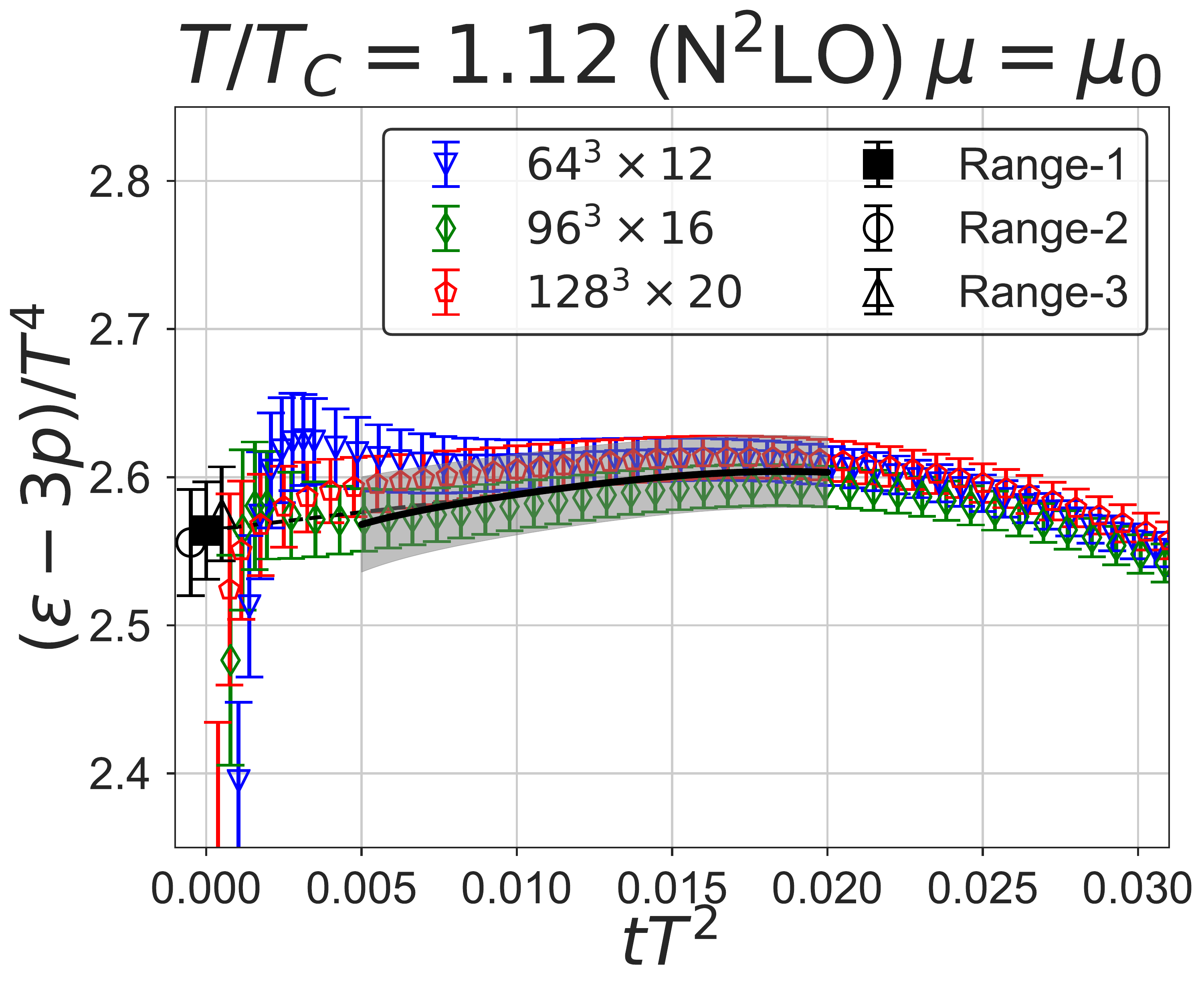}
\caption{}
\label{}
\end{subfigure}
\hspace{10mm}
\begin{subfigure}{0.35\columnwidth}
\centering
\includegraphics[width=\columnwidth]{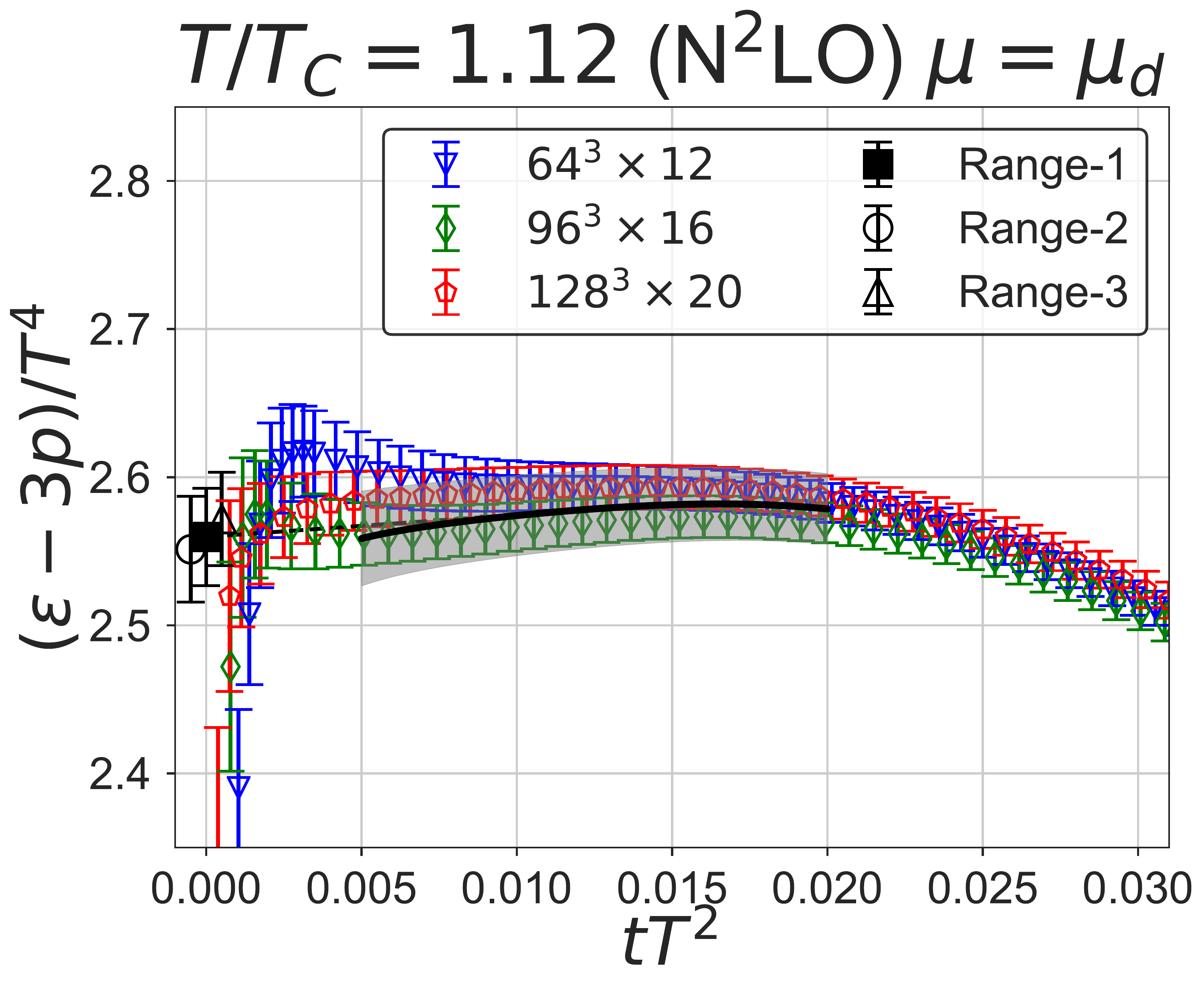}
\caption{}
\label{}
\end{subfigure}
\begin{subfigure}{0.35\columnwidth}
\centering
\includegraphics[width=\columnwidth]{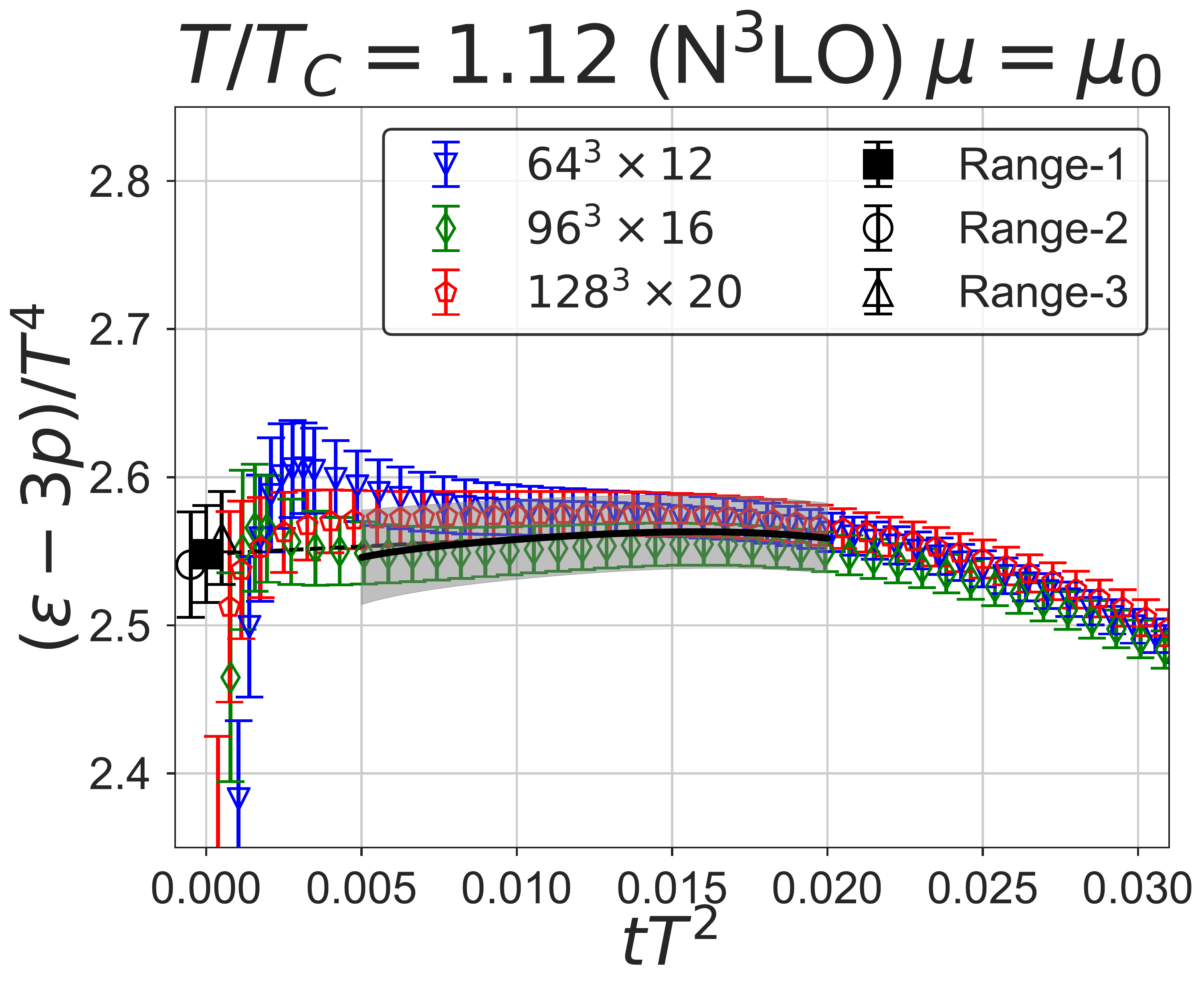}
\caption{}
\label{}
\end{subfigure}
\hspace{10mm}
\begin{subfigure}{0.35\columnwidth}
\centering
\includegraphics[width=\columnwidth]{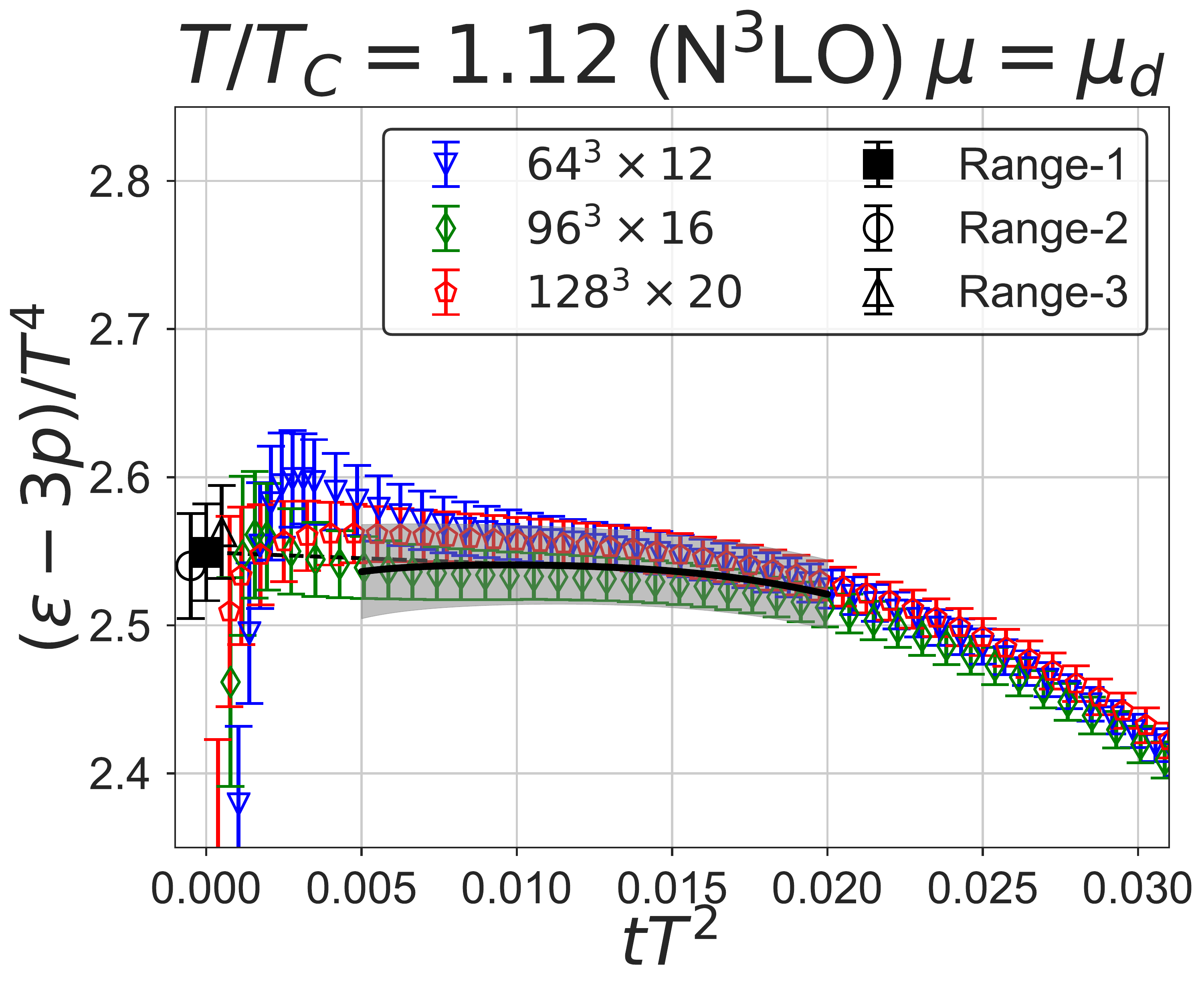}
\caption{}
\label{}
\end{subfigure}
\caption{Same as~Fig.~\ref{fig:3}. $T/T_c=1.12$.}
\label{fig:A10}
\end{figure}

\begin{figure}[htbp]
\centering
\begin{subfigure}{0.35\columnwidth}
\centering
\includegraphics[width=\columnwidth]{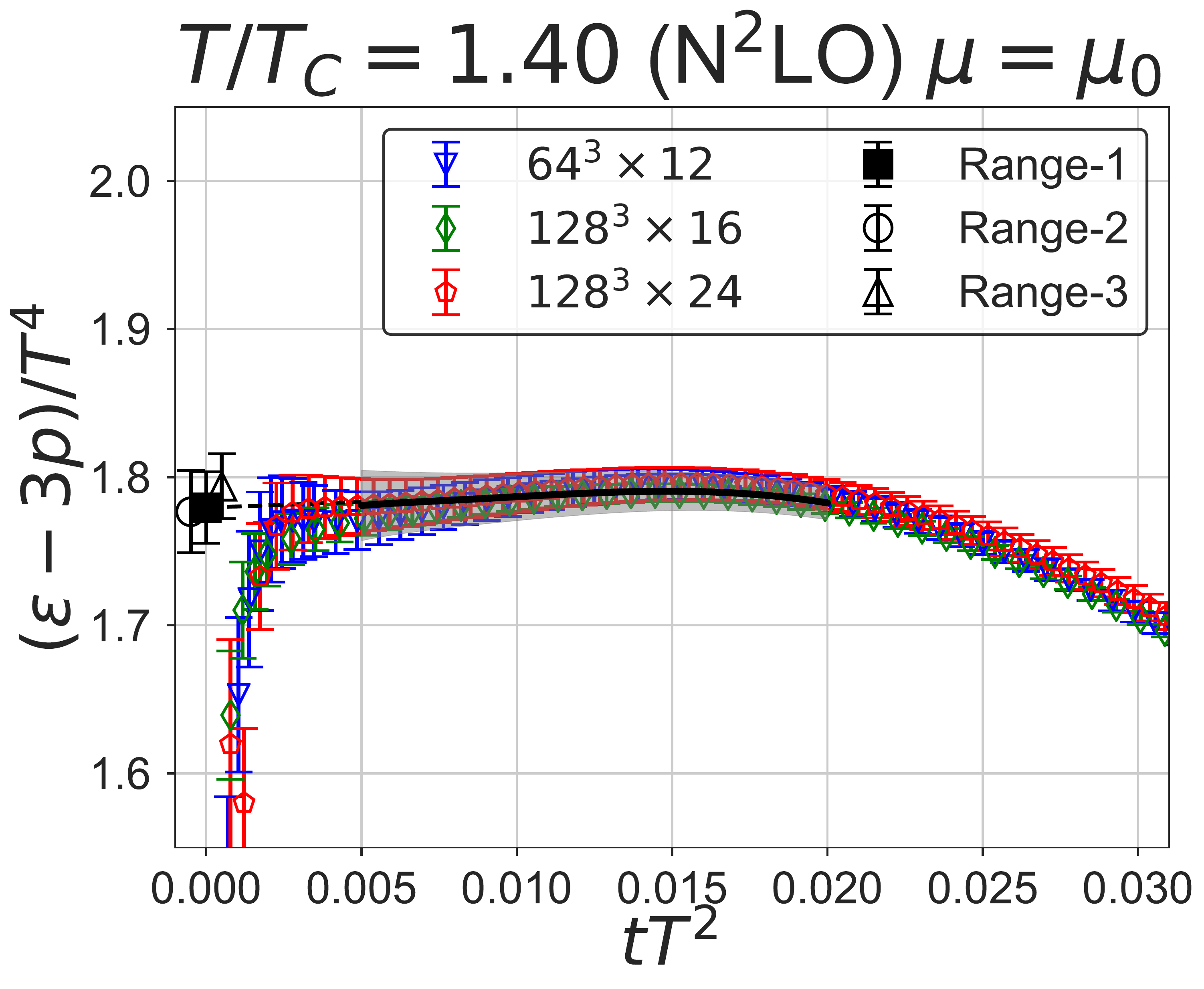}
\caption{}
\label{}
\end{subfigure}
\hspace{10mm}
\begin{subfigure}{0.35\columnwidth}
\centering
\includegraphics[width=\columnwidth]{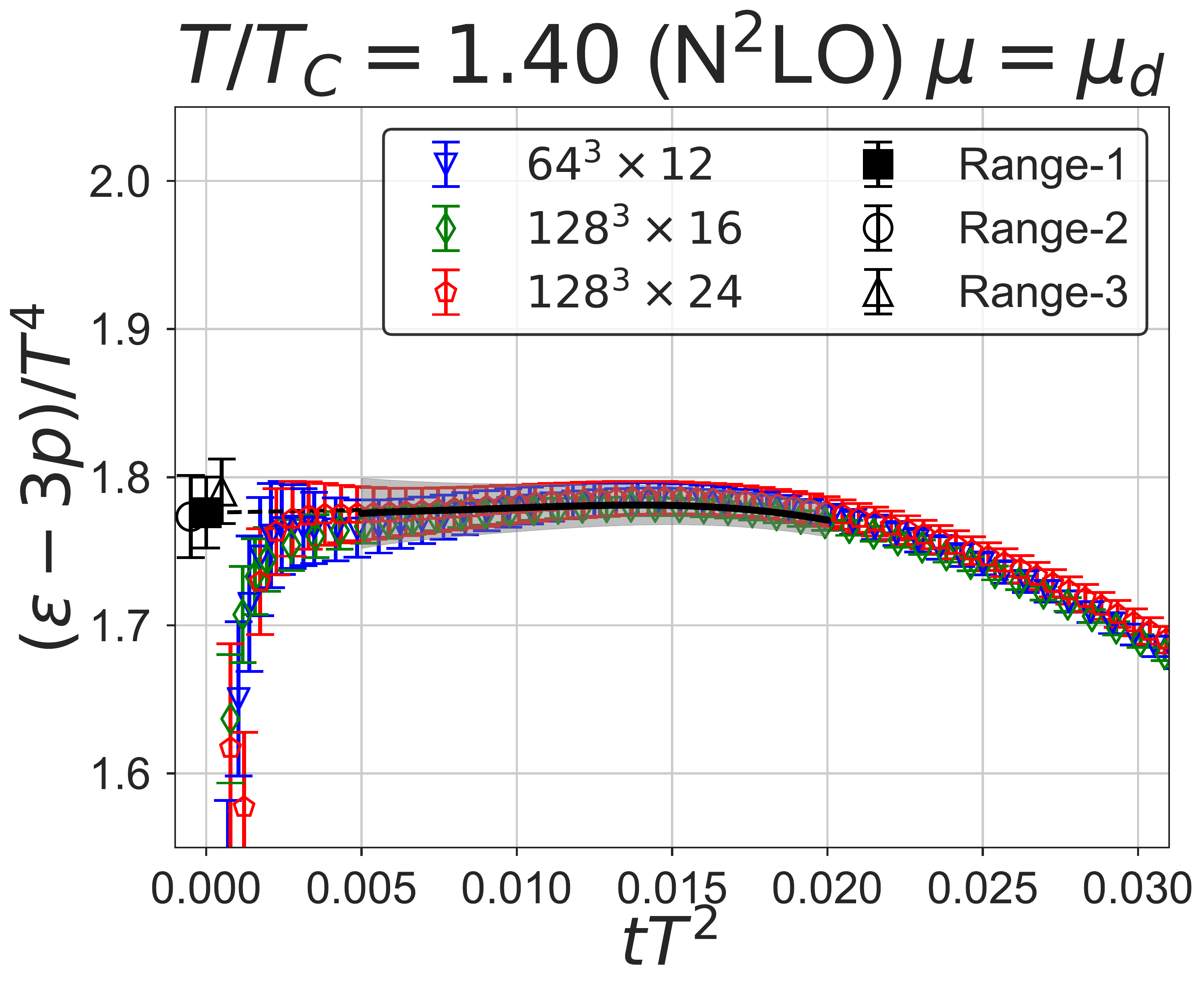}
\caption{}
\label{}
\end{subfigure}
\begin{subfigure}{0.35\columnwidth}
\centering
\includegraphics[width=\columnwidth]{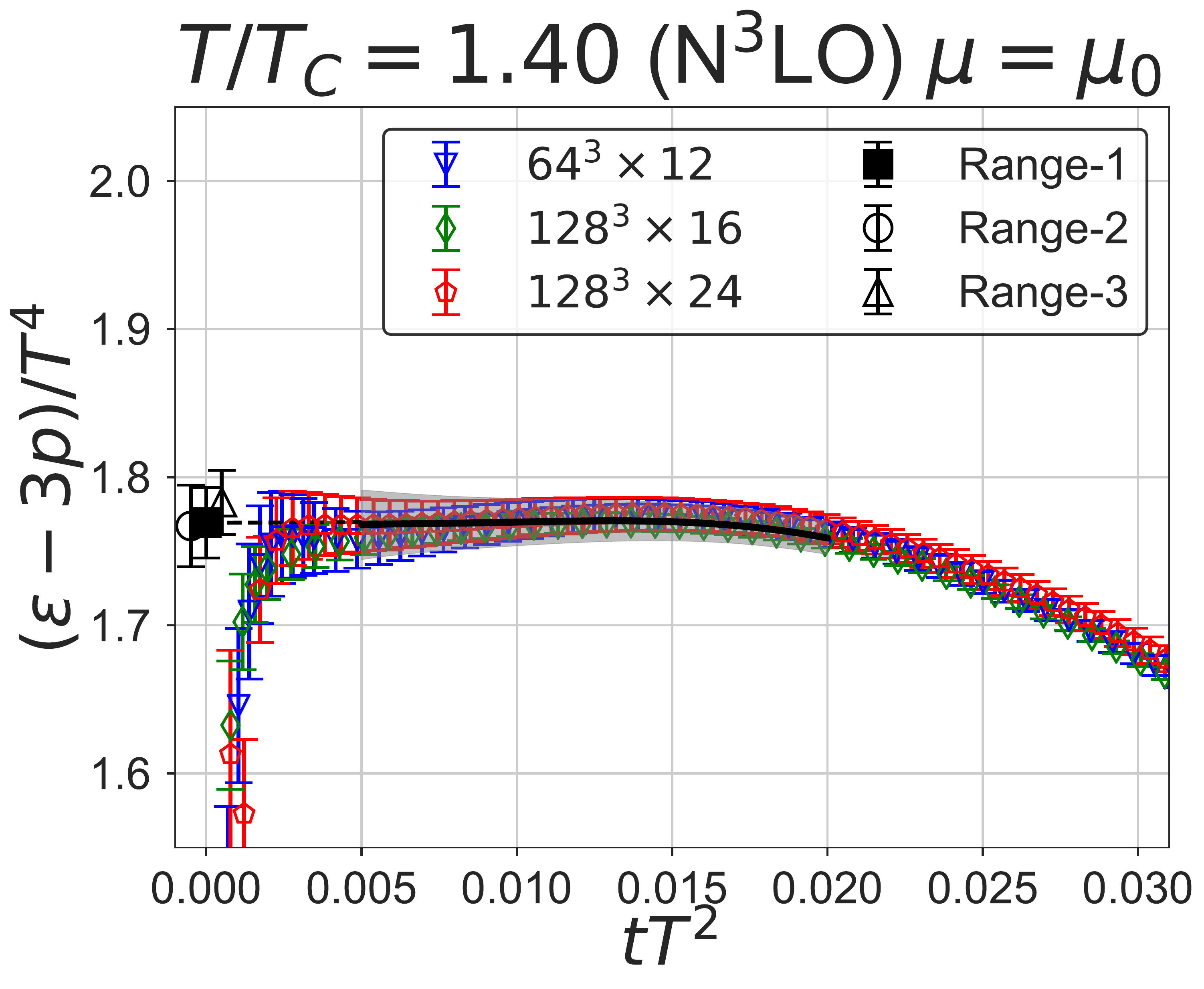}
\caption{}
\label{}
\end{subfigure}
\hspace{10mm}
\begin{subfigure}{0.35\columnwidth}
\centering
\includegraphics[width=\columnwidth]{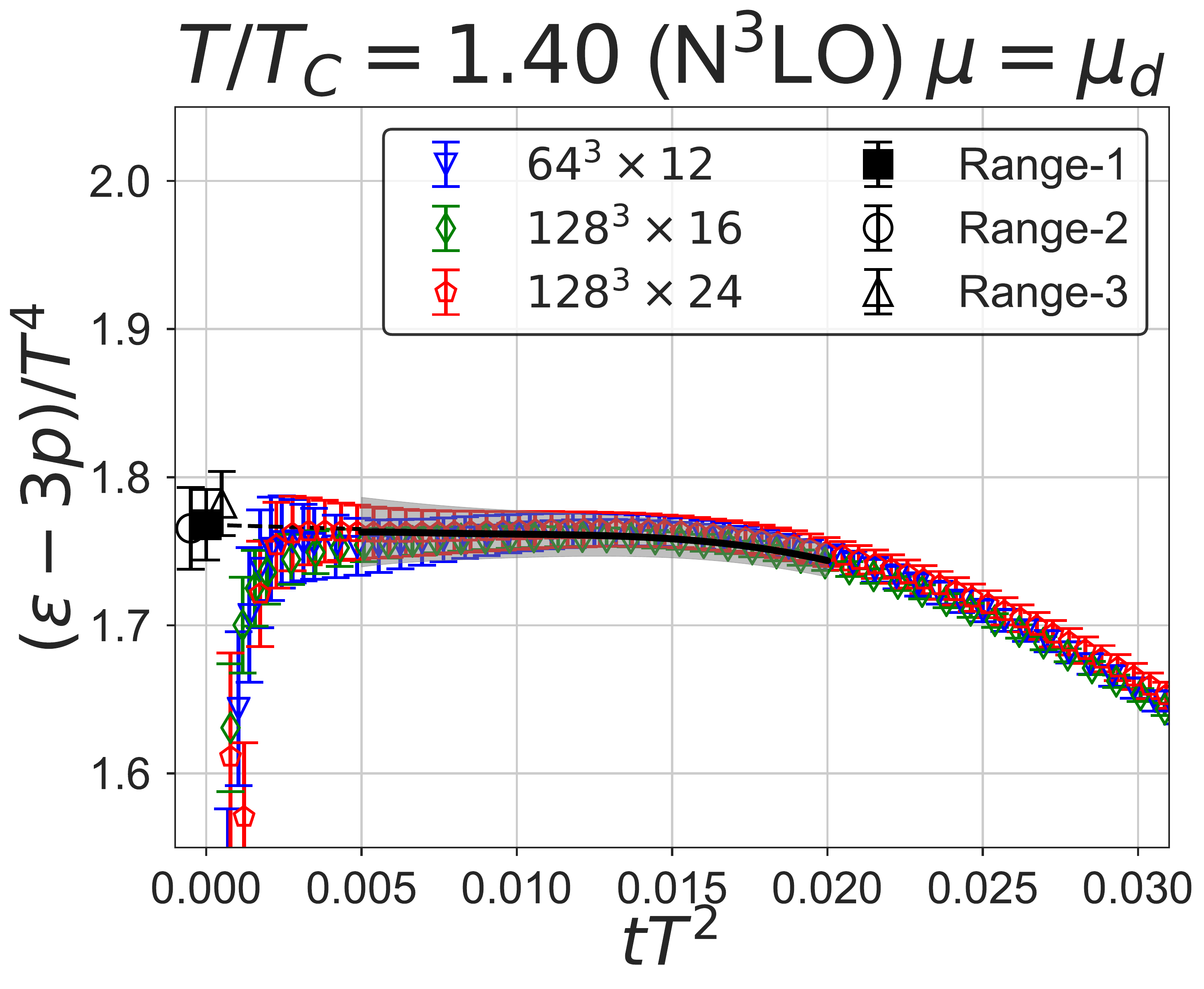}
\caption{}
\label{}
\end{subfigure}
\caption{Same as~Fig.~\ref{fig:3}. $T/T_c=1.40$.}
\label{fig:A11}
\end{figure}

In this appendix, we show the plots of Eqs.~\eqref{eq:(3.1)}
and~\eqref{eq:(3.2)} for temperatures other than~$T/T_c=1.68$
in~Figs.~\ref{fig:A1}--\ref{fig:A11}.

\end{document}